\documentclass[aps, prd, twocolumn, preprintnumbers, superscriptaddress, nofootinbib, floatfix]{revtex4-1}
\usepackage{amsmath}
\usepackage{bm}
\usepackage{amsfonts}
\usepackage{graphicx}
\usepackage{epstopdf}
\usepackage{hyperref}
\usepackage{array}
\usepackage{soul}
\usepackage{color}
\enlargethispage{\baselineskip}
\hypersetup{colorlinks = true, citecolor = blue}
\everymath{\displaystyle}

\newcommand{\mL}{\mathcal{L}}

\newcommand{\MP}{M_{\rm Pl}}
\newcommand{\MS}{M_{\rm susy}}

\renewcommand\({\left(}
\renewcommand\){\right)}

\newcommand{\be}{\begin{equation}}
\newcommand{\ee}{\end{equation}}
\newcommand{\bea}{\begin{eqnarray}}
\newcommand{\eea}{\end{eqnarray}}

\begin{document}

\title{Cosmological window onto the string axiverse and the supersymmetry breaking scale}

\newcommand{\FIRSTAFF}{\affiliation{The Oskar Klein Centre for Cosmoparticle Physics, Stockholm University, Roslagstullsbacken 21A, SE-106 91 Stockholm, Sweden}}
\newcommand{\SECONDAFF}{\affiliation{Nordita, KTH Royal Institute of Technology and Stockholm University, Roslagstullsbacken 23, SE-106 91 Stockholm, Sweden}}
\newcommand{\THIRDAFF}{\affiliation{Department of Physics and Astronomy, Uppsala University, L\"{a}gerhyddsv\"{a}gen 1, 75120 Uppsala, Sweden}}
\author{Luca Visinelli}\email[Electronic address: ]{luca.visinelli@physics.uu.se}\FIRSTAFF \SECONDAFF \THIRDAFF
\author{Sunny Vagnozzi}\email[Electronic address: ]{sunny.vagnozzi@fysik.su.se}\FIRSTAFF \SECONDAFF

\date{\today}
\preprint{NORDITA-2018-051}

\begin{abstract}
In the simplest picture, the masses of string axions populating the axiverse depend on two parameters: the supersymmetry breaking scale $M_{\rm susy}$ and the action $S$ of the string instantons responsible for breaking the axion shift symmetry. In this work, we explore whether cosmological data can be used to probe these two parameters. Adopting string-inspired flat priors on $\log_{10}M_{\rm susy}$ and $S$, and imposing that $M_{\rm susy}$ be sub-Planckian, we find $S=198\pm28$. These bounds suggest that cosmological data complemented with string-inspired priors select a quite narrow axion mass range within the axiverse, $\log_{10}\(m_a/{\rm eV}\) = -21.5^{+1.3}_{-2.3}$. We find that $M_{\rm susy}$ remains unconstrained due to a fundamental parameter degeneracy with $S$. We explore the significant impact of other choices of priors on the results, and we comment on similar findings in recent previous literature.
\end{abstract}

\maketitle

\section{Introduction}
\label{intro}

The theory of Quantum ChromoDynamics (QCD) allows for a currently unobserved CP-violating interaction~\cite{Belavin:1975fg, tHooft:1976rip, Jackiw:1976pf, Callan:1976je}. A plausible solution to the so-called ``strong-CP problem''~\cite{Peccei:1977hh, Peccei:1977ur} predicts the existence of the QCD axion $a=a(x)$~\cite{Wilczek:1977pj, Weinberg:1977ma}, a pseudo-scalar Goldstone boson that couples to the number density of QCD instantons via
\begin{eqnarray}
	S_a = \frac{1}{32\pi^2 f_a} \int d^4x \,a\, \epsilon^{\mu\nu\lambda\sigma}\,{\rm \bf Tr}\,G_{\mu\nu}G_{\lambda\sigma}.
\end{eqnarray}
Here, the trace is taken in the three-dimensional representation of SU(3) and $f_a$ is the spontaneous symmetry breaking scale (or axion decay scale). In viable ``invisible'' axion models, stellar object cooling considerations provide a bound $f_a \gtrsim 10^9\,{\rm GeV}$~\cite{Kim:1979if, Dine:1981rt, Zhitnitsky:1980tq, Shifman1980493}, with the exact value of the bound depending on the axion model considered. The shift symmetry $a \to a + $const, which holds at the classical level, is explicitly broken by the same QCD instanton effects which are also responsible for generating the axion potential. Owing to these non-perturbative effects, the QCD axion acquires a mass $m_a = \Lambda_{\rm QCD}^2/f_a$, where $\Lambda_{\rm QCD} \simeq 75.5\,{\rm MeV}$~\cite{Wilczek:1977pj, Weinberg:1977ma, Gross:53.43}. It is well known that the axion can constitute the Cold Dark Matter (CDM)~\cite{Preskill:1982cy, Abbott:1982af, Dine:1982ah}, with the axion decay constant being as large as the Grand Unification Theory (GUT) energy scale in the so-called ``anthropic'' axion window~\cite{Linde:1987bx, Linde:1991km, Turner:1991, Wilczek:2004cr, Tegmark:2005dy, Hertzberg:2008wr, Freivogel:2008qc, Mack:2009hv, Visinelli:2009zm, Visinelli:2009kt, Acharya:2010zx, Visinelli:2014twa, Visinelli:2017ooc, Visinelli:2018zif, Visinelli:2018wza}.

Along with the QCD axion, other ``axion-like'' particles (ALPs) arise either from the breaking of ``accidental'' symmetries~\cite{Choi:2006qj, Choi:2009jt, Dias:2014osa, Higaki:2014pja, Kim:2015yna, Redi:2016esr, DiLuzio:2017tjx} or from manifold compactification within string theory~\cite{Witten:1984, Fox:2004kb, Svrcek:2006yi, Arvanitaki:2009fg, Acharya:2010zx, Dine:2010cr, Ringwald:2012cu, Kim:2013fga, Bachlechner:2014gfa, Halverson:2017deq, Stott:2017hvl}. Both these scenarios feature a symmetry-breaking scale $\Lambda_a$ and an ALP decay constant $f_a$, with the axion field acquiring the mass $m_a = \Lambda_a^2/f_a$. However, unlike the case for the QCD axion, the ALP energy scale $\Lambda_a$ is not tied to the QCD energy scale, so that the mass $m_a$ and the decay scale $f_a$ can effectively be treated as two independent parameters. Although somewhat fundamentally less motivated than the QCD axion, ALPs are potentially suitable dark matter candidates~\cite{Arias:2012az, Hui:2016ltb, Visinelli:2017imh, Diez-Tejedor:2017ivd}. Of particular interest are ultra-light axions (ULAs)~\cite{Baldeschi:1983, Membrado:1989bqo, Press:1990, Sin:1994, Ji:1994, Lee:1996, Guzman:2000, Sahni:2000, Peebles:2000yy, Goodman:2000, Matos:2000, Hu:2000ke}, whose mass resides in the range $m_a \in [10^{-27}; 10^{-18}]\,{\rm eV}$. These ULAs manifest their wave-like behavior by suppressing the matter power spectrum at astrophysical scales. It has been argued that this suppression of power could be the key to address a number of controversies (e.g. the ``missing satellites" and the ``cusp-core" problems) arising in the standard $\Lambda$CDM cosmology on galactic and subgalactic scales~(see Ref.~\cite{Weinberg:2013aya} for a review). Hereafter we shall refer to axion-like particles interchangeably as ``ALPs'' or ``axions'' when we are not interested in their origin (e.g. from the string axiverse), as opposed to ``string axions'' which originate specifically from the string axiverse (to be discussed later).

It has long been noticed that axions arise naturally within string theory compactifications as Kaluza-Klein zero-modes of antisymmetric tensor fields~\cite{Witten:1984, Svrcek:2006yi}. Zero-modes originate from non-contractable cycles on the compactified manifold. The number of zero-modes is fixed by the topology of the compactification manifold itself and is generally in the order of hundreds (for instance, for compactifications on Calabi-Yau manifolds the number of zero-modes is given by the Hodge number of the manifold itself). Notice that this assumes that the size of the extra-dimensions is finite, unlike e.g. the case of Randall-Sundrum models~\cite{Randall:1999vf, Caldwell:2001ja, Maartens:2010ar, Visinelli:2017bny}. A fraction of these zero-modes are expected to acquire a mass through non-perturbative string instanton effects~\cite{Vandoren:2008xg}, which can be characterised by the (dimensionless) action of the instantons $S$ (which scales with the volume of the corresponding cycles)~\cite{Svrcek:2006yi}, as well as a non-perturbative ultra-violet (UV) cut-off scale $\mu$. This non-perturbative scale is related in turn to the supersymmetry (SUSY) breaking scale $\MS$~\cite{Svrcek:2006yi, Arvanitaki:2009fg} (with SUSY almost inevitably appearing in any realistic string theory) as $\mu \propto \sqrt{\MS}$, if the axion potential arises from the superpotential generated by  string instantons (which is often the case). This suggests that a reasonable way of characterizing string axions is by exploring the $\mu$-$S$ (or equivalently $\MS$-$S$) parameter space.

Away from the swampland~(\cite{Vafa:2005ui}; see Refs.~\cite{Danielsson:2018ztv,Obied:2018sgi,Agrawal:2018own,Dvali:2018fqu,Andriot:2018wzk,Banerjee:2018qey,Achucarro:2018vey,Garg:2018reu,
Kehagias:2018uem,Dias:2018ngv,Denef:2018etk,Roupec:2018mbn,Andriot:2018ept,Matsui:2018bsy,Ben-Dayan:2018mhe,Heisenberg:2018yae,Damian:2018tlf,Conlon:2018eyr,Kinney:2018nny,Dasgupta:2018rtp,Cicoli:2018kdo,Kachru:2018aqn,Akrami:2018ylq,
Heisenberg:2018rdu,Murayama:2018lie,Marsh:2018kub,Brahma:2018hrd,Choi:2018rze} for recent developments), the landscape of string vacua~\cite{Fox:2004kb, Svrcek:2006yi} gives rise to a plethora of light axions, known in the literature as the string axiverse~\cite{Arvanitaki:2009fg}, along with various other massless modes~\cite{Arvanitaki:2009hb}. Since for the QCD axion the $\theta$-parameter is constrained to be smaller than $10^{-10}$, string corrections are negligible compared to those given by the QCD potential. On the contrary, the mass of lighter axions is primarily fixed by the non-perturbative string contributions. The exploration of the string axiverse scenario, wherein a multitude of axions populate various orders of magnitude in masses, is being pursued by a variety of searches, including a rotation in the polarisation of the Cosmic Microwave Background (CMB) radiation spectrum (for masses ranging from $10^{-33}{\rm \,eV}$ to $4\times10^{-28}{\rm \, eV}$), a suppression in the power spectrum of density perturbations at small scales (for masses ranging from $10^{-28}{\rm \,eV}$ to $10^{-18}{\rm \, eV}$), the altered dynamics of rotating black holes due to the effect of super-radiance (for masses ranging from $10^{-22}{\rm \,eV}$ to $10^{-10}{\rm \, eV}$)~\cite{Arvanitaki:2009fg}, and various laboratory searches (for masses larger than about $10^{-15}{\rm \,eV}$) like ABRACADABRA~\cite{Kahn:2016aff}, ADMX~\cite{Stern:2016bbw}, KLASH~\cite{Alesini:2017ifp}, QUAX~\cite{Barbieri:2016vwg}, X3~\cite{Brubaker:2016ktl}, CULTASK~\cite{Chung:2016ysi}, MADMAX~\cite{TheMADMAXWorkingGroup:2016hpc}, ARIADNE~\cite{Arvanitaki:2014dfa}, IAXO~\cite{Vogel:2013bta}, and CASPEr~\cite{Budker:2013hfa}, see e.g. Ref.~\cite{Irastorza:2018dyq} for a review. Axions lighter than the present Hubble rate $H_0 \sim 10^{-33}{\rm \,eV}$ are still frozen today and do not contribute to the present matter content of the universe.

In the present work, it is our aim to address the following question: ``\textit{What can cosmology tell us about the string axiverse and its parameters?}". As discussed previously, we address the question by focusing on the non-perturbative scale $\mu$ (or equivalently the SUSY breaking scale $M_{\rm susy}$) and the dimensionless symmetry-breaking instanton action $S$ as parameters characterising the string axiverse. Focusing for definiteness on the case where the axions are present during inflation, we will also consider the initial misalignment angle $\theta_i$ and the primordial isocurvature fraction $\beta$ as additional parameters. For an incomplete list of other works examining the axiverse, and especially its cosmology, see e.g.~\cite{Marsh:2011gr,Cicoli:2012sz,Marsh:2013taa,Tashiro:2013yea,Yoshino:2014wwa,Kamionkowski:2014zda,Obata:2014loa,Yoshino:2015nsa, Daido:2015bva,Acharya:2015zfk,Emami:2016mrt,Karwal:2016vyq,Gorbunov:2017ayg,Yoshida:2017cjl,Emami:2018rxq}. We do not necessarily assume that the axion is the totality of the dark matter, so we assume a mixed dark matter scenario~\cite{Bae:2014efa, Visinelli:2014xsa, Bae:2015rra, Baum:2016oow}.

We therefore characterise the string axiverse by the 4-dimensional parameter space spanned by the parameters $(\mu\,,S\,,\theta_i\,,\beta)$ [alternatively $(M_{\rm susy}\,,S\,,\theta_i\,,\beta)$] and explore how these parameters can be constrained by cosmological data. On this matter, a caveat/warning is in order at this point. As it is, to the best of our knowledge, the first time an attempt to constrain the parameters $M_{\rm susy}$ and $S$ is made, our goal is not to provide a full-fledged analysis utilising all available cosmological data (e.g. the full CMB temperature and polarisation anisotropy power spectra, or measurements of galaxy power spectra), but rather to get a feel for whether cosmology can actually provide information on $M_{\rm susy}$ and $S$ and, if so, which region of parameter space is selected and what is the physical motivation for such region being chosen. To this end, we include the following information/requirements from cosmology:
\begin{itemize}
\item constraints on the tensor-to-scalar ratio $r$ from the \textit{Planck} satellite;
\vspace{-0.2cm}
\item constraints on the primordial isocurvature fraction from the \textit{Planck} satellite;
\vspace{-0.2cm}
\item the requirement that the energy density in axions should not exceed the dark matter energy density measured by \textit{Planck}.
\end{itemize}

We obtain constraints on the model parameters $(M_{\rm susy}\,,S\,,\theta_i\,,\beta)$ by performing Bayesian parameter inference in light of the aforementioned cosmological data, with string-motivated priors for the model parameters (flat in $\log_{10}M_{\rm susy}$ and $S$), which we discuss in detail in Sec.~\ref{sec:analysis}. In this first analysis, we find $S=198 \pm 28$ at 68\% confidence level (C.L.). On the other hand, we find $\mu$ and hence $M_{\rm susy}$ to be poorly constrained, due to a fundamental parameter degeneracy with $S$. To break this degeneracy, we perform for purely illustrative purposes two additional analyses where we fix the instanton action to $S=198$ and $S=153$, corresponding to the central value and the $2\sigma$ lower bound on $S$ respectively. When fixing $S=198$, we find that the SUSY breaking scale is of the order of $M_{\rm susy} \sim 10^8\,{\rm TeV}$, whereas for $S=153$ we find that $M_{\rm susy} \lesssim 10^4\,{\rm TeV}$ at 95\%~C.L., which could lead to potentially interesting signatures at the proposed 100 TeV collider~\cite{Arkani-Hamed:2015vfh}.

Within the string axiverse, theoretical results make it possible to link the quantities $M_{\rm susy}$ and $S$ to the more familiar axion mass and decay constant, $m_a$ and $f_a$~\cite{Witten:1984, Fox:2004kb, Svrcek:2006yi}. We exploit these results to convert our previous constraints on $\MS$ and $S$ to constraints on $m_a$ and $f_a$ \textit{for the string axiverse case} (hence adopting string-inspired priors on the parameters). Perhaps surprisingly, we find that the string axiverse results select a rather tight range of axion masses, around $m_a \approx 10^{-22}\,{\rm eV}$, thus favouring the ULA interpretation. Interestingly, this appears to be in agreement with theoretical work which suggests that most of the axions originating from the string axiverse would be of the ULA type~\cite{Arvanitaki:2009fg}.

\begin{figure}[bt]
\begin{center}
	\includegraphics[width=\linewidth]{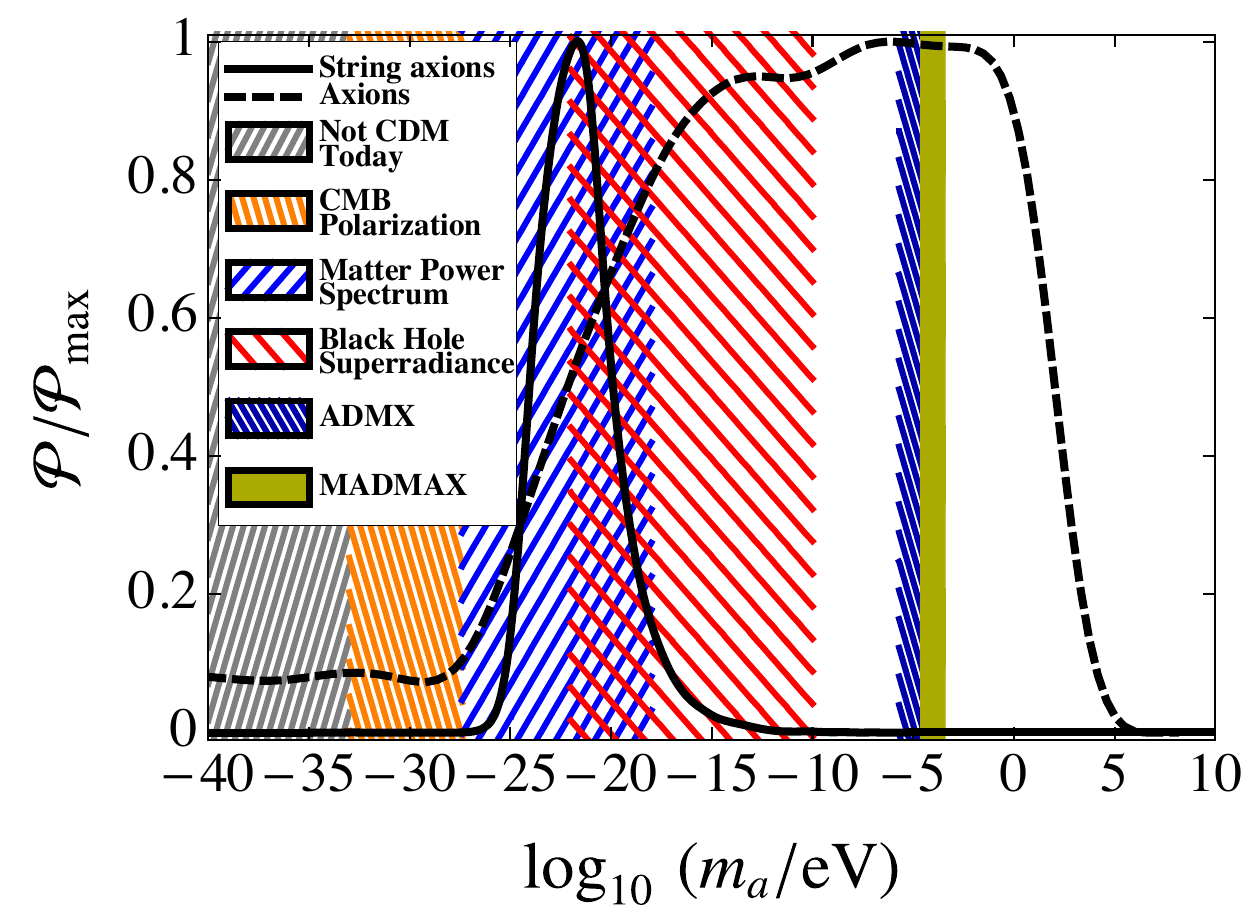}
	\caption{Marginalized posterior probability distributions for $\log_{10}(m_a/{\rm eV})$ (with $m_a$ the axion mass) normalized to their maximum values, for two of the analyses considered in our work. In the first analysis (solid black curve) we take string-inspired flat priors on $\log_{10}\MS$ and $S$, with $\MS$ the supersymmetry breaking scale and $S$ the dimensionless action of the instantons responsible for breaking the axion shift symmetry and generating the axion mass, see Sec.~\ref{string}. In the latter analysis (dashed black curve) we assume phenomenology-inspired flat priors on $\log_{10}m_a$ and $\log_{10}f_a$, with $m_a$ the ALP mass and $f_a$ the axion decay scale. In both cases we apply the bounds on the parameter space described in Sec.~\ref{sec:axion_cosmology}. Hashed areas represent the various detection techniques for different mass ranges of the string axiverse.}
	\label{fig:Plotaxiverse_logmass}
\end{center}
\end{figure}

The previous result, selecting a very specific range for $m_a$ in the string axiverse case, raises the question: what if we were to repeat the analysis without focusing on string axions, hence without adopting string-inspired priors for $\mu$ and $S$? We address this question by performing a fourth and final analysis where we consider the parameter space spanned by the parameters $(m_a\,,f_a\,,\theta_i\,,\beta)$, as per previous discussions, with phenomenology-inspired priors on $m_a$ and $f_a$ (flat in $\log_{10}m_a$ and $\log_{10}f_a$). In this case, we find that our analysis selects a rather broad region in the $m_a$ parameter space, with $\log_{10}(m_a/{\rm eV}) \approx -11^{+11}_{-6}$ at 68\%~C.L. and $\approx -11^{+14}_{-15}$ at 95\%~C.L., slightly preferring the heavier mass region, albeit at a very mild significance. This suggests that the choice of prior distributions (string-inspired versus non-string-inspired) plays a relevant role in our results, a fact worth keeping in mind when reading our paper. Our main results are shown in Fig.~\ref{fig:Plotaxiverse_logmass}, where we plot the posterior probability distributions we obtain for the string axion (first analysis, solid black line) and the ALP (fourth analysis, dashed black line) cases, with detection techniques for axions of various mass ranges overlain on the plot. The results of the four analyses we perform are briefly summarised in Tab.~\ref{table_coefficients}.
\begin{table}[bt]
\begin{center}
\scriptsize
{\renewcommand{\arraystretch}{1.4}
\begin{tabular}{|l|l|l|l|}
\hline
	Analysis & Parameters & $\log_{10}\!\(m_a\! /{\rm eV}\)\!$ & $\log_{10}\!\!\(\MS\!/\!{\rm TeV}\!\)\!$\\
\hline
	String axion & $\{\MS, S, \theta_i, \beta\}$ & $-21.5^{+1.3}_{-2.3} $ & Unconstrained\\
	String axion, $S \!=\! 198$ & $\{\MS, \theta_i, \beta\}$ & $-20.9^{+1.3}_{-2.3}$ & $7.8^{+1.4}_{-2.4}$\\
	String axion, $S \!=\! 153$ & $\{\MS, \theta_i, \beta\}$ & $<\! -14.9$ & $<\! 3.8$\\
	Axion (ALPs) & $\{m_a, f_a, \theta_i, \beta\}$ & $-11.1^{+11.5}_{-5.9}$ & Unconstrained\\
\hline
\end{tabular}}
\caption{Summary of the parameters used and inferred values for $\log_{10}(m_a /{\rm eV})$ and $\log_{10}(\MS/{\rm TeV})$ obtained for each analysis. Intervals of the form $\mu \pm \sigma$ are 68\%~C.L. intervals whereas quoted upper limits are 95\%~C.L. upper bounds. We consider a parameter to be unconstrained when its 95\%~C.L. interval is almost as wide as the parameter prior.}
\label{table_coefficients}
\normalsize
\end{center}
\end{table}
In principle, additional readily available datasets can be added to the analysis performed, to further constrain different parts of the axion parameter space range, including:
\begin{itemize}
	\item Limits on the axion-photon coupling from direct searches for the axion in the lab through light-shining-through-wall experiments, like OSQAR~\cite{Pugnat:2007nu, Ballou:2015cka} and ALPS~\cite{Ehret:2009sq, Ehret:2010mh};
	\item Cavity searches~\cite{Sikivie:1983ip, Sikivie:1985yu}, whose sensitivity depends on the local density of axion dark matter, like ADMX~\cite{Asztalos:2001tf, Stern:2016bbw} and YWL~\cite{Kenany:2016tta, Brubaker:2016ktl};
	\item Axion helioscopes~\cite{vanBibber:1988ge}, like CAST~\cite{Andriamonje:2007ew, Anastassopoulos:2017ftl};
	\item Astrophysical constraints on the axion-photon coupling from the supernova 1987A~\cite{Payez:2014xsa};
	\item Constraints on the axion-electron coupling from the cooling of white dwarfs~\cite{Giannotti:2015kwo, Giannotti:2017hny};
	\item Constraints on the couplings of the axion to electrons and photons from consideration on the branch of red giant stars~\cite{Viaux:2013lha};
	\item Searches for gravitational waves in relation to black hole super-radiance~\cite{Arvanitaki:2009fg};
	\item Constraints on the mass of ultra-light bosons in relation to the Lyman-$\alpha$ forest~\cite{Kobayashi:2017jcf, Desjacques:2017fmf, Poulin:2018dzj};
	\item The matter power spectrum~\cite{Hlozek:2014lca, Hlozek:2017zzf}.
\end{itemize}
A list of experiments and astrophysical observations have been recently used to constrain the parameter space of the QCD axion in Ref.~\cite{Hoof:2018ieb}, while precision cosmological data have been used to explore the axiverse in Refs.~\cite{Hlozek:2014lca, Hlozek:2017zzf}. We discuss these findings in comparison with our results in Sec.~\ref{conclusions}.

This paper is organised as follows. In Sec.~\ref{sec:axion_cosmology} we first review some basics of axion cosmology. We review the string theory axion and the string axiverse in Sec.~\ref{string}. In Sec.~\ref{sec:analysis}, we discuss in more detail the cosmological data used and the statistical analysis performed. In Sec.~\ref{sec:results}, we discuss the constraints we obtain on the relevant string parameters, $M_{\rm susy}$ and $S$. We then compare the constraints obtained on $m_a$ (axion mass) and $f_a$ (axion decay constant) in the string axion case (by converting the the bounds we obtained on $M_{\rm susy}$ and $S$ in the first analysis), as well as in the ALP case (fourth analysis). We also discuss the results obtained by fixing $S$ for purely illustrative purposes, in order to resolve the $M_{\rm susy}$-$S$ degeneracy. Finally, in Sec.~\ref{conclusions} we provide concluding remarks and discuss the implications of our results for future axion and supersymmetry searches.

\section{Axion cosmology}
\label{sec:axion_cosmology}

In this Section, we review axion cosmology without focusing on the possible string nature of the axion itself. See e.g.~\cite{Marsh:2015xka} for a recent comprehensive review. As stated previously, we consider for definiteness the case where the axion decay constant is larger than the Hubble rate during inflation $H_I$, implying that the axion is present during inflation. The initial value of the axion misalignment angle $\theta_i$ is drawn randomly from the unit circle, with the randomness arising from the Peccei-Quinn mechanism. The axion field is frozen in its initial configuration until the Universe's expansion rate has slowed down to a value comparable to the axion mass. At this point the axion field starts to oscillate about the minimum of its potential and the axion number density redshifts as expected for a non-relativistic matter component. This transition occurs when the scale factor takes a value $a_{\rm osc}$ obtained by requiring $m_a \approx 3H(a_{\rm osc})$, where the value of the Hubble rate as a function of the scale factor $a$ is given in terms of the Hubble parameter and scale factor at present time, $H_0$ and $a_0$, as follows:
\be
	H(a) = H_0\, \left [ \Omega_\Lambda + \Omega_m \(\frac{a_0}{a}\)^3 + \Omega_r \(\frac{a_0}{a}\)^4 \right ] ^{\frac{1}{2}}.
\ee
Here we denote by $\Omega_i \equiv \rho_i/\rho_{\rm crit}$ the ratio of the current energy density of species $i \in \{\Lambda, m, r\}$ (with $\Lambda$, $m$, and $r$ corresponding to dark energy, matter, and radiation respectively) to the current critical energy density $\rho_{\rm crit} = 3H^2_0\MP^2$, with $M_{\rm Pl}$ the reduced Planck mass. The present energy density of cold axions, $\rho_a = (1/2)\Lambda_a^4\,\theta_i^2\,\(a_{\rm osc}/a_0\)^{3}$, must not exceed the present CDM energy density measured by Planck $\rho_{\rm CDM} \sim 10^{-47}{\rm \,GeV^4}$~\cite{Ade:2015xua}.

As for any nearly massless scalar field, axions inherit quantum fluctuations from the inflationary period, with a standard deviation $\sigma_\theta$. Primordial quantum fluctuations later develop into isocurvature perturbations~\cite{Kobayashi:2013nva}, which modify the number density of axions, since the gauge invariant entropy perturbation $\mathcal{S}_a$ is non-zero~\cite{Axenides:1983, Linde:1985yf, Seckel:1985}. Cold axions that spectated inflation differ from thermally-produced WIMPs because of these imprints from isocurvature fluctuations, whose amplitude is related to the energy scale of inflation. In this work, we focus on single-field, slow-roll inflation for which the amplitude of the bispectrum is related to the spectral tilt of the power spectrum~\cite{Maldacena:2002vr}, and a relation between the Hubble rate at the end of inflation $H_I$ and the tensor-to-scalar ratio $r$ emerges~\cite{Lyth:1984, Lyth:1990, Lyth:1992yy}. In this framework, the standard deviation of the axion field in units of the decay constant is $\sigma_\theta = H_I/2\pi f_a$. In the following, we assume that there are no couplings between the axion and the inflaton field other than gravity. Other scenarios have been discussed in Ref.~\cite{Ballesteros:2016xej} in relation to axion isocurvature fluctuations. Isocurvature fluctuations can also be suppressed by coupling the axion to a hidden sector~\cite{Kitajima:2014xla}, which we do not take into consideration here.

We have parametrised the power spectrum of the isocurvature fluctuations at the scale $k_0$ according to~\cite{Crotty:2003rz, Beltran:2005xd, Beltran:2006sq}
\begin{equation}
	\Delta^2_{A}(k_0) \!\equiv\! \langle \mathcal{S}_a^2 \rangle \!=\!\! \(\frac{\partial \ln \Omega_a}{\partial \theta_i}\)^{\!\!2}\!\!\(\frac{\Omega_a}{\Omega_{\rm CDM}}\)^{\!\!2}\!\sigma_\theta^2 \!=\! \Delta^2_{\mathcal{R}}(k_0)\frac{\beta}{1\!-\!\beta}.
	\label{eq:axionisocurvaturebound}
\end{equation}

The \textit{Planck} collaboration~\cite{Ade:2013zuv, Planck:2013jfk, Barkats:2013jfa, Ade:2015tva, Ade:2015lrj} constrains both the primordial isocurvature fraction $\beta$ and the tensor-to-scalar ratio $r$, while the curvature power spectrum is measured to be $\Delta^2_{\mathcal{R}}(k_0) \approx 2.2 \times 10^{-9}$~\cite{Ade:2015xua}. In terms of the axion physics quantities, the ratio of the axion energy density to the CDM energy density today, $\omega$, and the tensor-to-scalar ratio, $r$, read
\bea
	\omega &\equiv& \frac{\rho_a}{\rho_{\rm CDM}} = \frac{m_a^2\,f_a^2\,\theta_i^2}{2\rho_{\rm CDM}}\,\(\frac{a_{\rm osc}}{a_0}\)^{3} \leq 1 \,,
	\label{eq:energydensity}\\
	r &=& \frac{2}{\omega^2}\(\frac{f_a\theta_i}{\MP}\)^2 \,\frac{\beta}{1-\beta}\,.
	\label{eq:tensor_to_scalar_ratio}
\eea

An important caveat is in order at this point. In the following, we will conservatively require $\omega \leq 1$ so that the current axion energy density does not exceed the current CDM energy density. In reality, this is an approximate requirement for two reasons. The first is that, for $m_a \lesssim 10^{-27}\,{\rm eV}$, the resulting ULA actually has a dark energy-like rather than dark matter-like behaviour, and so a more appropriate requirement would be $\rho_a/\rho_{\Lambda}<1$. Moreover, an earlier analysis using precision cosmological data (including measurements of the CMB temperature and polarization anisotropy power spectra, galaxy power spectrum, and Baryon Acoustic Oscillations) showed that, in the region where $10^{-32} {\rm \,eV} \lesssim m_a \lesssim 10^{-26} {\rm \,eV}$, $\omega \lesssim 0.05$ is required~\cite{Hlozek:2014lca} (see also~\cite{Poulin:2018dzj,Hlozek:2016lzm}).

In our work, we conservatively choose to only require $\omega \leq 1$ instead, for a variety of reasons. Firstly, a posteriori we do not expect the choice of setting $\omega \lesssim 0.05$ for $10^{-32} {\rm \,eV} \lesssim m_a \lesssim 10^{-26} {\rm \,eV}$ to affect our bounds substantially, since the posterior distributions in the region of $m_a$ parameter space in question are already quite suppressed for both the string axion and the ALP analyses (see Fig.~\ref{fig:Plotaxiverse_logmass}). Second, our focus in this paper is on the string axion case, and it is worth noting that the mathematical relations connecting $M_{\rm susy}$ and $S$ to $m_a$ and $f_a$ [Eqs.~(\ref{mamu},\ref{fs}) to be discussed later] are uncertain to a factor of a few (especially the relation between $f_a$ and $S$), so actually worrying about modelling the exact constraints on $\omega$ when there are these other uncertainties at play appears to some extent incongruous. We also remind the reader that our goal is not to provide a full-fledged analysis with precision cosmological data, but to get a feel for whether cosmology can actually provide information on $M_{\rm susy}$ and $S$: we believe that to this end, modelling the exact constraints on $\omega$ is not essential. Of course, for future work aiming to obtain more robust constraints on the string axiverse from precision cosmology data, dealing with the aforementioned problem will be of utmost importance and we plan to return to this in a follow-up work using the modified Boltzmann solver \texttt{axionCAMB}~\cite{Hlozek:2014lca}.

\section{Axions in string theory}
\label{string}

In this Section, we consider axions originating from string theory. Starting from the ten-dimensional low energy Lagrangian of the heterotic string~\cite{polchinski_1998}, and reducing to four dimensions by compactifying a six-dimensional manifold $Z$ whose volume is $V_Z$, an effective Lagrangian describing the field $\phi = a/f_a$ is found to be~\cite{Witten:1984, Svrcek:2006yi, Conlon:2006tq}:
\be
	\mL = \frac{f_a^2}{2}\(\partial^\mu\phi\)\(\partial_\mu \phi\) - \frac{\Lambda_a^4}{2}\phi^2,
\ee
where, in various string compactification models~\cite{Svrcek:2006yi, Svrcek:2006hf}, the axion decay constant satisfies $f_a \lesssim x\MP/S$, with $x$ a factor of ${\cal O}(1)$~\footnote{For example, $x\!=\!\sqrt{2}/2$ in the context of the model-independent heterotic string.} and $S$ is the (dimensionless) action of the instantons that break the axion shift symmetry, generating the axion potential. As stated previously, we expect $S \approx {\cal O}(200)$. The dependence on the type of string theory (the string scale $\ell_s$, the asymptotic expansion parameter $g_s$, and the volume of $Z$) enters only through the string definition of the reduced Planck scale $\MP = \sqrt{4\pi V_Z}/g_s\ell_s^4$ and $S = 2\pi V_Z/g_s^2\ell_s^6$. For instance, non-perturbative world-sheet or membrane instantons violate the shift symmetry~\cite{Svrcek:2006yi, Svrcek:2006hf}, leading to $\Lambda_a^4 = \mu^4e^{-S}$, where $\mu$ is an UV non-perturbative scale which can be as high as the Planck energy scale.

To connect to more physical quantities, the resulting axion mass is then given by $m_a = \Lambda_a^2/f_a$. Supersymmetry, which almost inevitably appears in any realistic string theory, suppresses the UV non-perturbative energy scale by a factor $\(\MS/\MP\)^2$, where $\MS$ is the SUSY breaking scale. We then expect $\mu^2 \approx \MS \MP$~\cite{Beasley:2004ys}. In the following, we assume $f_a = x\MP/S$ with $x = 1$. Since the axion decay constant enters Eqs.~(\ref{eq:energydensity},\ref{eq:tensor_to_scalar_ratio}) only through the combination $f_a\theta_i$, choosing lower values $x < 1$ should not alter our results significantly since the factor $x$ would be re-absorbed into a different value of the misalignment angle $\theta_i$ (which we are not particularly interested in). Admittedly, this does nonetheless introduce an uncertainty of order unity when converting from $S$ to $f_a$ which should be kept in mind. When performing the analysis, we convert from the parameters $\{\mu, S\}$ (or alternatively $\{\MS, S\}$) to the parameters $\{m_a, f_a\}$ through
\begin{eqnarray}
	m_a & \approx & \frac{\mu^2 e^{-\frac{S}{2}}S}{\MP} \approx \MS e^{-\frac{S}{2}}S \,, \label{mamu} \\
	f_a & \approx & \frac{\MP}{S} \label{fs} \,.
\end{eqnarray}
In terms of the string parameters, Eqs.~(\ref{eq:energydensity},\ref{eq:tensor_to_scalar_ratio}) read
\bea
	\omega &\equiv& \frac{\Omega_a}{\Omega_{\rm CDM}} \approx \frac{\mu^4\,\theta_i^2\,e^{-S}}{2\rho_{\rm CDM}}\,\(\frac{a_{\rm osc}}{a_0}\)^{3} \leq 1 \,,
	\label{eq:energydensity_str}\\
	r &\approx& \frac{2}{\omega^2}\(\frac{\theta_i}{S}\)^2 \,\frac{\beta}{1-\beta}\,.
	\label{eq:tensor_to_scalar_ratio_str}
\eea

\section{Analysis}
\label{sec:analysis}

In this Section, we describe the method used to obtain observational constraints on the string axiverse parameters. Our aim is to obtain observational constraints on the four parameters $\mu$, $S$, $\theta_i$, and $\beta$, jointly denoted by $\boldsymbol{\Theta_1}$, in light of observational data $\boldsymbol{d}$. In order to do so, we perform a Bayesian analysis, for which we need to specify priors for the parameters $\boldsymbol{\Theta_1}$.

We begin by discussing our choice of priors. For the initial misalignment angle $\theta_i$ we choose a uniform prior over the region $[-\pi; \pi]$, consistent with the parameter being drawn randomly from the unit circle. Our prior on $\beta$ is given by the posterior distribution for the same parameter obtained by the \textit{Planck} collaboration analysing the \textit{Planck}~\textrm{TT,TE,EE+lowP} dataset (including temperature as well as large-scale and small-scale polarisation data) at the wavenumber $k_0 = 0.002{\rm \,Mpc^{-1}}$, and provided in~Ref.~\cite{Ade:2015lrj}. The fractional primordial contribution of isocurvature modes at the comoving wavenumber considered is constrained as $\beta < 0.02$ at 95\% C.L. The distribution of $\beta$ peaks roughly at $\beta \approx 0$, i.e. it is consistent with an upper limit and not a detection. We have also tested the results against a different prior based on the~\textrm{TT+lowP} dataset, obtaining that results are not sensitive to the different choice on the prior on $\beta$. We choose a flat prior for $S \in [50; 450]$. The range is chosen to match the theoretical expectation $S \approx 200$, and we verify a posteriori that it is broad enough to not cut the posterior where the latter is significantly non-zero. Finally, we impose a flat prior for $\log_{10}(\mu/{\rm TeV}) \in [7;16]$, where the choice for the range considered corresponds roughly to $\MS \approx \mu^2/\MP \in [100\,{\rm GeV};\MP]$, i.e. to the SUSY breaking scale lying between the electroweak scale and the Planck scale. The upper limit in the choice on the range for $\MS$ conforms to theoretical expectations from realistic string theories, whereas the lower limit is consistent with the non-observation of supersymmetric partners at colliders.

On top of the priors we discussed, we include one further prior on the set of parameters $\boldsymbol{\Theta_1}$, conditioned on the value of the tensor-to-scalar ratio $r$ computed from $\boldsymbol{\Theta_1} \equiv \{\mu\,,S\,,\theta_i\,,\beta\}$ through Eq.~(\ref{eq:tensor_to_scalar_ratio_str}). We choose the prior in such a way that it reflects constraints on the tensor-to-scalar ratio obtained by the \textit{Planck} collaboration by analyzing the \textit{Planck}~\textrm{TT,TE,EE+lowP} dataset. Operationally, for each point in parameter space $\boldsymbol{\hat{\Theta}_1}$ selected by our Markov Chain Monte Carlo algorithm (to be discussed briefly later), we first compute $\hat{r}$ using Eq.~(\ref{eq:tensor_to_scalar_ratio_str}); then, to this point in parameter space, we assign a prior probability whose value is numerically equivalent to the posterior probability distribution for the tensor-to-scalar ratio $r$ determined by the \textit{Planck} collaboration and evaluated at $\hat{r}$ (or, equivalently, we reweigh each point in our Markov chain by this value).~\footnote{Technically speaking, it would perhaps have been more appropriate to use the posterior distribution for $r$ \textit{conditioned} to the given value of $\hat{\beta}$, i.e. $P(r=\hat{r} \vert \hat{\beta})$. However, this procedure would make a substantial difference only if $\beta$ and $r$ were strongly correlated, which is not the case since the results in Ref.~\cite{Ade:2015lrj} suggest that the correlation between $\beta$ and $r$ is mild. In fact, we read on the bottom of Page~50 of Ref.~\cite{Ade:2015lrj} that ``[cold dark matter isocurvature perturbations] hardly affect the determination of $r$, and allowing for tensor perturbations hardly affects the determination of the non-adiabaticity parameters". From this statement we conclude that $P(r=\hat{r} \vert \hat{\beta}) \approx P(r=\hat{r})$, explaining the simplification made in our choice of likelihood.} This distribution peaks at $r \approx 0$ and constrains $r<0.12$ at 95\%~C.L., \textit{i.e.} the distribution is consistent with a non-detection, similarly to what discussed for the prior on $\beta$.

On top of these priors, we further restrict the available parameter space by requiring that:
\begin{itemize}
	\item the axion decay rate into two photons be smaller that the present expansion rate of the universe: this imposes further cuts in the $m_a$-$f_a$ subspace, so in turn on the $\mu$-$S$ subspace through Eqs.~(\ref{mamu},\ref{fs});

	\item the axion be present during inflation, so we demand $f_a > H_I/2\pi$, with $H_I$ the Hubble rate at the end of inflation. Since $H_I$ can be expressed as a function of $r$ and hence of all four parameters in $\boldsymbol{\Theta_1}$, this condition imposes further cuts over the whole parameter space;

	\item the current axion energy density not exceed the CDM energy density, i.e. that $\omega\leq1$, with $\omega$ given by Eq.~(\ref{eq:energydensity_str}). Since $\omega$ can be expressed as a function of $\mu$, $S$, and $\theta_i$, this condition restricts the available $\mu$-$S$-$\theta_i$ subspace, without affecting $\beta$.
\end{itemize}
%

Our discussion so far was concerned with string axions, and on the parameters $\boldsymbol{\Theta_1} \equiv \{\mu\,,S\,,\theta_i\,,\beta\}$. As discussed in the introduction, after converting the resulting bounds on $\mu$ and $S$ to bounds on $m_a$ and $f_a$ using Eqs.~(\ref{mamu},\ref{fs}), we are a posteriori brought to also consider a more generic analysis where we sample directly on the latter two parameters, with phenomenology-inspired priors on the two. We therefore perform a separate analysis wherein we consider the parameter space spanned by $\boldsymbol{\Theta_2} \equiv \{X\,,Y\,,\theta_i\,,\beta\}$, with $X \equiv \log_{10}(m_a/{\rm eV})$ and $Y \equiv \log_{10}(f_a/M_{\rm Pl})$. This choice is driven by our expectation that $m_a$ and $f_a$ evenly span various orders of magnitude~\cite{Arvanitaki:2009fg}, hence it is more appropriate to work with their logarithms (however see~\cite{Easther:2005zr}). We impose uniform priors on $X$ and $Y$ within the ranges $[-40;8]$ and $[-10;0]$ respectively. We also further impose the same bounds as discussed for the string axion analysis (i.e. the bounds concerning the axion decay rate into two photons, the axion spectating inflation, and the axion energy density not exceeding the CDM energy density today). The range chosen for $X$ is very broad and conservative and allows for values of $m_a<H_0\sim 10^{-33}\,$eV, for which the axion has yet to begin oscillating. The range for $Y$ is chosen such that the decay constant $f_a$ is sub-Planckian, and is bounded from below by the negative results of the CAST searches~\cite{Anastassopoulos:2017ftl}. As in Sec.~\ref{sec:analysis}, we again construct our likelihood as:
\begin{eqnarray}
\hskip -0.5 cm {\cal L}(\boldsymbol{d} \vert \hat{X}\,,\hat{Y}\,,\hat{\theta}_i\,,\hat{\beta}) = {\rm Pr}_{\rm Planck} \( r = \hat{r}(\hat{X}\,,\hat{Y}\,,\hat{\theta}_i\,,\hat{\beta}) \)\,,
\end{eqnarray}
with $\hat{r}(\hat{X}\,,\hat{Y}\,,\hat{\theta}_i\,,\hat{\beta})$ given by Eq.~\eqref{eq:tensor_to_scalar_ratio}.

To sample the posterior distribution we use Markov Chain Monte Carlo (MCMC) methods. We use the Metropolis-Hastings sampler implemented in the cosmological MCMC package \texttt{Montepython}~\cite{Audren:2013}, which we configure to act as a generic sampler. From the generated MCMC chains we compute the joint and marginalized posterior probability distributions of the four parameters and, in particular, of the axion mass and the axion decay constant. From now on, we quote credible regions for the parameters at 68\%~C.L. unless otherwise stated, whereas all upper limits are quoted at 95\%~C.L. to conform to standard practices.

\section{Results}
\label{sec:results}

In this Section, we describe the results obtained using the methodology outlined previously. We begin by considering the string axion case, with parameter space described by $\boldsymbol{\Theta_1} \equiv \{\mu\,,S\,,\theta_i\,,\beta\}$ (or equivalently $\{ M_{\rm susy}\,,S\,,\theta_i\,,\beta \}$). Our first analysis yields a poor constraint on $\mu$ (and hence $\MS$), which remains basically unconstrained due to the strong degeneracy between $\mu$ and $S$ which data is unable to lift, while we find $S=198 \pm 28$. One might at first glance be surprised by the fact that $S$ is quite well constrained whereas $\mu$ is not. However, an explanation for this puzzling observation is readily found by examining Eqs.~(\ref{eq:energydensity_str},\ref{eq:tensor_to_scalar_ratio_str}), where we see that $S$ enters exponentially in the observables, and hence it is not possible to vary $S$ \textit{too much} without then spoiling agreement with observations. In particular, high values of $S$ are excluded by the limits on $r$, since $r \propto e^{2S}$, whereas low values are excluded by the requirement that $\omega \leq 1$, since $\omega \propto e^{-S}$.

\begin{figure}[!h]
\begin{center}
	\includegraphics[width=\linewidth]{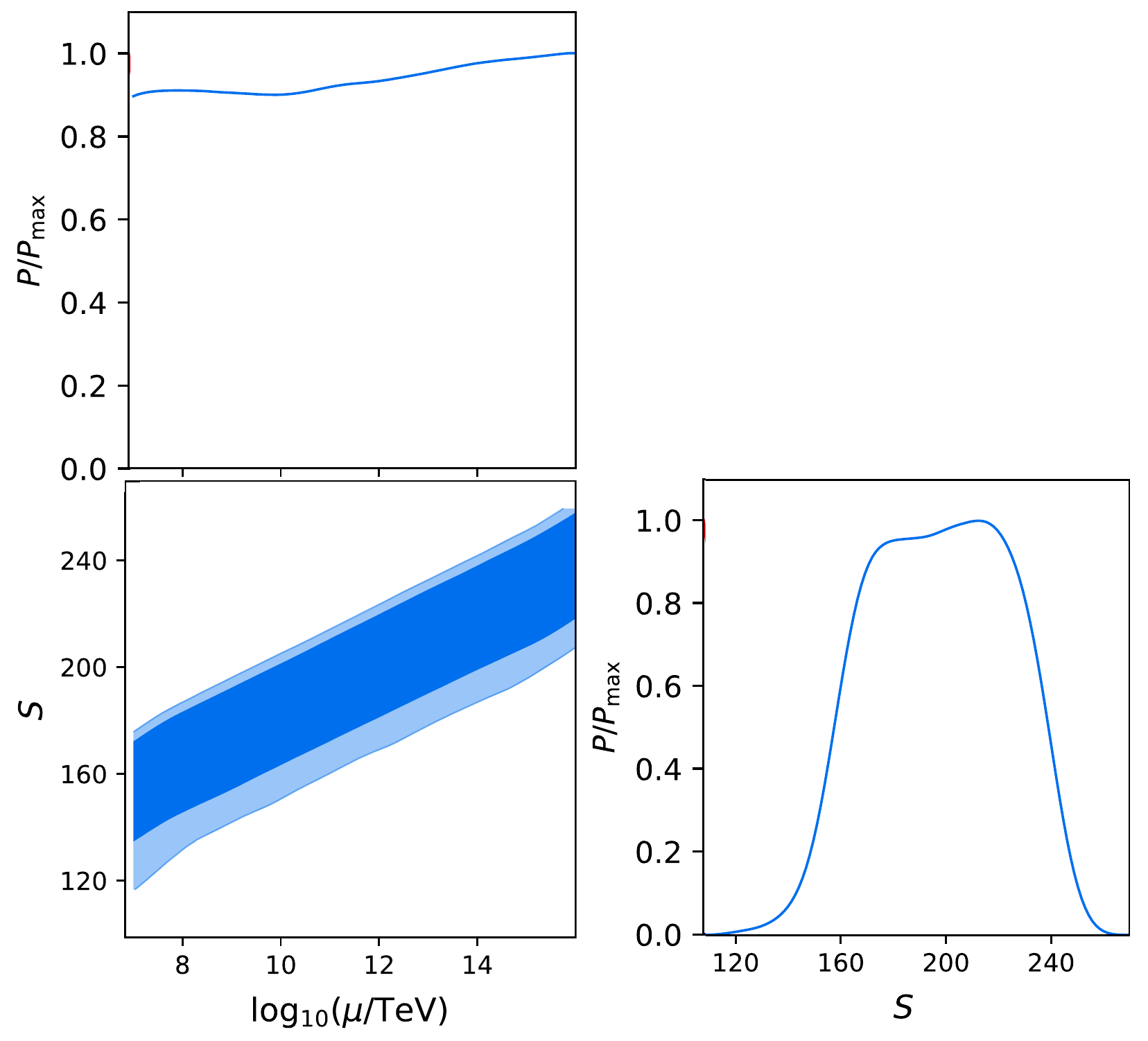}
	\caption{Triangular plot showing the joint and marginalised posterior distributions for $\log_{10}\(\mu/{\rm TeV}\)$ and $S$, with $\mu \approx \sqrt{\MS \MP}$ an ultraviolet non-perturbative scale and $S$ the dimensionless action of the instantons responsible for generating the axion mass, obtained from our string-inspired axion analysis. The bottom left panel shows the 2D joint posterior distribution, whereas the plots along the diagonal show the marginalised posterior distributions normalized to their maximum values.}
	\label{fig:axiverse_mus_tri}
\end{center}
\end{figure}

The degeneracy between $\mu$ and $S$ is easy to understand if we glance at Eqs.~(\ref{eq:energydensity_str},\ref{eq:tensor_to_scalar_ratio_str}). There, we see that $\mu$ and $S$ enter the observables through the combination $\mu^4e^{-S}$, and the data is unable to break this degeneracy (this is somewhat similar to the case of the CMB temperature power spectrum, where the overall amplitude depends on the combination $A_se^{-2\tau}$, with $A_s$ and $\tau$ thus strongly correlated and the degeneracy only being partially broken by including polarisation data, see e.g. discussion in~\cite{Vagnozzi:2017ovm}). We thus expect a strong correlation between the two parameters. This is confirmed by our analysis, wherein we find a correlation coefficient of $0.93$ between the two parameters, which are thus close to being perfectly correlated. In Fig.~\ref{fig:axiverse_mus_tri} we show a triangular plot in the $\log_{10}\mu$-$S$ space which clearly shows this degeneracy.

Using Eqs.~(\ref{mamu},\ref{fs}), we translate the obtained bounds on $\mu$ and $S$ into bounds on $m_a$ and $f_a$. We do so by using the aforementioned relations and converting each $\{\mu\,,S\}$ sample in our MCMC chains into an $\{m_a\,,f_a\}$ sample, so that the latter two are effectively derived parameters. Notice that doing so retains the information on the priors we used on $\mu$ and $S$. The resulting marginalised posterior probability distribution for $m_a$, normalised to its maximum value, is given by the solid black line in Fig.~\ref{fig:Plotaxiverse_logmass}. We find $\log_{10}(m_a/{\rm eV}) = -21.5^{+1.3}_{-2.3}$ and $f_a = (5.1^{+0.5}_{-0.9})\times 10^{-3}\,\MP$. The reason for $m_a$ and $f_a$ being quite well constrained despite $\mu$ being unconstrained is due to the fact that $S$ is quite well constrained. Note that the values obtained for $m_a$ and $f_a$ are centred around what was originally proposed for ULAs~\cite{Hu:2000ke}, corresponding to an axion decay constant of the order of the GUT scale and with a Compton wavelength of order $m_a^{-1} \approx 0.1\,{\rm pc}$, thus with a de Broglie wavelength of galactic size. We therefore conclude that cosmological data complemented with our choice of string-inspired priors favour the region of the string axiverse that comprises an ULA of mass $m_a \approx 10^{-22}\,{\rm eV}$. This result is perhaps somewhat artificial, as it depends on the choice of priors and in particular on our choice of having a sub-Planckian $M_{\rm susy}$. We will discuss in much more detail the impact of this choice on our results in the final paragraphs of this Section.

\begin{figure}[!h]
\begin{center}
	\includegraphics[width=\linewidth]{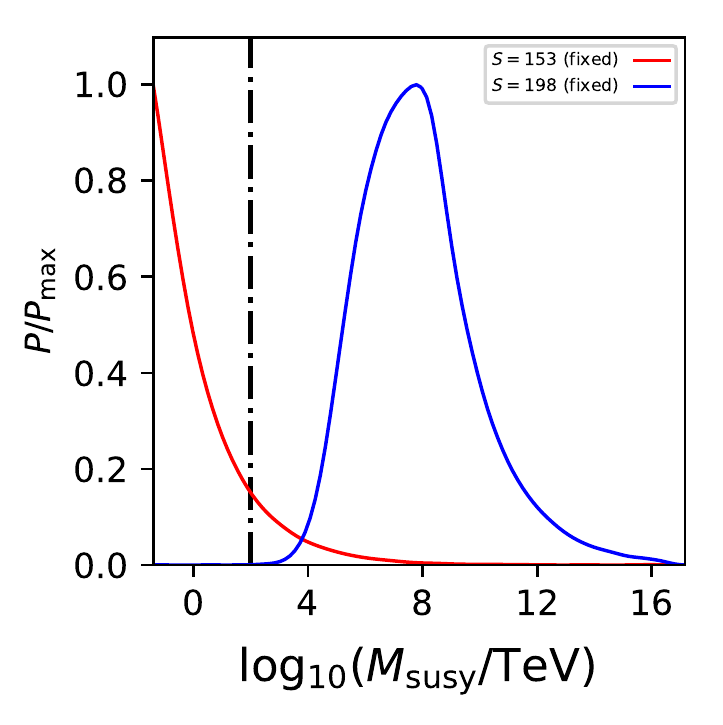}
	\caption{Marginalized posterior probability distribution for the supersymmetry breaking scale, normalised to its maximum value, obtained from our string axion analysis, when fixing $S=198$ (blue curve) and $S=153$ (red curve), respectively the central value and $2\sigma$ lower limit on $S$ obtained in our analysis. The vertical dot-dashed line sets the scale of a proposed 100 TeV collider.}
	\label{fig:axiverse_mus_fixed_msusy}
\end{center}
\end{figure}

The strong degeneracy between $\mu$ (or $\MS$) and $S$ suggests that an improved determination of the former parameter(s) would be greatly enhanced by an improved determination of the latter. For purely pedagogical purposes, we explore how the constraints on $\MS$ (currently unconstrained) would change if we were to fix $S$ to a selected value. We choose two representative values of $S$, namely $S=198$ (corresponding to the inferred central value of $S$) and $S=153$ (corresponding to the $2\sigma$ lower limit on $S$). One could for instance imagine considering a specific string realisation wherein the value of $S$ is so well determined from theoretical considerations that it is for all intents and purposes fixed. In the former case, we find $\log_{10}(\MS/{\rm TeV}) = 7.8^{+1.4}_{-2.3}$, whereas in the latter case we find $\MS \lesssim 8 \times 10^3\,{\rm TeV}$ at 95\% C.L., indicating that within this scenario SUSY could possibly be within reach of the proposed 100 TeV collider~\cite{Arkani-Hamed:2015vfh}. We show the marginalised posterior distribution, normalised to its maximum value, of $\log_{10}(\MS/{\rm TeV})$ when $S$ is fixed to $198$ and $153$ in Fig.~\ref{fig:axiverse_mus_fixed_msusy}, with the dashed vertical line corresponding to $\MS=100\,{\rm TeV}$.

We now perform our fourth analysis, where we consider ALPs (thus disregarding their fundamental origin) with the parameter space described by $\boldsymbol{\Theta_2} \equiv \{m_a\,,f_a\,,\theta_i\,,\beta\}$ and flat priors on $\log_{10}m_a$ and $\log_{10}f_a$. The marginalised posterior probability distribution for $m_a$, normalised to its maximum value, is given by the dashed black line in Fig.~\ref{fig:Plotaxiverse_logmass}. In particular, we find $X = \log_{10}(m_a/{\rm eV}) = -11.1^{+11.5}_{-5.9}$, mildly favouring higher values for $m_a$ and mildly disfavouring ULAs, although not at a statistically significant level. Two notable features are discernible from the posterior distribution. The first is the complete loss of sensitivity (flat posterior) at very low values of $m_a \lesssim H_0$, due to the fact that axions which are extremely light have yet to begin oscillating. In the very high mass range ($m_a \gtrsim {\cal O}({\rm eV})$), instead, the posterior distribution is sharply cut by the requirement that the axion be present during inflation, hence $f_a \gtrsim H_I$. The requirements that $\omega\leq1$ and that $r \ll 1$ also cut the posterior distribution at low and high masses respectively, as evident from Eqs.~(\ref{eq:energydensity},\ref{eq:tensor_to_scalar_ratio}).

\begin{figure}[tb]
\begin{center}
	\includegraphics[width=\linewidth]{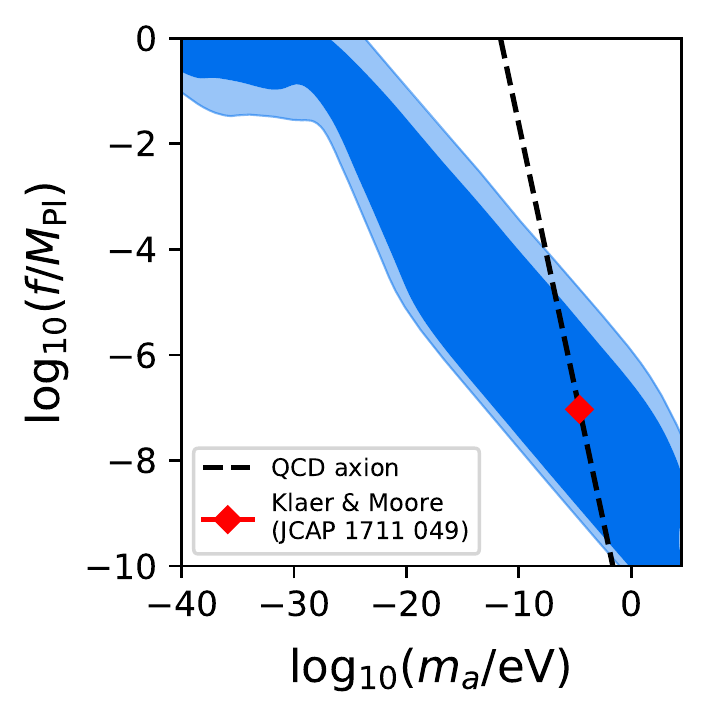}
	\caption{2D joint posterior distribution in $\log_{10}\(m_a/{\rm eV}\)$-$\log_{10}(f_a/\MP)$ parameter space, obtained in our axion analysis with flat priors on $\log_{10}m_a$ and $\log_{10}f_a$, with 68\%~C.L. and 95\%~C.L. credible regions corresponding to the dark and light blue regions respectively. Also shown is the relation defining the QCD axion (dashed black line), and the value of the axion mass predicted by Klaer and Moore~\cite{Klaer:2017ond} assuming that the axion makes up the totality of the dark matter (red diamond), $m_a = (26.2 \pm 3.4)\,\mu{\rm eV}$.}
	\label{fig:Plotaxiverse_Run1_mf}
\end{center}
\end{figure}

For the axion decay constant, we find $Y = \log_{10}(f_a/\MP) = -5.5^{+1.8}_{-3.0}$. The central value corresponds to $f_a \approx 10^{13}{\rm \,GeV}$ which is slightly higher than what expected for the QCD axion to be the CDM particle. Moreover, we find that $X$ and $Y$ are strongly anti-correlated, with a correlation coefficient of $-0.92$. We show this Fig.~\ref{fig:Plotaxiverse_Run1_mf}, where we plot the joint 2D posterior distribution in the $X$-$Y$ plane along with the the 68\% C.L. (dark blue) and the 95\% C.L. (light blue) contours. The degeneracy is approximately along the axis defined by the relation $m_af_a^2 \approx {\rm const}$. This can be understood by combining Eqs.~(\ref{eq:energydensity},\ref{eq:tensor_to_scalar_ratio}) for a radiation-dominated universe ($m_a \gtrsim 10^{-27}{\rm \,eV}$), for which we obtain $r \propto \(m_a f_a^2 \theta_i^2\)^{-1}\beta$. Overlain on the same figure is the $m_a$-$f_a$ relation for the QCD axion (black dashed line), which instead lies along the axis defined by $m_af_a = \Lambda_{\rm QCD}^2$, as well as the central value of the analysis by Klaer and Moore (which for the first time included large short-distance contributions to the axionic string tension in their numerical simulations) which yields $m_a = \(26.2 \pm 3.4\){\rm \, \mu eV}$~\cite{Klaer:2017ond}.

If we interpret the results obtained in Fig.~\ref{fig:Plotaxiverse_Run1_mf} in terms of the string axiverse parameters $\mu$ and $S$, the preferred value for the decay constant forces $S \sim \MP/f_a \approx 10^6$. This is far from the theoretically preferred value $S \approx {\cal O}(200)$. Moreover, it leads to a trans-Planckian value of $\MS = m_a e^{S/2}/S$ for any reasonable value of the axion mass, due to the exponential sensitivity of the SUSY scale to $S$. In fact, demanding that $\MS$ be at most equal to $\MP$, we find that the size of the cycle is at most $S = 145$ ($S = 240$) for $m_a =10^{-2}{\rm \,eV}$ ($m_a =10^{-22}{\rm \,eV}$), a value which is consistent with ${\cal O}(200)$~\footnote{Although this results seems to be in agreement with what obtained in Ref.~\cite{Svrcek:2006yi}, the model for the axion mass differs since we do not account for the contribution from the QCD instantons to the axion mass.}.

In fact, the consistency of the relation between $m_a$, $M_{\rm susy}$, and $S$ in Eq.~\eqref{mamu} with the theoretical prior $M_{\rm susy}<M_{\rm Pl}$ leads to $S \sim 200$ to within an uncertainty of about $\sim 50$. The role of the data is that of further shrinking the error bar, leading to the result $S=198\pm28$ we have reported. The role of the data in further improving the determination of $S$ relies on the fact that the upper limits on $\omega$ and $r$ squeeze the allowed region of parameter space from opposite directions, see Eqs.~\eqref{eq:energydensity_str}-\eqref{eq:tensor_to_scalar_ratio_str}, with the limit on $\omega$ excluding the tail at low values of $S$, whereas the bounds on $r$ exclude higher values of $S$. 

\section{Discussions and conclusions}
\label{conclusions}

In this work, we have for the first time attempted to constrain fundamental parameters describing the string axiverse, using cosmology. We have in particular focused on the SUSY breaking scale $M_{\rm susy}$ and the dimensionless action of the string instantons responsible for breaking the axion shift symmetry, $S$. Imposing string-inspired uniform priors on $\log_{10}M_{\rm susy}$ and $S$, and using current observational bounds on the tensor-to-scalar ratio, the primordial isocurvature fraction, and the energy density of dark matter, we have performed a Bayesian inference analysis to constrain the parameters characterising the string axiverse.

We have found that $M_{\rm susy}$ is essentially unconstrained (due to a strong parameter degeneracy with $S$), while we find $S=198\pm28$ at 68\%~C.L. which is partly due to consistency with the theoretical priors as discussed in the previous Section. When interpreting these results in terms of the more familiar axion mass and decay constant, $m_a$ and $f_a$, through Eqs.~(\ref{mamu},\ref{fs}), we find that the lower range of $m_a$ is somewhat artificially favoured, with $\log_{10}(m_a/{\rm eV}) = -21.5^{+1.3}_{-2.3}$ and $f_a = (5.1^{+0.5}_{-0.9})\times 10^{-3}\,\MP$ at 68\%~C.L.. These values lie within the ultra-light axion range, see the solid black line in Fig.~\ref{fig:Plotaxiverse_logmass}.

In order to break the $M_{\rm susy}$-$S$ degeneracy, for purely pedagogical purposes we explore the impact of fixing $S = 198$ or $S = 153$, respectively corresponding to the central value and $2\sigma$ lower limit from the first analysis. This respectively gives $\log_{10}(\MS/{\rm TeV}) = 7.8^{+1.4}_{-2.3}$ or $\MS \lesssim 8 \times 10^3\,{\rm TeV}$. The latter bound is particularly interesting, since it suggests that such a scenario could hypothetically be probed in a future 100 TeV collider~\cite{Arkani-Hamed:2015vfh}.

We have then explored the impact of our choice of string-inspired priors on the inferred values of $m_a$ and $f_a$. We have done so by performing a different analysis where we directly sample the $m_a$-$f_a$ parameter space, with flat priors in $\log_{10}m_a$ and $\log_{10}f_a$. We have found $\log_{10} \(m_a/{\rm eV}\) = -11.1^{+11.5}_{-5.9}$ and $\log_{10} \(f_a/\MP\) = -5.5^{+1.8}_{-3.0}$ at 68\%~C.L., which mildly disfavours the lighter end of the axion mass spectrum and is consistent with the value of the mass inferred by Klaer and Moore assuming that the axion makes up all the dark matter~\cite{Klaer:2017ond}, see the dashed black line in Fig.~\ref{fig:Plotaxiverse_logmass}. Clearly, the difference between the two analyses is at least partly due to the different choices of priors, which however we argued arise quite naturally when considering the two different theoretical or phenomenological approaches. In the string-inspired analysis, our results are consistent with theoretical works suggesting that axions originating from string theory preferentially populate the low-mass end of the string axiverse~\cite{Arvanitaki:2009fg}, although we find that only a relatively narrow band around the value $m_a \approx 10^{-22}\,$eV is consistent with a combination of theoretical priors and data, see the solid black line in Fig.~\ref{fig:Plotaxiverse_logmass}. Ultra-light axions with mass of order $10^{-22}\,$eV have been ruled out as the dark matter component by various methods: matching the profile of the Fornax dwarf~\cite{Schive:2014dra}, Jeans analysis on numerically simulated dark matter halos~\cite{Schive:2015kza, Chen:2016unw, Schive:2017biq}, Lyman-$\alpha$ forest~\cite{Kobayashi:2017jcf, Desjacques:2017fmf}.

Recently, the work in Ref.~\cite{Hoof:2018ieb} appeared where a global fit that uses a Bayesian analysis technique to explore the parameter space of the QCD axion based on the code GAMBIT~\cite{Athron:2017ard} and its module DarkBit~\cite{Workgroup:2017lvb} is presented. In particular, Ref.~\cite{Hoof:2018ieb} considers the scenario in which the Peccei-Quinn symmetry breaks during a period of inflation although, at this stage, a fit that accounts for the inflation module has not been implemented yet so the analysis reported does not take into account bounds from isocurvature fluctuations. A further difference with our analysis lies in the relation $m_a f_a = \Lambda_{\rm QCD}^2 = \(75.5{\rm \,MeV}\)^2$ which holds for the QCD axion. The likelihood analysis with GAMBIT takes into account various results, including laboratory experiments from light-shining-through-wall, helioscopes, and cavity searches, as well as astrophysical observations for the distortions of gamma-ray spectra, supernovae, horizontal branch stars and the hint from the cooling of white dwarfs. Given these differences, the marginalised posterior distribution obtained in Ref.~\cite{Hoof:2018ieb} when demanding that the totality of dark matter is in axions gives the range $0.12{\rm \,\mu eV} \leq m_a \leq 0.15{\rm \,meV}$ at the 95\% equal-tailed confidence interval. Their results are also dependent on the choice for the prior on $f_a$, as well as those on the axion-to-electron coupling and the anomaly ratio.

In Refs.~\cite{Hlozek:2014lca, Hlozek:2017zzf}, a Bayesian technique using a MCMC sampling has been performed using the cosmological parameter space $\{A_s, n_s, \tau, \Omega_bh^2, h , \Omega_ch^2, \Omega_ah^2, m_a, H_I\}$, where the amplitude of scalar fluctuations $A_s$ and the scalar spectral index $n_s$ are defined at the pivotal scale $k_0 = 0.05{\rm \,Mpc}^{-1}$, $\tau$ is the optical depth to reionisation, $\Omega_bh^2$ is the baryon density, and $\Omega_ch^2$ is the density in dark matter other than axions, so that $\Omega_{\rm CDM} = \Omega_a + \Omega_c$. The analysis in the paper is performed using the code \texttt{axionCAMB} code~\cite{Hlozek:2014lca} and reveals no evidence for an axion component in the mass range $10^{-33}{\rm \, eV} \leq m_a \leq 10^{-24}{\rm \, eV}$, extending similar previous findings~\cite{Amendola:2005ad}.

In conclusion, in this work we provide for the first time a new window into the string axiverse and the SUSY breaking scale from cosmology. There are plenty of avenues for follow-up work. Firstly, it would be worth performing a full-fledged analysis carefully taking into account precision cosmology data, such as data from the CMB temperature and polarisation anisotropy spectra, as well as from galaxy surveys (galaxy power spectrum and/or Baryon Acoustic Oscillations) and weak lensing surveys. Such an analysis requires a suitable modification of the \texttt{axionCAMB} code~\cite{Hlozek:2014lca}, and would be especially relevant given that our analysis of the string-inspired axion constrains the axion mass to reside within a narrow region where the axion affects the matter power spectrum at small, but still cosmologically relevant, scales. In addition, it would be useful to revisit the theoretical uncertainties entering Eqs.~(\ref{mamu},\ref{fs}) relating $M_{\rm susy}$ and $S$ to $m_a$ and $f_a$, in order to strengthen our analysis. We plan to return to these and other issues in future work.

\begin{acknowledgments}
We thank Paolo di Vecchia for a careful reading of the manuscript and useful explanations on string compactifications. We thank Martina Gerbino, Massimiliano Lattanzi, and Frank Wilczek for useful suggestions that improved the paper. We acknowledge support by the Vetenskapsr\r{a}det (Swedish Research Council) through contract No. 638-2013-8993 and the Oskar Klein Centre for Cosmoparticle Physics.
\end{acknowledgments}


\begin{thebibliography}{188}%
\makeatletter
\providecommand \@ifxundefined [1]{%
 \@ifx{#1\undefined}
}%
\providecommand \@ifnum [1]{%
 \ifnum #1\expandafter \@firstoftwo
 \else \expandafter \@secondoftwo
 \fi
}%
\providecommand \@ifx [1]{%
 \ifx #1\expandafter \@firstoftwo
 \else \expandafter \@secondoftwo
 \fi
}%
\providecommand \natexlab [1]{#1}%
\providecommand \enquote  [1]{``#1''}%
\providecommand \bibnamefont  [1]{#1}%
\providecommand \bibfnamefont [1]{#1}%
\providecommand \citenamefont [1]{#1}%
\providecommand \href@noop [0]{\@secondoftwo}%
\providecommand \href [0]{\begingroup \@sanitize@url \@href}%
\providecommand \@href[1]{\@@startlink{#1}\@@href}%
\providecommand \@@href[1]{\endgroup#1\@@endlink}%
\providecommand \@sanitize@url [0]{\catcode `\\12\catcode `\$12\catcode
  `\&12\catcode `\#12\catcode `\^12\catcode `\_12\catcode `\%12\relax}%
\providecommand \@@startlink[1]{}%
\providecommand \@@endlink[0]{}%
\providecommand \url  [0]{\begingroup\@sanitize@url \@url }%
\providecommand \@url [1]{\endgroup\@href {#1}{\urlprefix }}%
\providecommand \urlprefix  [0]{URL }%
\providecommand \Eprint [0]{\href }%
\providecommand \doibase [0]{http://dx.doi.org/}%
\providecommand \selectlanguage [0]{\@gobble}%
\providecommand \bibinfo  [0]{\@secondoftwo}%
\providecommand \bibfield  [0]{\@secondoftwo}%
\providecommand \translation [1]{[#1]}%
\providecommand \BibitemOpen [0]{}%
\providecommand \bibitemStop [0]{}%
\providecommand \bibitemNoStop [0]{.\EOS\space}%
\providecommand \EOS [0]{\spacefactor3000\relax}%
\providecommand \BibitemShut  [1]{\csname bibitem#1\endcsname}%
\let\auto@bib@innerbib\@empty
\bibitem [{\citenamefont {Belavin}\ \emph {et~al.}(1975)\citenamefont
  {Belavin}, \citenamefont {Polyakov}, \citenamefont {Schwartz},\ and\
  \citenamefont {Tyupkin}}]{Belavin:1975fg}%
  \BibitemOpen
  \bibfield  {author} {\bibinfo {author} {\bibfnamefont {A.~A.}\ \bibnamefont
  {Belavin}}, \bibinfo {author} {\bibfnamefont {A.~M.}\ \bibnamefont
  {Polyakov}}, \bibinfo {author} {\bibfnamefont {A.~S.}\ \bibnamefont
  {Schwartz}}, \ and\ \bibinfo {author} {\bibfnamefont {{\relax Yu}.~S.}\
  \bibnamefont {Tyupkin}},\ }\href {\doibase 10.1016/0370-2693(75)90163-X}
  {\bibfield  {journal} {\bibinfo  {journal} {Phys. Lett.}\ }\textbf {\bibinfo
  {volume} {B59}},\ \bibinfo {pages} {85} (\bibinfo {year} {1975})},\ \bibinfo
  {note} {[,350(1975)]}\BibitemShut {NoStop}%
\bibitem [{\citenamefont {'t~Hooft}(1976)}]{tHooft:1976rip}%
  \BibitemOpen
  \bibfield  {author} {\bibinfo {author} {\bibfnamefont {G.}~\bibnamefont
  {'t~Hooft}},\ }\href {\doibase 10.1103/PhysRevLett.37.8} {\bibfield
  {journal} {\bibinfo  {journal} {Phys. Rev. Lett.}\ }\textbf {\bibinfo
  {volume} {37}},\ \bibinfo {pages} {8} (\bibinfo {year} {1976})},\ \bibinfo
  {note} {[,226(1976)]}\BibitemShut {NoStop}%
\bibitem [{\citenamefont {Jackiw}\ and\ \citenamefont
  {Rebbi}(1976)}]{Jackiw:1976pf}%
  \BibitemOpen
  \bibfield  {author} {\bibinfo {author} {\bibfnamefont {R.}~\bibnamefont
  {Jackiw}}\ and\ \bibinfo {author} {\bibfnamefont {C.}~\bibnamefont {Rebbi}},\
  }\href@noop {} {\bibfield  {journal} {\bibinfo  {journal} {Phys. Rev. Lett.}\
  }\textbf {\bibinfo {volume} {37}},\ \bibinfo {pages} {172} (\bibinfo {year}
  {1976})}\BibitemShut {NoStop}%
\bibitem [{\citenamefont {Callan}\ \emph {et~al.}(1976)\citenamefont {Callan},
  \citenamefont {Dashen},\ and\ \citenamefont {Gross}}]{Callan:1976je}%
  \BibitemOpen
  \bibfield  {author} {\bibinfo {author} {\bibfnamefont {C.~G.}\ \bibnamefont
  {Callan}, \bibfnamefont {Jr.}}, \bibinfo {author} {\bibfnamefont {R.~F.}\
  \bibnamefont {Dashen}}, \ and\ \bibinfo {author} {\bibfnamefont {D.~J.}\
  \bibnamefont {Gross}},\ }\href {\doibase 10.1016/0370-2693(76)90277-X}
  {\bibfield  {journal} {\bibinfo  {journal} {Phys. Lett.}\ }\textbf {\bibinfo
  {volume} {B63}},\ \bibinfo {pages} {334} (\bibinfo {year} {1976})},\ \bibinfo
  {note} {[,357(1976)]}\BibitemShut {NoStop}%
\bibitem [{\citenamefont {Peccei}\ and\ \citenamefont
  {Quinn}(1977{\natexlab{a}})}]{Peccei:1977hh}%
  \BibitemOpen
  \bibfield  {author} {\bibinfo {author} {\bibfnamefont {R.~D.}\ \bibnamefont
  {Peccei}}\ and\ \bibinfo {author} {\bibfnamefont {H.~R.}\ \bibnamefont
  {Quinn}},\ }\href {\doibase 10.1103/PhysRevLett.38.1440} {\bibfield
  {journal} {\bibinfo  {journal} {Phys. Rev. Lett.}\ }\textbf {\bibinfo
  {volume} {38}},\ \bibinfo {pages} {1440} (\bibinfo {year}
  {1977}{\natexlab{a}})}\BibitemShut {NoStop}%
\bibitem [{\citenamefont {Peccei}\ and\ \citenamefont
  {Quinn}(1977{\natexlab{b}})}]{Peccei:1977ur}%
  \BibitemOpen
  \bibfield  {author} {\bibinfo {author} {\bibfnamefont {R.~D.}\ \bibnamefont
  {Peccei}}\ and\ \bibinfo {author} {\bibfnamefont {H.~R.}\ \bibnamefont
  {Quinn}},\ }\href {\doibase 10.1103/PhysRevD.16.1791} {\bibfield  {journal}
  {\bibinfo  {journal} {Phys. Rev.}\ }\textbf {\bibinfo {volume} {D16}},\
  \bibinfo {pages} {1791} (\bibinfo {year} {1977}{\natexlab{b}})}\BibitemShut
  {NoStop}%
\bibitem [{\citenamefont {Wilczek}(1978)}]{Wilczek:1977pj}%
  \BibitemOpen
  \bibfield  {author} {\bibinfo {author} {\bibfnamefont {F.}~\bibnamefont
  {Wilczek}},\ }\href {\doibase 10.1103/PhysRevLett.40.279} {\bibfield
  {journal} {\bibinfo  {journal} {Phys. Rev. Lett.}\ }\textbf {\bibinfo
  {volume} {40}},\ \bibinfo {pages} {279} (\bibinfo {year} {1978})}\BibitemShut
  {NoStop}%
\bibitem [{\citenamefont {Weinberg}(1978)}]{Weinberg:1977ma}%
  \BibitemOpen
  \bibfield  {author} {\bibinfo {author} {\bibfnamefont {S.}~\bibnamefont
  {Weinberg}},\ }\href {\doibase 10.1103/PhysRevLett.40.223} {\bibfield
  {journal} {\bibinfo  {journal} {Phys. Rev. Lett.}\ }\textbf {\bibinfo
  {volume} {40}},\ \bibinfo {pages} {223} (\bibinfo {year} {1978})}\BibitemShut
  {NoStop}%
\bibitem [{\citenamefont {Kim}(1979)}]{Kim:1979if}%
  \BibitemOpen
  \bibfield  {author} {\bibinfo {author} {\bibfnamefont {J.~E.}\ \bibnamefont
  {Kim}},\ }\href {\doibase 10.1103/PhysRevLett.43.103} {\bibfield  {journal}
  {\bibinfo  {journal} {Phys. Rev. Lett.}\ }\textbf {\bibinfo {volume} {43}},\
  \bibinfo {pages} {103} (\bibinfo {year} {1979})}\BibitemShut {NoStop}%
\bibitem [{\citenamefont {Dine}\ \emph {et~al.}(1981)\citenamefont {Dine},
  \citenamefont {Fischler},\ and\ \citenamefont {Srednicki}}]{Dine:1981rt}%
  \BibitemOpen
  \bibfield  {author} {\bibinfo {author} {\bibfnamefont {M.}~\bibnamefont
  {Dine}}, \bibinfo {author} {\bibfnamefont {W.}~\bibnamefont {Fischler}}, \
  and\ \bibinfo {author} {\bibfnamefont {M.}~\bibnamefont {Srednicki}},\ }\href
  {\doibase 10.1016/0370-2693(81)90590-6} {\bibfield  {journal} {\bibinfo
  {journal} {Phys. Lett.}\ }\textbf {\bibinfo {volume} {B104}},\ \bibinfo
  {pages} {199} (\bibinfo {year} {1981})}\BibitemShut {NoStop}%
\bibitem [{\citenamefont {Zhitnitsky}(1980)}]{Zhitnitsky:1980tq}%
  \BibitemOpen
  \bibfield  {author} {\bibinfo {author} {\bibfnamefont {A.~R.}\ \bibnamefont
  {Zhitnitsky}},\ }\href@noop {} {\bibfield  {journal} {\bibinfo  {journal}
  {Sov. J. Nucl. Phys.}\ }\textbf {\bibinfo {volume} {31}},\ \bibinfo {pages}
  {260} (\bibinfo {year} {1980})}\BibitemShut {NoStop}%
\bibitem [{\citenamefont {Shifman}\ \emph {et~al.}(1980)\citenamefont
  {Shifman}, \citenamefont {Vainshtein},\ and\ \citenamefont
  {Zakharov}}]{Shifman1980493}%
  \BibitemOpen
  \bibfield  {author} {\bibinfo {author} {\bibfnamefont {M.}~\bibnamefont
  {Shifman}}, \bibinfo {author} {\bibfnamefont {A.}~\bibnamefont {Vainshtein}},
  \ and\ \bibinfo {author} {\bibfnamefont {V.}~\bibnamefont {Zakharov}},\
  }\href {\doibase http://dx.doi.org/10.1016/0550-3213(80)90209-6} {\bibfield
  {journal} {\bibinfo  {journal} {Nuclear Physics B}\ }\textbf {\bibinfo
  {volume} {166}},\ \bibinfo {pages} {493 } (\bibinfo {year}
  {1980})}\BibitemShut {NoStop}%
\bibitem [{\citenamefont {Gross}\ \emph {et~al.}(1981)\citenamefont {Gross},
  \citenamefont {Pisarski},\ and\ \citenamefont {Yaffe}}]{Gross:53.43}%
  \BibitemOpen
  \bibfield  {author} {\bibinfo {author} {\bibfnamefont {D.~J.}\ \bibnamefont
  {Gross}}, \bibinfo {author} {\bibfnamefont {R.~D.}\ \bibnamefont {Pisarski}},
  \ and\ \bibinfo {author} {\bibfnamefont {L.~G.}\ \bibnamefont {Yaffe}},\
  }\href {\doibase 10.1103/RevModPhys.53.43} {\bibfield  {journal} {\bibinfo
  {journal} {Rev. Mod. Phys.}\ }\textbf {\bibinfo {volume} {53}},\ \bibinfo
  {pages} {43} (\bibinfo {year} {1981})}\BibitemShut {NoStop}%
\bibitem [{\citenamefont {Preskill}\ \emph {et~al.}(1983)\citenamefont
  {Preskill}, \citenamefont {Wise},\ and\ \citenamefont
  {Wilczek}}]{Preskill:1982cy}%
  \BibitemOpen
  \bibfield  {author} {\bibinfo {author} {\bibfnamefont {J.}~\bibnamefont
  {Preskill}}, \bibinfo {author} {\bibfnamefont {M.~B.}\ \bibnamefont {Wise}},
  \ and\ \bibinfo {author} {\bibfnamefont {F.}~\bibnamefont {Wilczek}},\ }\href
  {\doibase 10.1016/0370-2693(83)90637-8} {\bibfield  {journal} {\bibinfo
  {journal} {Phys. Lett.}\ }\textbf {\bibinfo {volume} {B120}},\ \bibinfo
  {pages} {127} (\bibinfo {year} {1983})}\BibitemShut {NoStop}%
\bibitem [{\citenamefont {Abbott}\ and\ \citenamefont
  {Sikivie}(1983)}]{Abbott:1982af}%
  \BibitemOpen
  \bibfield  {author} {\bibinfo {author} {\bibfnamefont {L.~F.}\ \bibnamefont
  {Abbott}}\ and\ \bibinfo {author} {\bibfnamefont {P.}~\bibnamefont
  {Sikivie}},\ }\href {\doibase 10.1016/0370-2693(83)90638-X} {\bibfield
  {journal} {\bibinfo  {journal} {Phys. Lett.}\ }\textbf {\bibinfo {volume}
  {B120}},\ \bibinfo {pages} {133} (\bibinfo {year} {1983})}\BibitemShut
  {NoStop}%
\bibitem [{\citenamefont {Dine}\ and\ \citenamefont
  {Fischler}(1983)}]{Dine:1982ah}%
  \BibitemOpen
  \bibfield  {author} {\bibinfo {author} {\bibfnamefont {M.}~\bibnamefont
  {Dine}}\ and\ \bibinfo {author} {\bibfnamefont {W.}~\bibnamefont
  {Fischler}},\ }\href {\doibase 10.1016/0370-2693(83)90639-1} {\bibfield
  {journal} {\bibinfo  {journal} {Phys. Lett.}\ }\textbf {\bibinfo {volume}
  {B120}},\ \bibinfo {pages} {137} (\bibinfo {year} {1983})}\BibitemShut
  {NoStop}%
\bibitem [{\citenamefont {Linde}(1988)}]{Linde:1987bx}%
  \BibitemOpen
  \bibfield  {author} {\bibinfo {author} {\bibfnamefont {A.~D.}\ \bibnamefont
  {Linde}},\ }\href {\doibase 10.1016/0370-2693(88)90597-7} {\bibfield
  {journal} {\bibinfo  {journal} {Phys. Lett.}\ }\textbf {\bibinfo {volume}
  {B201}},\ \bibinfo {pages} {437} (\bibinfo {year} {1988})}\BibitemShut
  {NoStop}%
\bibitem [{\citenamefont {Linde}(1991)}]{Linde:1991km}%
  \BibitemOpen
  \bibfield  {author} {\bibinfo {author} {\bibfnamefont {A.~D.}\ \bibnamefont
  {Linde}},\ }\href {\doibase 10.1016/0370-2693(91)90130-I} {\bibfield
  {journal} {\bibinfo  {journal} {Phys. Lett.}\ }\textbf {\bibinfo {volume}
  {B259}},\ \bibinfo {pages} {38} (\bibinfo {year} {1991})}\BibitemShut
  {NoStop}%
\bibitem [{\citenamefont {Turner}\ and\ \citenamefont
  {Wilczek}(1991)}]{Turner:1991}%
  \BibitemOpen
  \bibfield  {author} {\bibinfo {author} {\bibfnamefont {M.~S.}\ \bibnamefont
  {Turner}}\ and\ \bibinfo {author} {\bibfnamefont {F.}~\bibnamefont
  {Wilczek}},\ }\href {\doibase 10.1103/PhysRevLett.66.5} {\bibfield  {journal}
  {\bibinfo  {journal} {Phys. Rev. Lett.}\ }\textbf {\bibinfo {volume} {66}},\
  \bibinfo {pages} {5} (\bibinfo {year} {1991})}\BibitemShut {NoStop}%
\bibitem [{\citenamefont {Wilczek}(2004)}]{Wilczek:2004cr}%
  \BibitemOpen
  \bibfield  {author} {\bibinfo {author} {\bibfnamefont {F.}~\bibnamefont
  {Wilczek}},\ }\href@noop {} {\  (\bibinfo {year} {2004})},\ \Eprint
  {http://arxiv.org/abs/hep-ph/0408167} {arXiv:hep-ph/0408167 [hep-ph]}
  \BibitemShut {NoStop}%
\bibitem [{\citenamefont {Tegmark}\ \emph {et~al.}(2006)\citenamefont
  {Tegmark}, \citenamefont {Aguirre}, \citenamefont {Rees},\ and\ \citenamefont
  {Wilczek}}]{Tegmark:2005dy}%
  \BibitemOpen
  \bibfield  {author} {\bibinfo {author} {\bibfnamefont {M.}~\bibnamefont
  {Tegmark}}, \bibinfo {author} {\bibfnamefont {A.}~\bibnamefont {Aguirre}},
  \bibinfo {author} {\bibfnamefont {M.}~\bibnamefont {Rees}}, \ and\ \bibinfo
  {author} {\bibfnamefont {F.}~\bibnamefont {Wilczek}},\ }\href {\doibase
  10.1103/PhysRevD.73.023505} {\bibfield  {journal} {\bibinfo  {journal} {Phys.
  Rev.}\ }\textbf {\bibinfo {volume} {D73}},\ \bibinfo {pages} {023505}
  (\bibinfo {year} {2006})},\ \Eprint {http://arxiv.org/abs/astro-ph/0511774}
  {arXiv:astro-ph/0511774 [astro-ph]} \BibitemShut {NoStop}%
\bibitem [{\citenamefont {Hertzberg}\ \emph {et~al.}(2008)\citenamefont
  {Hertzberg}, \citenamefont {Tegmark},\ and\ \citenamefont
  {Wilczek}}]{Hertzberg:2008wr}%
  \BibitemOpen
  \bibfield  {author} {\bibinfo {author} {\bibfnamefont {M.~P.}\ \bibnamefont
  {Hertzberg}}, \bibinfo {author} {\bibfnamefont {M.}~\bibnamefont {Tegmark}},
  \ and\ \bibinfo {author} {\bibfnamefont {F.}~\bibnamefont {Wilczek}},\ }\href
  {\doibase 10.1103/PhysRevD.78.083507} {\bibfield  {journal} {\bibinfo
  {journal} {Phys. Rev.}\ }\textbf {\bibinfo {volume} {D78}},\ \bibinfo {pages}
  {083507} (\bibinfo {year} {2008})},\ \Eprint {http://arxiv.org/abs/0807.1726}
  {arXiv:0807.1726 [astro-ph]} \BibitemShut {NoStop}%
\bibitem [{\citenamefont {Freivogel}(2010)}]{Freivogel:2008qc}%
  \BibitemOpen
  \bibfield  {author} {\bibinfo {author} {\bibfnamefont {B.}~\bibnamefont
  {Freivogel}},\ }\href {\doibase 10.1088/1475-7516/2010/03/021} {\bibfield
  {journal} {\bibinfo  {journal} {JCAP}\ }\textbf {\bibinfo {volume} {1003}},\
  \bibinfo {pages} {021} (\bibinfo {year} {2010})},\ \Eprint
  {http://arxiv.org/abs/0810.0703} {arXiv:0810.0703 [hep-th]} \BibitemShut
  {NoStop}%
\bibitem [{\citenamefont {Mack}(2011)}]{Mack:2009hv}%
  \BibitemOpen
  \bibfield  {author} {\bibinfo {author} {\bibfnamefont {K.~J.}\ \bibnamefont
  {Mack}},\ }\href {\doibase 10.1088/1475-7516/2011/07/021} {\bibfield
  {journal} {\bibinfo  {journal} {JCAP}\ }\textbf {\bibinfo {volume} {1107}},\
  \bibinfo {pages} {021} (\bibinfo {year} {2011})},\ \Eprint
  {http://arxiv.org/abs/0911.0421} {arXiv:0911.0421 [astro-ph.CO]} \BibitemShut
  {NoStop}%
\bibitem [{\citenamefont {Visinelli}\ and\ \citenamefont
  {Gondolo}(2009)}]{Visinelli:2009zm}%
  \BibitemOpen
  \bibfield  {author} {\bibinfo {author} {\bibfnamefont {L.}~\bibnamefont
  {Visinelli}}\ and\ \bibinfo {author} {\bibfnamefont {P.}~\bibnamefont
  {Gondolo}},\ }\href {\doibase 10.1103/PhysRevD.80.035024} {\bibfield
  {journal} {\bibinfo  {journal} {Phys. Rev.}\ }\textbf {\bibinfo {volume}
  {D80}},\ \bibinfo {pages} {035024} (\bibinfo {year} {2009})},\ \Eprint
  {http://arxiv.org/abs/0903.4377} {arXiv:0903.4377 [astro-ph.CO]} \BibitemShut
  {NoStop}%
\bibitem [{\citenamefont {Visinelli}\ and\ \citenamefont
  {Gondolo}(2010)}]{Visinelli:2009kt}%
  \BibitemOpen
  \bibfield  {author} {\bibinfo {author} {\bibfnamefont {L.}~\bibnamefont
  {Visinelli}}\ and\ \bibinfo {author} {\bibfnamefont {P.}~\bibnamefont
  {Gondolo}},\ }\href {\doibase 10.1103/PhysRevD.81.063508} {\bibfield
  {journal} {\bibinfo  {journal} {Phys. Rev.}\ }\textbf {\bibinfo {volume}
  {D81}},\ \bibinfo {pages} {063508} (\bibinfo {year} {2010})},\ \Eprint
  {http://arxiv.org/abs/0912.0015} {arXiv:0912.0015 [astro-ph.CO]} \BibitemShut
  {NoStop}%
\bibitem [{\citenamefont {Acharya}\ \emph {et~al.}(2010)\citenamefont
  {Acharya}, \citenamefont {Bobkov},\ and\ \citenamefont
  {Kumar}}]{Acharya:2010zx}%
  \BibitemOpen
  \bibfield  {author} {\bibinfo {author} {\bibfnamefont {B.~S.}\ \bibnamefont
  {Acharya}}, \bibinfo {author} {\bibfnamefont {K.}~\bibnamefont {Bobkov}}, \
  and\ \bibinfo {author} {\bibfnamefont {P.}~\bibnamefont {Kumar}},\ }\href
  {\doibase 10.1007/JHEP11(2010)105} {\bibfield  {journal} {\bibinfo  {journal}
  {JHEP}\ }\textbf {\bibinfo {volume} {11}},\ \bibinfo {pages} {105} (\bibinfo
  {year} {2010})},\ \Eprint {http://arxiv.org/abs/1004.5138} {arXiv:1004.5138
  [hep-th]} \BibitemShut {NoStop}%
\bibitem [{\citenamefont {Gondolo}\ and\ \citenamefont
  {Visinelli}(2014)}]{Visinelli:2014twa}%
  \BibitemOpen
  \bibfield  {author} {\bibinfo {author} {\bibfnamefont {P.}~\bibnamefont
  {Gondolo}}\ and\ \bibinfo {author} {\bibfnamefont {L.}~\bibnamefont
  {Visinelli}},\ }\href {\doibase 10.1103/PhysRevLett.113.011802} {\bibfield
  {journal} {\bibinfo  {journal} {Phys. Rev. Lett.}\ }\textbf {\bibinfo
  {volume} {113}},\ \bibinfo {pages} {011802} (\bibinfo {year} {2014})},\
  \Eprint {http://arxiv.org/abs/1403.4594} {arXiv:1403.4594 [hep-ph]}
  \BibitemShut {NoStop}%
\bibitem [{\citenamefont {Visinelli}\ \emph
  {et~al.}(2018{\natexlab{a}})\citenamefont {Visinelli}, \citenamefont {Baum},
  \citenamefont {Redondo}, \citenamefont {Freese},\ and\ \citenamefont
  {Wilczek}}]{Visinelli:2017ooc}%
  \BibitemOpen
  \bibfield  {author} {\bibinfo {author} {\bibfnamefont {L.}~\bibnamefont
  {Visinelli}}, \bibinfo {author} {\bibfnamefont {S.}~\bibnamefont {Baum}},
  \bibinfo {author} {\bibfnamefont {J.}~\bibnamefont {Redondo}}, \bibinfo
  {author} {\bibfnamefont {K.}~\bibnamefont {Freese}}, \ and\ \bibinfo {author}
  {\bibfnamefont {F.}~\bibnamefont {Wilczek}},\ }\href {\doibase
  10.1016/j.physletb.2017.12.010} {\bibfield  {journal} {\bibinfo  {journal}
  {Phys. Lett.}\ }\textbf {\bibinfo {volume} {B777}},\ \bibinfo {pages} {64}
  (\bibinfo {year} {2018}{\natexlab{a}})},\ \Eprint
  {http://arxiv.org/abs/1710.08910} {arXiv:1710.08910 [astro-ph.CO]}
  \BibitemShut {NoStop}%
\bibitem [{\citenamefont {Visinelli}\ and\ \citenamefont {Ter{\c
  c}as}(2018)}]{Visinelli:2018zif}%
  \BibitemOpen
  \bibfield  {author} {\bibinfo {author} {\bibfnamefont {L.}~\bibnamefont
  {Visinelli}}\ and\ \bibinfo {author} {\bibfnamefont {H.}~\bibnamefont {Ter{\c
  c}as}},\ }\href@noop {} {\  (\bibinfo {year} {2018})},\ \Eprint
  {http://arxiv.org/abs/1807.06828} {arXiv:1807.06828 [hep-ph]} \BibitemShut
  {NoStop}%
\bibitem [{\citenamefont {Visinelli}\ and\ \citenamefont
  {Redondo}(2018)}]{Visinelli:2018wza}%
  \BibitemOpen
  \bibfield  {author} {\bibinfo {author} {\bibfnamefont {L.}~\bibnamefont
  {Visinelli}}\ and\ \bibinfo {author} {\bibfnamefont {J.}~\bibnamefont
  {Redondo}},\ }\href@noop {} {\  (\bibinfo {year} {2018})},\ \Eprint
  {http://arxiv.org/abs/1808.01879} {arXiv:1808.01879 [astro-ph.CO]}
  \BibitemShut {NoStop}%
\bibitem [{\citenamefont {Choi}\ \emph {et~al.}(2007)\citenamefont {Choi},
  \citenamefont {Kim},\ and\ \citenamefont {Kim}}]{Choi:2006qj}%
  \BibitemOpen
  \bibfield  {author} {\bibinfo {author} {\bibfnamefont {K.-S.}\ \bibnamefont
  {Choi}}, \bibinfo {author} {\bibfnamefont {I.-W.}\ \bibnamefont {Kim}}, \
  and\ \bibinfo {author} {\bibfnamefont {J.~E.}\ \bibnamefont {Kim}},\ }\href
  {\doibase 10.1088/1126-6708/2007/03/116} {\bibfield  {journal} {\bibinfo
  {journal} {JHEP}\ }\textbf {\bibinfo {volume} {03}},\ \bibinfo {pages} {116}
  (\bibinfo {year} {2007})},\ \Eprint {http://arxiv.org/abs/hep-ph/0612107}
  {arXiv:hep-ph/0612107 [hep-ph]} \BibitemShut {NoStop}%
\bibitem [{\citenamefont {Choi}\ \emph {et~al.}(2009)\citenamefont {Choi},
  \citenamefont {Nilles}, \citenamefont {Ramos-Sanchez},\ and\ \citenamefont
  {Vaudrevange}}]{Choi:2009jt}%
  \BibitemOpen
  \bibfield  {author} {\bibinfo {author} {\bibfnamefont {K.-S.}\ \bibnamefont
  {Choi}}, \bibinfo {author} {\bibfnamefont {H.~P.}\ \bibnamefont {Nilles}},
  \bibinfo {author} {\bibfnamefont {S.}~\bibnamefont {Ramos-Sanchez}}, \ and\
  \bibinfo {author} {\bibfnamefont {P.~K.~S.}\ \bibnamefont {Vaudrevange}},\
  }\href {\doibase 10.1016/j.physletb.2009.04.028} {\bibfield  {journal}
  {\bibinfo  {journal} {Phys. Lett.}\ }\textbf {\bibinfo {volume} {B675}},\
  \bibinfo {pages} {381} (\bibinfo {year} {2009})},\ \Eprint
  {http://arxiv.org/abs/0902.3070} {arXiv:0902.3070 [hep-th]} \BibitemShut
  {NoStop}%
\bibitem [{\citenamefont {Dias}\ \emph {et~al.}(2014)\citenamefont {Dias},
  \citenamefont {Machado}, \citenamefont {Nishi}, \citenamefont {Ringwald},\
  and\ \citenamefont {Vaudrevange}}]{Dias:2014osa}%
  \BibitemOpen
  \bibfield  {author} {\bibinfo {author} {\bibfnamefont {A.~G.}\ \bibnamefont
  {Dias}}, \bibinfo {author} {\bibfnamefont {A.~C.~B.}\ \bibnamefont
  {Machado}}, \bibinfo {author} {\bibfnamefont {C.~C.}\ \bibnamefont {Nishi}},
  \bibinfo {author} {\bibfnamefont {A.}~\bibnamefont {Ringwald}}, \ and\
  \bibinfo {author} {\bibfnamefont {P.}~\bibnamefont {Vaudrevange}},\ }\href
  {\doibase 10.1007/JHEP06(2014)037} {\bibfield  {journal} {\bibinfo  {journal}
  {JHEP}\ }\textbf {\bibinfo {volume} {06}},\ \bibinfo {pages} {037} (\bibinfo
  {year} {2014})},\ \Eprint {http://arxiv.org/abs/1403.5760} {arXiv:1403.5760
  [hep-ph]} \BibitemShut {NoStop}%
\bibitem [{\citenamefont {Higaki}\ and\ \citenamefont
  {Takahashi}(2014)}]{Higaki:2014pja}%
  \BibitemOpen
  \bibfield  {author} {\bibinfo {author} {\bibfnamefont {T.}~\bibnamefont
  {Higaki}}\ and\ \bibinfo {author} {\bibfnamefont {F.}~\bibnamefont
  {Takahashi}},\ }\href {\doibase 10.1007/JHEP07(2014)074} {\bibfield
  {journal} {\bibinfo  {journal} {JHEP}\ }\textbf {\bibinfo {volume} {07}},\
  \bibinfo {pages} {074} (\bibinfo {year} {2014})},\ \Eprint
  {http://arxiv.org/abs/1404.6923} {arXiv:1404.6923 [hep-th]} \BibitemShut
  {NoStop}%
\bibitem [{\citenamefont {Kim}\ and\ \citenamefont
  {Marsh}(2016)}]{Kim:2015yna}%
  \BibitemOpen
  \bibfield  {author} {\bibinfo {author} {\bibfnamefont {J.~E.}\ \bibnamefont
  {Kim}}\ and\ \bibinfo {author} {\bibfnamefont {D.~J.~E.}\ \bibnamefont
  {Marsh}},\ }\href {\doibase 10.1103/PhysRevD.93.025027} {\bibfield  {journal}
  {\bibinfo  {journal} {Phys. Rev.}\ }\textbf {\bibinfo {volume} {D93}},\
  \bibinfo {pages} {025027} (\bibinfo {year} {2016})},\ \Eprint
  {http://arxiv.org/abs/1510.01701} {arXiv:1510.01701 [hep-ph]} \BibitemShut
  {NoStop}%
\bibitem [{\citenamefont {Redi}\ and\ \citenamefont
  {Sato}(2016)}]{Redi:2016esr}%
  \BibitemOpen
  \bibfield  {author} {\bibinfo {author} {\bibfnamefont {M.}~\bibnamefont
  {Redi}}\ and\ \bibinfo {author} {\bibfnamefont {R.}~\bibnamefont {Sato}},\
  }\href {\doibase 10.1007/JHEP05(2016)104} {\bibfield  {journal} {\bibinfo
  {journal} {JHEP}\ }\textbf {\bibinfo {volume} {05}},\ \bibinfo {pages} {104}
  (\bibinfo {year} {2016})},\ \Eprint {http://arxiv.org/abs/1602.05427}
  {arXiv:1602.05427 [hep-ph]} \BibitemShut {NoStop}%
\bibitem [{\citenamefont {Di~Luzio}\ \emph {et~al.}(2017)\citenamefont
  {Di~Luzio}, \citenamefont {Nardi},\ and\ \citenamefont
  {Ubaldi}}]{DiLuzio:2017tjx}%
  \BibitemOpen
  \bibfield  {author} {\bibinfo {author} {\bibfnamefont {L.}~\bibnamefont
  {Di~Luzio}}, \bibinfo {author} {\bibfnamefont {E.}~\bibnamefont {Nardi}}, \
  and\ \bibinfo {author} {\bibfnamefont {L.}~\bibnamefont {Ubaldi}},\
  }\href@noop {} {\  (\bibinfo {year} {2017})},\ \Eprint
  {http://arxiv.org/abs/1704.01122} {arXiv:1704.01122 [hep-ph]} \BibitemShut
  {NoStop}%
\bibitem [{\citenamefont {Witten}(1984)}]{Witten:1984}%
  \BibitemOpen
  \bibfield  {author} {\bibinfo {author} {\bibfnamefont {E.}~\bibnamefont
  {Witten}},\ }\href {\doibase https://doi.org/10.1016/0370-2693(84)90422-2}
  {\bibfield  {journal} {\bibinfo  {journal} {Physics Letters B}\ }\textbf
  {\bibinfo {volume} {149}},\ \bibinfo {pages} {351 } (\bibinfo {year}
  {1984})}\BibitemShut {NoStop}%
\bibitem [{\citenamefont {Fox}\ \emph {et~al.}(2004)\citenamefont {Fox},
  \citenamefont {Pierce},\ and\ \citenamefont {Thomas}}]{Fox:2004kb}%
  \BibitemOpen
  \bibfield  {author} {\bibinfo {author} {\bibfnamefont {P.}~\bibnamefont
  {Fox}}, \bibinfo {author} {\bibfnamefont {A.}~\bibnamefont {Pierce}}, \ and\
  \bibinfo {author} {\bibfnamefont {S.~D.}\ \bibnamefont {Thomas}},\
  }\href@noop {} {\  (\bibinfo {year} {2004})},\ \Eprint
  {http://arxiv.org/abs/hep-th/0409059} {arXiv:hep-th/0409059 [hep-th]}
  \BibitemShut {NoStop}%
\bibitem [{\citenamefont {Svrcek}\ and\ \citenamefont
  {Witten}(2006)}]{Svrcek:2006yi}%
  \BibitemOpen
  \bibfield  {author} {\bibinfo {author} {\bibfnamefont {P.}~\bibnamefont
  {Svrcek}}\ and\ \bibinfo {author} {\bibfnamefont {E.}~\bibnamefont
  {Witten}},\ }\href {\doibase 10.1088/1126-6708/2006/06/051} {\bibfield
  {journal} {\bibinfo  {journal} {JHEP}\ }\textbf {\bibinfo {volume} {06}},\
  \bibinfo {pages} {051} (\bibinfo {year} {2006})},\ \Eprint
  {http://arxiv.org/abs/hep-th/0605206} {arXiv:hep-th/0605206 [hep-th]}
  \BibitemShut {NoStop}%
\bibitem [{\citenamefont {Arvanitaki}\ \emph
  {et~al.}(2010{\natexlab{a}})\citenamefont {Arvanitaki}, \citenamefont
  {Dimopoulos}, \citenamefont {Dubovsky}, \citenamefont {Kaloper},\ and\
  \citenamefont {March-Russell}}]{Arvanitaki:2009fg}%
  \BibitemOpen
  \bibfield  {author} {\bibinfo {author} {\bibfnamefont {A.}~\bibnamefont
  {Arvanitaki}}, \bibinfo {author} {\bibfnamefont {S.}~\bibnamefont
  {Dimopoulos}}, \bibinfo {author} {\bibfnamefont {S.}~\bibnamefont
  {Dubovsky}}, \bibinfo {author} {\bibfnamefont {N.}~\bibnamefont {Kaloper}}, \
  and\ \bibinfo {author} {\bibfnamefont {J.}~\bibnamefont {March-Russell}},\
  }\href {\doibase 10.1103/PhysRevD.81.123530} {\bibfield  {journal} {\bibinfo
  {journal} {Phys. Rev.}\ }\textbf {\bibinfo {volume} {D81}},\ \bibinfo {pages}
  {123530} (\bibinfo {year} {2010}{\natexlab{a}})},\ \Eprint
  {http://arxiv.org/abs/0905.4720} {arXiv:0905.4720 [hep-th]} \BibitemShut
  {NoStop}%
\bibitem [{\citenamefont {Dine}\ \emph {et~al.}(2011)\citenamefont {Dine},
  \citenamefont {Festuccia}, \citenamefont {Kehayias},\ and\ \citenamefont
  {Wu}}]{Dine:2010cr}%
  \BibitemOpen
  \bibfield  {author} {\bibinfo {author} {\bibfnamefont {M.}~\bibnamefont
  {Dine}}, \bibinfo {author} {\bibfnamefont {G.}~\bibnamefont {Festuccia}},
  \bibinfo {author} {\bibfnamefont {J.}~\bibnamefont {Kehayias}}, \ and\
  \bibinfo {author} {\bibfnamefont {W.}~\bibnamefont {Wu}},\ }\href {\doibase
  10.1007/JHEP01(2011)012} {\bibfield  {journal} {\bibinfo  {journal} {JHEP}\
  }\textbf {\bibinfo {volume} {01}},\ \bibinfo {pages} {012} (\bibinfo {year}
  {2011})},\ \Eprint {http://arxiv.org/abs/1010.4803} {arXiv:1010.4803
  [hep-th]} \BibitemShut {NoStop}%
\bibitem [{\citenamefont {Ringwald}(2014)}]{Ringwald:2012cu}%
  \BibitemOpen
  \bibfield  {author} {\bibinfo {author} {\bibfnamefont {A.}~\bibnamefont
  {Ringwald}},\ }\bibfield  {booktitle} {\emph {\bibinfo {booktitle}
  {{Proceedings, 18th International Symposium on Particles, Strings and
  Cosmology (PASCOS 2012): Merida, Yucatan, Mexico, June 3-8, 2012}}},\ }\href
  {\doibase 10.1088/1742-6596/485/1/012013} {\bibfield  {journal} {\bibinfo
  {journal} {J. Phys. Conf. Ser.}\ }\textbf {\bibinfo {volume} {485}},\
  \bibinfo {pages} {012013} (\bibinfo {year} {2014})},\ \Eprint
  {http://arxiv.org/abs/1209.2299} {arXiv:1209.2299 [hep-ph]} \BibitemShut
  {NoStop}%
\bibitem [{\citenamefont {Kim}(2013)}]{Kim:2013fga}%
  \BibitemOpen
  \bibfield  {author} {\bibinfo {author} {\bibfnamefont {J.~E.}\ \bibnamefont
  {Kim}},\ }\href {\doibase 10.1103/PhysRevLett.111.031801} {\bibfield
  {journal} {\bibinfo  {journal} {Phys. Rev. Lett.}\ }\textbf {\bibinfo
  {volume} {111}},\ \bibinfo {pages} {031801} (\bibinfo {year} {2013})},\
  \Eprint {http://arxiv.org/abs/1303.1822} {arXiv:1303.1822 [hep-ph]}
  \BibitemShut {NoStop}%
\bibitem [{\citenamefont {Bachlechner}\ \emph {et~al.}(2015)\citenamefont
  {Bachlechner}, \citenamefont {Long},\ and\ \citenamefont
  {McAllister}}]{Bachlechner:2014gfa}%
  \BibitemOpen
  \bibfield  {author} {\bibinfo {author} {\bibfnamefont {T.~C.}\ \bibnamefont
  {Bachlechner}}, \bibinfo {author} {\bibfnamefont {C.}~\bibnamefont {Long}}, \
  and\ \bibinfo {author} {\bibfnamefont {L.}~\bibnamefont {McAllister}},\
  }\href {\doibase 10.1007/JHEP12(2015)042} {\bibfield  {journal} {\bibinfo
  {journal} {JHEP}\ }\textbf {\bibinfo {volume} {12}},\ \bibinfo {pages} {042}
  (\bibinfo {year} {2015})},\ \Eprint {http://arxiv.org/abs/1412.1093}
  {arXiv:1412.1093 [hep-th]} \BibitemShut {NoStop}%
\bibitem [{\citenamefont {Halverson}\ \emph {et~al.}(2017)\citenamefont
  {Halverson}, \citenamefont {Long},\ and\ \citenamefont
  {Nath}}]{Halverson:2017deq}%
  \BibitemOpen
  \bibfield  {author} {\bibinfo {author} {\bibfnamefont {J.}~\bibnamefont
  {Halverson}}, \bibinfo {author} {\bibfnamefont {C.}~\bibnamefont {Long}}, \
  and\ \bibinfo {author} {\bibfnamefont {P.}~\bibnamefont {Nath}},\ }\href@noop
  {} {\  (\bibinfo {year} {2017})},\ \Eprint {http://arxiv.org/abs/1703.07779}
  {arXiv:1703.07779 [hep-ph]} \BibitemShut {NoStop}%
\bibitem [{\citenamefont {Stott}\ \emph {et~al.}(2017)\citenamefont {Stott},
  \citenamefont {Marsh}, \citenamefont {Pongkitivanichkul}, \citenamefont
  {Price},\ and\ \citenamefont {Acharya}}]{Stott:2017hvl}%
  \BibitemOpen
  \bibfield  {author} {\bibinfo {author} {\bibfnamefont {M.~J.}\ \bibnamefont
  {Stott}}, \bibinfo {author} {\bibfnamefont {D.~J.~E.}\ \bibnamefont {Marsh}},
  \bibinfo {author} {\bibfnamefont {C.}~\bibnamefont {Pongkitivanichkul}},
  \bibinfo {author} {\bibfnamefont {L.~C.}\ \bibnamefont {Price}}, \ and\
  \bibinfo {author} {\bibfnamefont {B.~S.}\ \bibnamefont {Acharya}},\
  }\href@noop {} {\  (\bibinfo {year} {2017})},\ \Eprint
  {http://arxiv.org/abs/1706.03236} {arXiv:1706.03236 [astro-ph.CO]}
  \BibitemShut {NoStop}%
\bibitem [{\citenamefont {Arias}\ \emph {et~al.}(2012)\citenamefont {Arias},
  \citenamefont {Cadamuro}, \citenamefont {Goodsell}, \citenamefont {Jaeckel},
  \citenamefont {Redondo},\ and\ \citenamefont {Ringwald}}]{Arias:2012az}%
  \BibitemOpen
  \bibfield  {author} {\bibinfo {author} {\bibfnamefont {P.}~\bibnamefont
  {Arias}}, \bibinfo {author} {\bibfnamefont {D.}~\bibnamefont {Cadamuro}},
  \bibinfo {author} {\bibfnamefont {M.}~\bibnamefont {Goodsell}}, \bibinfo
  {author} {\bibfnamefont {J.}~\bibnamefont {Jaeckel}}, \bibinfo {author}
  {\bibfnamefont {J.}~\bibnamefont {Redondo}}, \ and\ \bibinfo {author}
  {\bibfnamefont {A.}~\bibnamefont {Ringwald}},\ }\href {\doibase
  10.1088/1475-7516/2012/06/013} {\bibfield  {journal} {\bibinfo  {journal}
  {JCAP}\ }\textbf {\bibinfo {volume} {1206}},\ \bibinfo {pages} {013}
  (\bibinfo {year} {2012})},\ \Eprint {http://arxiv.org/abs/1201.5902}
  {arXiv:1201.5902 [hep-ph]} \BibitemShut {NoStop}%
\bibitem [{\citenamefont {Hui}\ \emph {et~al.}(2016)\citenamefont {Hui},
  \citenamefont {Ostriker}, \citenamefont {Tremaine},\ and\ \citenamefont
  {Witten}}]{Hui:2016ltb}%
  \BibitemOpen
  \bibfield  {author} {\bibinfo {author} {\bibfnamefont {L.}~\bibnamefont
  {Hui}}, \bibinfo {author} {\bibfnamefont {J.~P.}\ \bibnamefont {Ostriker}},
  \bibinfo {author} {\bibfnamefont {S.}~\bibnamefont {Tremaine}}, \ and\
  \bibinfo {author} {\bibfnamefont {E.}~\bibnamefont {Witten}},\ }\href@noop {}
  {\  (\bibinfo {year} {2016})},\ \Eprint {http://arxiv.org/abs/1610.08297}
  {arXiv:1610.08297 [astro-ph.CO]} \BibitemShut {NoStop}%
\bibitem [{\citenamefont {Visinelli}(2017)}]{Visinelli:2017imh}%
  \BibitemOpen
  \bibfield  {author} {\bibinfo {author} {\bibfnamefont {L.}~\bibnamefont
  {Visinelli}},\ }\href {\doibase 10.1103/PhysRevD.96.023013} {\bibfield
  {journal} {\bibinfo  {journal} {Phys. Rev.}\ }\textbf {\bibinfo {volume}
  {D96}},\ \bibinfo {pages} {023013} (\bibinfo {year} {2017})},\ \Eprint
  {http://arxiv.org/abs/1703.08798} {arXiv:1703.08798 [astro-ph.CO]}
  \BibitemShut {NoStop}%
\bibitem [{\citenamefont {Diez-Tejedor}\ and\ \citenamefont
  {Marsh}(2017)}]{Diez-Tejedor:2017ivd}%
  \BibitemOpen
  \bibfield  {author} {\bibinfo {author} {\bibfnamefont {A.}~\bibnamefont
  {Diez-Tejedor}}\ and\ \bibinfo {author} {\bibfnamefont {D.~J.~E.}\
  \bibnamefont {Marsh}},\ }\href@noop {} {\  (\bibinfo {year} {2017})},\
  \Eprint {http://arxiv.org/abs/1702.02116} {arXiv:1702.02116 [hep-ph]}
  \BibitemShut {NoStop}%
\bibitem [{\citenamefont {Baldeschi}\ \emph {et~al.}(1983)\citenamefont
  {Baldeschi}, \citenamefont {Gelmini},\ and\ \citenamefont
  {Ruffini}}]{Baldeschi:1983}%
  \BibitemOpen
  \bibfield  {author} {\bibinfo {author} {\bibfnamefont {M.}~\bibnamefont
  {Baldeschi}}, \bibinfo {author} {\bibfnamefont {G.}~\bibnamefont {Gelmini}},
  \ and\ \bibinfo {author} {\bibfnamefont {R.}~\bibnamefont {Ruffini}},\ }\href
  {\doibase http://dx.doi.org/10.1016/0370-2693(83)90688-3} {\bibfield
  {journal} {\bibinfo  {journal} {Physics Letters B}\ }\textbf {\bibinfo
  {volume} {122}},\ \bibinfo {pages} {221 } (\bibinfo {year}
  {1983})}\BibitemShut {NoStop}%
\bibitem [{\citenamefont {Membrado}\ \emph {et~al.}(1989)\citenamefont
  {Membrado}, \citenamefont {Pacheco},\ and\ \citenamefont
  {Sa{\~n}udo}}]{Membrado:1989bqo}%
  \BibitemOpen
  \bibfield  {author} {\bibinfo {author} {\bibfnamefont {M.}~\bibnamefont
  {Membrado}}, \bibinfo {author} {\bibfnamefont {A.~F.}\ \bibnamefont
  {Pacheco}}, \ and\ \bibinfo {author} {\bibfnamefont {J.}~\bibnamefont
  {Sa{\~n}udo}},\ }\href {\doibase 10.1103/PhysRevA.39.4207} {\bibfield
  {journal} {\bibinfo  {journal} {Phys. Rev.}\ }\textbf {\bibinfo {volume}
  {A39}},\ \bibinfo {pages} {4207} (\bibinfo {year} {1989})}\BibitemShut
  {NoStop}%
\bibitem [{\citenamefont {Press}\ \emph {et~al.}(1990)\citenamefont {Press},
  \citenamefont {Ryden},\ and\ \citenamefont {Spergel}}]{Press:1990}%
  \BibitemOpen
  \bibfield  {author} {\bibinfo {author} {\bibfnamefont {W.~H.}\ \bibnamefont
  {Press}}, \bibinfo {author} {\bibfnamefont {B.~S.}\ \bibnamefont {Ryden}}, \
  and\ \bibinfo {author} {\bibfnamefont {D.~N.}\ \bibnamefont {Spergel}},\
  }\href {\doibase 10.1103/PhysRevLett.64.1084} {\bibfield  {journal} {\bibinfo
   {journal} {Phys. Rev. Lett.}\ }\textbf {\bibinfo {volume} {64}},\ \bibinfo
  {pages} {1084} (\bibinfo {year} {1990})}\BibitemShut {NoStop}%
\bibitem [{\citenamefont {Sin}(1994)}]{Sin:1994}%
  \BibitemOpen
  \bibfield  {author} {\bibinfo {author} {\bibfnamefont {S.-J.}\ \bibnamefont
  {Sin}},\ }\href {\doibase 10.1103/PhysRevD.50.3650} {\bibfield  {journal}
  {\bibinfo  {journal} {Phys. Rev. D}\ }\textbf {\bibinfo {volume} {50}},\
  \bibinfo {pages} {3650} (\bibinfo {year} {1994})}\BibitemShut {NoStop}%
\bibitem [{\citenamefont {Ji}\ and\ \citenamefont {Sin}(1994)}]{Ji:1994}%
  \BibitemOpen
  \bibfield  {author} {\bibinfo {author} {\bibfnamefont {S.~U.}\ \bibnamefont
  {Ji}}\ and\ \bibinfo {author} {\bibfnamefont {S.~J.}\ \bibnamefont {Sin}},\
  }\href {\doibase 10.1103/PhysRevD.50.3655} {\bibfield  {journal} {\bibinfo
  {journal} {Phys. Rev. D}\ }\textbf {\bibinfo {volume} {50}},\ \bibinfo
  {pages} {3655} (\bibinfo {year} {1994})}\BibitemShut {NoStop}%
\bibitem [{\citenamefont {Lee}\ and\ \citenamefont {Koh}(1996)}]{Lee:1996}%
  \BibitemOpen
  \bibfield  {author} {\bibinfo {author} {\bibfnamefont {J.-w.}\ \bibnamefont
  {Lee}}\ and\ \bibinfo {author} {\bibfnamefont {I.~G.}\ \bibnamefont {Koh}},\
  }\href {\doibase 10.1103/PhysRevD.53.2236} {\bibfield  {journal} {\bibinfo
  {journal} {Phys. Rev. D}\ }\textbf {\bibinfo {volume} {53}},\ \bibinfo
  {pages} {2236} (\bibinfo {year} {1996})}\BibitemShut {NoStop}%
\bibitem [{\citenamefont {Guzm{\'a}n}\ and\ \citenamefont
  {Matos}(2000)}]{Guzman:2000}%
  \BibitemOpen
  \bibfield  {author} {\bibinfo {author} {\bibfnamefont {F.~S.}\ \bibnamefont
  {Guzm{\'a}n}}\ and\ \bibinfo {author} {\bibfnamefont {T.}~\bibnamefont
  {Matos}},\ }\href {http://stacks.iop.org/0264-9381/17/i=1/a=102} {\bibfield
  {journal} {\bibinfo  {journal} {Classical and Quantum Gravity}\ }\textbf
  {\bibinfo {volume} {17}},\ \bibinfo {pages} {L9} (\bibinfo {year}
  {2000})}\BibitemShut {NoStop}%
\bibitem [{\citenamefont {Sahni}\ and\ \citenamefont
  {Wang}(2000)}]{Sahni:2000}%
  \BibitemOpen
  \bibfield  {author} {\bibinfo {author} {\bibfnamefont {V.}~\bibnamefont
  {Sahni}}\ and\ \bibinfo {author} {\bibfnamefont {L.}~\bibnamefont {Wang}},\
  }\href {\doibase 10.1103/PhysRevD.62.103517} {\bibfield  {journal} {\bibinfo
  {journal} {Phys. Rev. D}\ }\textbf {\bibinfo {volume} {62}},\ \bibinfo
  {pages} {103517} (\bibinfo {year} {2000})}\BibitemShut {NoStop}%
\bibitem [{\citenamefont {Peebles}(2000)}]{Peebles:2000yy}%
  \BibitemOpen
  \bibfield  {author} {\bibinfo {author} {\bibfnamefont {P.~J.~E.}\
  \bibnamefont {Peebles}},\ }\href {\doibase 10.1086/312677} {\bibfield
  {journal} {\bibinfo  {journal} {Astrophys. J.}\ }\textbf {\bibinfo {volume}
  {534}},\ \bibinfo {pages} {L127} (\bibinfo {year} {2000})},\ \Eprint
  {http://arxiv.org/abs/astro-ph/0002495} {arXiv:astro-ph/0002495 [astro-ph]}
  \BibitemShut {NoStop}%
\bibitem [{\citenamefont {Goodman}(2000)}]{Goodman:2000}%
  \BibitemOpen
  \bibfield  {author} {\bibinfo {author} {\bibfnamefont {J.}~\bibnamefont
  {Goodman}},\ }\href {\doibase
  http://dx.doi.org/10.1016/S1384-1076(00)00015-4} {\bibfield  {journal}
  {\bibinfo  {journal} {New Astronomy}\ }\textbf {\bibinfo {volume} {5}},\
  \bibinfo {pages} {103 } (\bibinfo {year} {2000})}\BibitemShut {NoStop}%
\bibitem [{\citenamefont {Matos}\ and\ \citenamefont
  {Ure{\~n}a-L{\'o}pez}(2000)}]{Matos:2000}%
  \BibitemOpen
  \bibfield  {author} {\bibinfo {author} {\bibfnamefont {T.}~\bibnamefont
  {Matos}}\ and\ \bibinfo {author} {\bibfnamefont {L.~A.}\ \bibnamefont
  {Ure{\~n}a-L{\'o}pez}},\ }\href
  {http://stacks.iop.org/0264-9381/17/i=13/a=101} {\bibfield  {journal}
  {\bibinfo  {journal} {Classical and Quantum Gravity}\ }\textbf {\bibinfo
  {volume} {17}},\ \bibinfo {pages} {L75} (\bibinfo {year} {2000})}\BibitemShut
  {NoStop}%
\bibitem [{\citenamefont {Hu}\ \emph {et~al.}(2000)\citenamefont {Hu},
  \citenamefont {Barkana},\ and\ \citenamefont {Gruzinov}}]{Hu:2000ke}%
  \BibitemOpen
  \bibfield  {author} {\bibinfo {author} {\bibfnamefont {W.}~\bibnamefont
  {Hu}}, \bibinfo {author} {\bibfnamefont {R.}~\bibnamefont {Barkana}}, \ and\
  \bibinfo {author} {\bibfnamefont {A.}~\bibnamefont {Gruzinov}},\ }\href
  {\doibase 10.1103/PhysRevLett.85.1158} {\bibfield  {journal} {\bibinfo
  {journal} {Phys. Rev. Lett.}\ }\textbf {\bibinfo {volume} {85}},\ \bibinfo
  {pages} {1158} (\bibinfo {year} {2000})},\ \Eprint
  {http://arxiv.org/abs/astro-ph/0003365} {arXiv:astro-ph/0003365 [astro-ph]}
  \BibitemShut {NoStop}%
\bibitem [{\citenamefont {Weinberg}\ \emph {et~al.}(2014)\citenamefont
  {Weinberg}, \citenamefont {Bullock}, \citenamefont {Governato}, \citenamefont
  {Kuzio~de Naray},\ and\ \citenamefont {Peter}}]{Weinberg:2013aya}%
  \BibitemOpen
  \bibfield  {author} {\bibinfo {author} {\bibfnamefont {D.~H.}\ \bibnamefont
  {Weinberg}}, \bibinfo {author} {\bibfnamefont {J.~S.}\ \bibnamefont
  {Bullock}}, \bibinfo {author} {\bibfnamefont {F.}~\bibnamefont {Governato}},
  \bibinfo {author} {\bibfnamefont {R.}~\bibnamefont {Kuzio~de Naray}}, \ and\
  \bibinfo {author} {\bibfnamefont {A.~H.~G.}\ \bibnamefont {Peter}},\
  }\bibfield  {booktitle} {\emph {\bibinfo {booktitle} {{Sackler Colloquium:
  Dark Matter Universe: On the Threshhold of Discovery Irvine, USA, October
  18-20, 2012}}},\ }\href {\doibase 10.1073/pnas.1308716112} {\bibfield
  {journal} {\bibinfo  {journal} {Proc. Nat. Acad. Sci.}\ }\textbf {\bibinfo
  {volume} {112}},\ \bibinfo {pages} {12249} (\bibinfo {year} {2014})},\
  \Eprint {http://arxiv.org/abs/1306.0913} {arXiv:1306.0913 [astro-ph.CO]}
  \BibitemShut {NoStop}%
\bibitem [{\citenamefont {Randall}\ and\ \citenamefont
  {Sundrum}(1999)}]{Randall:1999vf}%
  \BibitemOpen
  \bibfield  {author} {\bibinfo {author} {\bibfnamefont {L.}~\bibnamefont
  {Randall}}\ and\ \bibinfo {author} {\bibfnamefont {R.}~\bibnamefont
  {Sundrum}},\ }\href {\doibase 10.1103/PhysRevLett.83.4690} {\bibfield
  {journal} {\bibinfo  {journal} {Phys. Rev. Lett.}\ }\textbf {\bibinfo
  {volume} {83}},\ \bibinfo {pages} {4690} (\bibinfo {year} {1999})},\ \Eprint
  {http://arxiv.org/abs/hep-th/9906064} {arXiv:hep-th/9906064 [hep-th]}
  \BibitemShut {NoStop}%
\bibitem [{\citenamefont {Caldwell}\ and\ \citenamefont
  {Langlois}(2001)}]{Caldwell:2001ja}%
  \BibitemOpen
  \bibfield  {author} {\bibinfo {author} {\bibfnamefont {R.~R.}\ \bibnamefont
  {Caldwell}}\ and\ \bibinfo {author} {\bibfnamefont {D.}~\bibnamefont
  {Langlois}},\ }\href {\doibase 10.1016/S0370-2693(01)00631-1} {\bibfield
  {journal} {\bibinfo  {journal} {Phys. Lett.}\ }\textbf {\bibinfo {volume}
  {B511}},\ \bibinfo {pages} {129} (\bibinfo {year} {2001})},\ \Eprint
  {http://arxiv.org/abs/gr-qc/0103070} {arXiv:gr-qc/0103070 [gr-qc]}
  \BibitemShut {NoStop}%
\bibitem [{\citenamefont {Maartens}\ and\ \citenamefont
  {Koyama}(2010)}]{Maartens:2010ar}%
  \BibitemOpen
  \bibfield  {author} {\bibinfo {author} {\bibfnamefont {R.}~\bibnamefont
  {Maartens}}\ and\ \bibinfo {author} {\bibfnamefont {K.}~\bibnamefont
  {Koyama}},\ }\href {\doibase 10.12942/lrr-2010-5} {\bibfield  {journal}
  {\bibinfo  {journal} {Living Rev. Rel.}\ }\textbf {\bibinfo {volume} {13}},\
  \bibinfo {pages} {5} (\bibinfo {year} {2010})},\ \Eprint
  {http://arxiv.org/abs/1004.3962} {arXiv:1004.3962 [hep-th]} \BibitemShut
  {NoStop}%
\bibitem [{\citenamefont {Visinelli}\ \emph
  {et~al.}(2018{\natexlab{b}})\citenamefont {Visinelli}, \citenamefont
  {Bolis},\ and\ \citenamefont {Vagnozzi}}]{Visinelli:2017bny}%
  \BibitemOpen
  \bibfield  {author} {\bibinfo {author} {\bibfnamefont {L.}~\bibnamefont
  {Visinelli}}, \bibinfo {author} {\bibfnamefont {N.}~\bibnamefont {Bolis}}, \
  and\ \bibinfo {author} {\bibfnamefont {S.}~\bibnamefont {Vagnozzi}},\ }\href
  {\doibase 10.1103/PhysRevD.97.064039} {\bibfield  {journal} {\bibinfo
  {journal} {Phys. Rev.}\ }\textbf {\bibinfo {volume} {D97}},\ \bibinfo {pages}
  {064039} (\bibinfo {year} {2018}{\natexlab{b}})},\ \Eprint
  {http://arxiv.org/abs/1711.06628} {arXiv:1711.06628 [gr-qc]} \BibitemShut
  {NoStop}%
\bibitem [{\citenamefont {Vandoren}\ and\ \citenamefont {van
  Nieuwenhuizen}(2008)}]{Vandoren:2008xg}%
  \BibitemOpen
  \bibfield  {author} {\bibinfo {author} {\bibfnamefont {S.}~\bibnamefont
  {Vandoren}}\ and\ \bibinfo {author} {\bibfnamefont {P.}~\bibnamefont {van
  Nieuwenhuizen}},\ }\href@noop {} {\  (\bibinfo {year} {2008})},\ \Eprint
  {http://arxiv.org/abs/0802.1862} {arXiv:0802.1862 [hep-th]} \BibitemShut
  {NoStop}%
\bibitem [{\citenamefont {Vafa}(2005)}]{Vafa:2005ui}%
  \BibitemOpen
  \bibfield  {author} {\bibinfo {author} {\bibfnamefont {C.}~\bibnamefont
  {Vafa}},\ }\href@noop {} {\  (\bibinfo {year} {2005})},\ \Eprint
  {http://arxiv.org/abs/hep-th/0509212} {arXiv:hep-th/0509212 [hep-th]}
  \BibitemShut {NoStop}%
\bibitem [{\citenamefont {Danielsson}\ and\ \citenamefont
  {Van~Riet}(2018)}]{Danielsson:2018ztv}%
  \BibitemOpen
  \bibfield  {author} {\bibinfo {author} {\bibfnamefont {U.~H.}\ \bibnamefont
  {Danielsson}}\ and\ \bibinfo {author} {\bibfnamefont {T.}~\bibnamefont
  {Van~Riet}},\ }\href {\doibase 10.1142/S0218271818300070} {\bibfield
  {journal} {\bibinfo  {journal} {Int. J. Mod. Phys.}\ }\textbf {\bibinfo
  {volume} {D27}},\ \bibinfo {pages} {1830007} (\bibinfo {year} {2018})},\
  \Eprint {http://arxiv.org/abs/1804.01120} {arXiv:1804.01120 [hep-th]}
  \BibitemShut {NoStop}%
\bibitem [{\citenamefont {Obied}\ \emph {et~al.}(2018)\citenamefont {Obied},
  \citenamefont {Ooguri}, \citenamefont {Spodyneiko},\ and\ \citenamefont
  {Vafa}}]{Obied:2018sgi}%
  \BibitemOpen
  \bibfield  {author} {\bibinfo {author} {\bibfnamefont {G.}~\bibnamefont
  {Obied}}, \bibinfo {author} {\bibfnamefont {H.}~\bibnamefont {Ooguri}},
  \bibinfo {author} {\bibfnamefont {L.}~\bibnamefont {Spodyneiko}}, \ and\
  \bibinfo {author} {\bibfnamefont {C.}~\bibnamefont {Vafa}},\ }\href@noop {}
  {\  (\bibinfo {year} {2018})},\ \Eprint {http://arxiv.org/abs/1806.08362}
  {arXiv:1806.08362 [hep-th]} \BibitemShut {NoStop}%
\bibitem [{\citenamefont {Agrawal}\ \emph {et~al.}(2018)\citenamefont
  {Agrawal}, \citenamefont {Obied}, \citenamefont {Steinhardt},\ and\
  \citenamefont {Vafa}}]{Agrawal:2018own}%
  \BibitemOpen
  \bibfield  {author} {\bibinfo {author} {\bibfnamefont {P.}~\bibnamefont
  {Agrawal}}, \bibinfo {author} {\bibfnamefont {G.}~\bibnamefont {Obied}},
  \bibinfo {author} {\bibfnamefont {P.~J.}\ \bibnamefont {Steinhardt}}, \ and\
  \bibinfo {author} {\bibfnamefont {C.}~\bibnamefont {Vafa}},\ }\href {\doibase
  10.1016/j.physletb.2018.07.040} {\bibfield  {journal} {\bibinfo  {journal}
  {Phys. Lett.}\ }\textbf {\bibinfo {volume} {B784}},\ \bibinfo {pages} {271}
  (\bibinfo {year} {2018})},\ \Eprint {http://arxiv.org/abs/1806.09718}
  {arXiv:1806.09718 [hep-th]} \BibitemShut {NoStop}%
\bibitem [{\citenamefont {Dvali}\ and\ \citenamefont
  {Gomez}(2018)}]{Dvali:2018fqu}%
  \BibitemOpen
  \bibfield  {author} {\bibinfo {author} {\bibfnamefont {G.}~\bibnamefont
  {Dvali}}\ and\ \bibinfo {author} {\bibfnamefont {C.}~\bibnamefont {Gomez}},\
  }\href@noop {} {\  (\bibinfo {year} {2018})},\ \Eprint
  {http://arxiv.org/abs/1806.10877} {arXiv:1806.10877 [hep-th]} \BibitemShut
  {NoStop}%
\bibitem [{\citenamefont {Andriot}(2018{\natexlab{a}})}]{Andriot:2018wzk}%
  \BibitemOpen
  \bibfield  {author} {\bibinfo {author} {\bibfnamefont {D.}~\bibnamefont
  {Andriot}},\ }\href@noop {} {\  (\bibinfo {year} {2018}{\natexlab{a}})},\
  \Eprint {http://arxiv.org/abs/1806.10999} {arXiv:1806.10999 [hep-th]}
  \BibitemShut {NoStop}%
\bibitem [{\citenamefont {Banerjee}\ \emph {et~al.}(2018)\citenamefont
  {Banerjee}, \citenamefont {Danielsson}, \citenamefont {Dibitetto},
  \citenamefont {Giri},\ and\ \citenamefont {Schillo}}]{Banerjee:2018qey}%
  \BibitemOpen
  \bibfield  {author} {\bibinfo {author} {\bibfnamefont {S.}~\bibnamefont
  {Banerjee}}, \bibinfo {author} {\bibfnamefont {U.}~\bibnamefont
  {Danielsson}}, \bibinfo {author} {\bibfnamefont {G.}~\bibnamefont
  {Dibitetto}}, \bibinfo {author} {\bibfnamefont {S.}~\bibnamefont {Giri}}, \
  and\ \bibinfo {author} {\bibfnamefont {M.}~\bibnamefont {Schillo}},\
  }\href@noop {} {\  (\bibinfo {year} {2018})},\ \Eprint
  {http://arxiv.org/abs/1807.01570} {arXiv:1807.01570 [hep-th]} \BibitemShut
  {NoStop}%
\bibitem [{\citenamefont {Ach{\'u}carro}\ and\ \citenamefont
  {Palma}(2018)}]{Achucarro:2018vey}%
  \BibitemOpen
  \bibfield  {author} {\bibinfo {author} {\bibfnamefont {A.}~\bibnamefont
  {Ach{\'u}carro}}\ and\ \bibinfo {author} {\bibfnamefont {G.~A.}\ \bibnamefont
  {Palma}},\ }\href@noop {} {\  (\bibinfo {year} {2018})},\ \Eprint
  {http://arxiv.org/abs/1807.04390} {arXiv:1807.04390 [hep-th]} \BibitemShut
  {NoStop}%
\bibitem [{\citenamefont {Garg}\ and\ \citenamefont
  {Krishnan}(2018)}]{Garg:2018reu}%
  \BibitemOpen
  \bibfield  {author} {\bibinfo {author} {\bibfnamefont {S.~K.}\ \bibnamefont
  {Garg}}\ and\ \bibinfo {author} {\bibfnamefont {C.}~\bibnamefont
  {Krishnan}},\ }\href@noop {} {\  (\bibinfo {year} {2018})},\ \Eprint
  {http://arxiv.org/abs/1807.05193} {arXiv:1807.05193 [hep-th]} \BibitemShut
  {NoStop}%
\bibitem [{\citenamefont {Kehagias}\ and\ \citenamefont
  {Riotto}(2018)}]{Kehagias:2018uem}%
  \BibitemOpen
  \bibfield  {author} {\bibinfo {author} {\bibfnamefont {A.}~\bibnamefont
  {Kehagias}}\ and\ \bibinfo {author} {\bibfnamefont {A.}~\bibnamefont
  {Riotto}},\ }\href@noop {} {\  (\bibinfo {year} {2018})},\ \Eprint
  {http://arxiv.org/abs/1807.05445} {arXiv:1807.05445 [hep-th]} \BibitemShut
  {NoStop}%
\bibitem [{\citenamefont {Dias}\ \emph {et~al.}(2018)\citenamefont {Dias},
  \citenamefont {Frazer}, \citenamefont {Retolaza},\ and\ \citenamefont
  {Westphal}}]{Dias:2018ngv}%
  \BibitemOpen
  \bibfield  {author} {\bibinfo {author} {\bibfnamefont {M.}~\bibnamefont
  {Dias}}, \bibinfo {author} {\bibfnamefont {J.}~\bibnamefont {Frazer}},
  \bibinfo {author} {\bibfnamefont {A.}~\bibnamefont {Retolaza}}, \ and\
  \bibinfo {author} {\bibfnamefont {A.}~\bibnamefont {Westphal}},\ }\href@noop
  {} {\  (\bibinfo {year} {2018})},\ \Eprint {http://arxiv.org/abs/1807.06579}
  {arXiv:1807.06579 [hep-th]} \BibitemShut {NoStop}%
\bibitem [{\citenamefont {Denef}\ \emph {et~al.}(2018)\citenamefont {Denef},
  \citenamefont {Hebecker},\ and\ \citenamefont {Wrase}}]{Denef:2018etk}%
  \BibitemOpen
  \bibfield  {author} {\bibinfo {author} {\bibfnamefont {F.}~\bibnamefont
  {Denef}}, \bibinfo {author} {\bibfnamefont {A.}~\bibnamefont {Hebecker}}, \
  and\ \bibinfo {author} {\bibfnamefont {T.}~\bibnamefont {Wrase}},\
  }\href@noop {} {\  (\bibinfo {year} {2018})},\ \Eprint
  {http://arxiv.org/abs/1807.06581} {arXiv:1807.06581 [hep-th]} \BibitemShut
  {NoStop}%
\bibitem [{\citenamefont {Roupec}\ and\ \citenamefont
  {Wrase}(2018)}]{Roupec:2018mbn}%
  \BibitemOpen
  \bibfield  {author} {\bibinfo {author} {\bibfnamefont {C.}~\bibnamefont
  {Roupec}}\ and\ \bibinfo {author} {\bibfnamefont {T.}~\bibnamefont {Wrase}},\
  }\href@noop {} {\  (\bibinfo {year} {2018})},\ \Eprint
  {http://arxiv.org/abs/1807.09538} {arXiv:1807.09538 [hep-th]} \BibitemShut
  {NoStop}%
\bibitem [{\citenamefont {Andriot}(2018{\natexlab{b}})}]{Andriot:2018ept}%
  \BibitemOpen
  \bibfield  {author} {\bibinfo {author} {\bibfnamefont {D.}~\bibnamefont
  {Andriot}},\ }\href@noop {} {\  (\bibinfo {year} {2018}{\natexlab{b}})},\
  \Eprint {http://arxiv.org/abs/1807.09698} {arXiv:1807.09698 [hep-th]}
  \BibitemShut {NoStop}%
\bibitem [{\citenamefont {Matsui}\ and\ \citenamefont
  {Takahashi}(2018)}]{Matsui:2018bsy}%
  \BibitemOpen
  \bibfield  {author} {\bibinfo {author} {\bibfnamefont {H.}~\bibnamefont
  {Matsui}}\ and\ \bibinfo {author} {\bibfnamefont {F.}~\bibnamefont
  {Takahashi}},\ }\href@noop {} {\  (\bibinfo {year} {2018})},\ \Eprint
  {http://arxiv.org/abs/1807.11938} {arXiv:1807.11938 [hep-th]} \BibitemShut
  {NoStop}%
\bibitem [{\citenamefont {Ben-Dayan}(2018)}]{Ben-Dayan:2018mhe}%
  \BibitemOpen
  \bibfield  {author} {\bibinfo {author} {\bibfnamefont {I.}~\bibnamefont
  {Ben-Dayan}},\ }\href@noop {} {\  (\bibinfo {year} {2018})},\ \Eprint
  {http://arxiv.org/abs/1808.01615} {arXiv:1808.01615 [hep-th]} \BibitemShut
  {NoStop}%
\bibitem [{\citenamefont {Heisenberg}\ \emph
  {et~al.}(2018{\natexlab{a}})\citenamefont {Heisenberg}, \citenamefont
  {Bartelmann}, \citenamefont {Brandenberger},\ and\ \citenamefont
  {Refregier}}]{Heisenberg:2018yae}%
  \BibitemOpen
  \bibfield  {author} {\bibinfo {author} {\bibfnamefont {L.}~\bibnamefont
  {Heisenberg}}, \bibinfo {author} {\bibfnamefont {M.}~\bibnamefont
  {Bartelmann}}, \bibinfo {author} {\bibfnamefont {R.}~\bibnamefont
  {Brandenberger}}, \ and\ \bibinfo {author} {\bibfnamefont {A.}~\bibnamefont
  {Refregier}},\ }\href@noop {} {\  (\bibinfo {year} {2018}{\natexlab{a}})},\
  \Eprint {http://arxiv.org/abs/1808.02877} {arXiv:1808.02877 [astro-ph.CO]}
  \BibitemShut {NoStop}%
\bibitem [{\citenamefont {Loaiza-Brito}\ and\ \citenamefont
  {Loaiza-Brito}(2018)}]{Damian:2018tlf}%
  \BibitemOpen
  \bibfield  {author} {\bibinfo {author} {\bibfnamefont {O.}~\bibnamefont
  {Loaiza-Brito}}\ and\ \bibinfo {author} {\bibfnamefont {O.}~\bibnamefont
  {Loaiza-Brito}},\ }\href@noop {} {\  (\bibinfo {year} {2018})},\ \Eprint
  {http://arxiv.org/abs/1808.03397} {arXiv:1808.03397 [hep-th]} \BibitemShut
  {NoStop}%
\bibitem [{\citenamefont {Conlon}(2018)}]{Conlon:2018eyr}%
  \BibitemOpen
  \bibfield  {author} {\bibinfo {author} {\bibfnamefont {J.~P.}\ \bibnamefont
  {Conlon}},\ }\href@noop {} {\  (\bibinfo {year} {2018})},\ \Eprint
  {http://arxiv.org/abs/1808.05040} {arXiv:1808.05040 [hep-th]} \BibitemShut
  {NoStop}%
\bibitem [{\citenamefont {Kinney}\ \emph {et~al.}(2018)\citenamefont {Kinney},
  \citenamefont {Vagnozzi},\ and\ \citenamefont {Visinelli}}]{Kinney:2018nny}%
  \BibitemOpen
  \bibfield  {author} {\bibinfo {author} {\bibfnamefont {W.~H.}\ \bibnamefont
  {Kinney}}, \bibinfo {author} {\bibfnamefont {S.}~\bibnamefont {Vagnozzi}}, \
  and\ \bibinfo {author} {\bibfnamefont {L.}~\bibnamefont {Visinelli}},\
  }\href@noop {} {\  (\bibinfo {year} {2018})},\ \Eprint
  {http://arxiv.org/abs/1808.06424} {arXiv:1808.06424 [astro-ph.CO]}
  \BibitemShut {NoStop}%
\bibitem [{\citenamefont {Dasgupta}\ \emph {et~al.}(2018)\citenamefont
  {Dasgupta}, \citenamefont {Emelin}, \citenamefont {McDonough},\ and\
  \citenamefont {Tatar}}]{Dasgupta:2018rtp}%
  \BibitemOpen
  \bibfield  {author} {\bibinfo {author} {\bibfnamefont {K.}~\bibnamefont
  {Dasgupta}}, \bibinfo {author} {\bibfnamefont {M.}~\bibnamefont {Emelin}},
  \bibinfo {author} {\bibfnamefont {E.}~\bibnamefont {McDonough}}, \ and\
  \bibinfo {author} {\bibfnamefont {R.}~\bibnamefont {Tatar}},\ }\href@noop {}
  {\  (\bibinfo {year} {2018})},\ \Eprint {http://arxiv.org/abs/1808.07498}
  {arXiv:1808.07498 [hep-th]} \BibitemShut {NoStop}%
\bibitem [{\citenamefont {Cicoli}\ \emph {et~al.}(2018)\citenamefont {Cicoli},
  \citenamefont {de~Alwis}, \citenamefont {Maharana}, \citenamefont {Muia},\
  and\ \citenamefont {Quevedo}}]{Cicoli:2018kdo}%
  \BibitemOpen
  \bibfield  {author} {\bibinfo {author} {\bibfnamefont {M.}~\bibnamefont
  {Cicoli}}, \bibinfo {author} {\bibfnamefont {S.}~\bibnamefont {de~Alwis}},
  \bibinfo {author} {\bibfnamefont {A.}~\bibnamefont {Maharana}}, \bibinfo
  {author} {\bibfnamefont {F.}~\bibnamefont {Muia}}, \ and\ \bibinfo {author}
  {\bibfnamefont {F.}~\bibnamefont {Quevedo}},\ }\href@noop {} {\  (\bibinfo
  {year} {2018})},\ \Eprint {http://arxiv.org/abs/1808.08967} {arXiv:1808.08967
  [hep-th]} \BibitemShut {NoStop}%
\bibitem [{\citenamefont {Kachru}\ and\ \citenamefont
  {Trivedi}(2018)}]{Kachru:2018aqn}%
  \BibitemOpen
  \bibfield  {author} {\bibinfo {author} {\bibfnamefont {S.}~\bibnamefont
  {Kachru}}\ and\ \bibinfo {author} {\bibfnamefont {S.}~\bibnamefont
  {Trivedi}},\ }\href@noop {} {\  (\bibinfo {year} {2018})},\ \Eprint
  {http://arxiv.org/abs/1808.08971} {arXiv:1808.08971 [hep-th]} \BibitemShut
  {NoStop}%
\bibitem [{\citenamefont {Akrami}\ \emph {et~al.}(2018)\citenamefont {Akrami},
  \citenamefont {Kallosh}, \citenamefont {Linde},\ and\ \citenamefont
  {Vardanyan}}]{Akrami:2018ylq}%
  \BibitemOpen
  \bibfield  {author} {\bibinfo {author} {\bibfnamefont {Y.}~\bibnamefont
  {Akrami}}, \bibinfo {author} {\bibfnamefont {R.}~\bibnamefont {Kallosh}},
  \bibinfo {author} {\bibfnamefont {A.}~\bibnamefont {Linde}}, \ and\ \bibinfo
  {author} {\bibfnamefont {V.}~\bibnamefont {Vardanyan}},\ }\href@noop {} {\
  (\bibinfo {year} {2018})},\ \Eprint {http://arxiv.org/abs/1808.09440}
  {arXiv:1808.09440 [hep-th]} \BibitemShut {NoStop}%
\bibitem [{\citenamefont {Heisenberg}\ \emph
  {et~al.}(2018{\natexlab{b}})\citenamefont {Heisenberg}, \citenamefont
  {Bartelmann}, \citenamefont {Brandenberger},\ and\ \citenamefont
  {Refregier}}]{Heisenberg:2018rdu}%
  \BibitemOpen
  \bibfield  {author} {\bibinfo {author} {\bibfnamefont {L.}~\bibnamefont
  {Heisenberg}}, \bibinfo {author} {\bibfnamefont {M.}~\bibnamefont
  {Bartelmann}}, \bibinfo {author} {\bibfnamefont {R.}~\bibnamefont
  {Brandenberger}}, \ and\ \bibinfo {author} {\bibfnamefont {A.}~\bibnamefont
  {Refregier}},\ }\href@noop {} {\  (\bibinfo {year} {2018}{\natexlab{b}})},\
  \Eprint {http://arxiv.org/abs/1809.00154} {arXiv:1809.00154 [astro-ph.CO]}
  \BibitemShut {NoStop}%
\bibitem [{\citenamefont {Murayama}\ \emph {et~al.}(2018)\citenamefont
  {Murayama}, \citenamefont {Yamazaki},\ and\ \citenamefont
  {Yanagida}}]{Murayama:2018lie}%
  \BibitemOpen
  \bibfield  {author} {\bibinfo {author} {\bibfnamefont {H.}~\bibnamefont
  {Murayama}}, \bibinfo {author} {\bibfnamefont {M.}~\bibnamefont {Yamazaki}},
  \ and\ \bibinfo {author} {\bibfnamefont {T.~T.}\ \bibnamefont {Yanagida}},\
  }\href@noop {} {\  (\bibinfo {year} {2018})},\ \Eprint
  {http://arxiv.org/abs/1809.00478} {arXiv:1809.00478 [hep-th]} \BibitemShut
  {NoStop}%
\bibitem [{\citenamefont {Marsh}(2018)}]{Marsh:2018kub}%
  \BibitemOpen
  \bibfield  {author} {\bibinfo {author} {\bibfnamefont {M.~C.~D.}\
  \bibnamefont {Marsh}},\ }\href@noop {} {\  (\bibinfo {year} {2018})},\
  \Eprint {http://arxiv.org/abs/1809.00726} {arXiv:1809.00726 [hep-th]}
  \BibitemShut {NoStop}%
\bibitem [{\citenamefont {Brahma}\ and\ \citenamefont
  {Hossain}(2018)}]{Brahma:2018hrd}%
  \BibitemOpen
  \bibfield  {author} {\bibinfo {author} {\bibfnamefont {S.}~\bibnamefont
  {Brahma}}\ and\ \bibinfo {author} {\bibfnamefont {M.~W.}\ \bibnamefont
  {Hossain}},\ }\href@noop {} {\  (\bibinfo {year} {2018})},\ \Eprint
  {http://arxiv.org/abs/1809.01277} {arXiv:1809.01277 [hep-th]} \BibitemShut
  {NoStop}%
\bibitem [{\citenamefont {Choi}\ \emph {et~al.}(2018)\citenamefont {Choi},
  \citenamefont {Chway},\ and\ \citenamefont {Shin}}]{Choi:2018rze}%
  \BibitemOpen
  \bibfield  {author} {\bibinfo {author} {\bibfnamefont {K.}~\bibnamefont
  {Choi}}, \bibinfo {author} {\bibfnamefont {D.}~\bibnamefont {Chway}}, \ and\
  \bibinfo {author} {\bibfnamefont {C.~S.}\ \bibnamefont {Shin}},\ }\href@noop
  {} {\  (\bibinfo {year} {2018})},\ \Eprint {http://arxiv.org/abs/1809.01475}
  {arXiv:1809.01475 [hep-th]} \BibitemShut {NoStop}%
\bibitem [{\citenamefont {Arvanitaki}\ \emph
  {et~al.}(2010{\natexlab{b}})\citenamefont {Arvanitaki}, \citenamefont
  {Craig}, \citenamefont {Dimopoulos}, \citenamefont {Dubovsky},\ and\
  \citenamefont {March-Russell}}]{Arvanitaki:2009hb}%
  \BibitemOpen
  \bibfield  {author} {\bibinfo {author} {\bibfnamefont {A.}~\bibnamefont
  {Arvanitaki}}, \bibinfo {author} {\bibfnamefont {N.}~\bibnamefont {Craig}},
  \bibinfo {author} {\bibfnamefont {S.}~\bibnamefont {Dimopoulos}}, \bibinfo
  {author} {\bibfnamefont {S.}~\bibnamefont {Dubovsky}}, \ and\ \bibinfo
  {author} {\bibfnamefont {J.}~\bibnamefont {March-Russell}},\ }\href {\doibase
  10.1103/PhysRevD.81.075018} {\bibfield  {journal} {\bibinfo  {journal} {Phys.
  Rev.}\ }\textbf {\bibinfo {volume} {D81}},\ \bibinfo {pages} {075018}
  (\bibinfo {year} {2010}{\natexlab{b}})},\ \Eprint
  {http://arxiv.org/abs/0909.5440} {arXiv:0909.5440 [hep-ph]} \BibitemShut
  {NoStop}%
\bibitem [{\citenamefont {Kahn}\ \emph {et~al.}(2016)\citenamefont {Kahn},
  \citenamefont {Safdi},\ and\ \citenamefont {Thaler}}]{Kahn:2016aff}%
  \BibitemOpen
  \bibfield  {author} {\bibinfo {author} {\bibfnamefont {Y.}~\bibnamefont
  {Kahn}}, \bibinfo {author} {\bibfnamefont {B.~R.}\ \bibnamefont {Safdi}}, \
  and\ \bibinfo {author} {\bibfnamefont {J.}~\bibnamefont {Thaler}},\ }\href
  {\doibase 10.1103/PhysRevLett.117.141801} {\bibfield  {journal} {\bibinfo
  {journal} {Phys. Rev. Lett.}\ }\textbf {\bibinfo {volume} {117}},\ \bibinfo
  {pages} {141801} (\bibinfo {year} {2016})},\ \Eprint
  {http://arxiv.org/abs/1602.01086} {arXiv:1602.01086 [hep-ph]} \BibitemShut
  {NoStop}%
\bibitem [{\citenamefont {Stern}(2016)}]{Stern:2016bbw}%
  \BibitemOpen
  \bibfield  {author} {\bibinfo {author} {\bibfnamefont {I.}~\bibnamefont
  {Stern}},\ }\bibfield  {booktitle} {\emph {\bibinfo {booktitle}
  {{Proceedings, 38th International Conference on High Energy Physics (ICHEP
  2016): Chicago, IL, USA, August 3-10, 2016}}},\ }\href@noop {} {\bibfield
  {journal} {\bibinfo  {journal} {PoS}\ }\textbf {\bibinfo {volume}
  {ICHEP2016}},\ \bibinfo {pages} {198} (\bibinfo {year} {2016})},\ \Eprint
  {http://arxiv.org/abs/1612.08296} {arXiv:1612.08296 [physics.ins-det]}
  \BibitemShut {NoStop}%
\bibitem [{\citenamefont {Alesini}\ \emph {et~al.}(2017)\citenamefont
  {Alesini}, \citenamefont {Babusci}, \citenamefont {Di~Gioacchino},
  \citenamefont {Gatti}, \citenamefont {Lamanna},\ and\ \citenamefont
  {Ligi}}]{Alesini:2017ifp}%
  \BibitemOpen
  \bibfield  {author} {\bibinfo {author} {\bibfnamefont {D.}~\bibnamefont
  {Alesini}}, \bibinfo {author} {\bibfnamefont {D.}~\bibnamefont {Babusci}},
  \bibinfo {author} {\bibfnamefont {D.}~\bibnamefont {Di~Gioacchino}}, \bibinfo
  {author} {\bibfnamefont {C.}~\bibnamefont {Gatti}}, \bibinfo {author}
  {\bibfnamefont {G.}~\bibnamefont {Lamanna}}, \ and\ \bibinfo {author}
  {\bibfnamefont {C.}~\bibnamefont {Ligi}},\ }\href@noop {} {\  (\bibinfo
  {year} {2017})},\ \Eprint {http://arxiv.org/abs/1707.06010} {arXiv:1707.06010
  [physics.ins-det]} \BibitemShut {NoStop}%
\bibitem [{\citenamefont {Barbieri}\ \emph {et~al.}(2017)\citenamefont
  {Barbieri}, \citenamefont {Braggio}, \citenamefont {Carugno}, \citenamefont
  {Gallo}, \citenamefont {Lombardi}, \citenamefont {Ortolan}, \citenamefont
  {Pengo}, \citenamefont {Ruoso},\ and\ \citenamefont
  {Speake}}]{Barbieri:2016vwg}%
  \BibitemOpen
  \bibfield  {author} {\bibinfo {author} {\bibfnamefont {R.}~\bibnamefont
  {Barbieri}}, \bibinfo {author} {\bibfnamefont {C.}~\bibnamefont {Braggio}},
  \bibinfo {author} {\bibfnamefont {G.}~\bibnamefont {Carugno}}, \bibinfo
  {author} {\bibfnamefont {C.~S.}\ \bibnamefont {Gallo}}, \bibinfo {author}
  {\bibfnamefont {A.}~\bibnamefont {Lombardi}}, \bibinfo {author}
  {\bibfnamefont {A.}~\bibnamefont {Ortolan}}, \bibinfo {author} {\bibfnamefont
  {R.}~\bibnamefont {Pengo}}, \bibinfo {author} {\bibfnamefont
  {G.}~\bibnamefont {Ruoso}}, \ and\ \bibinfo {author} {\bibfnamefont {C.~C.}\
  \bibnamefont {Speake}},\ }\href {\doibase 10.1016/j.dark.2017.01.003}
  {\bibfield  {journal} {\bibinfo  {journal} {Phys. Dark Univ.}\ }\textbf
  {\bibinfo {volume} {15}},\ \bibinfo {pages} {135} (\bibinfo {year} {2017})},\
  \Eprint {http://arxiv.org/abs/1606.02201} {arXiv:1606.02201 [hep-ph]}
  \BibitemShut {NoStop}%
\bibitem [{\citenamefont {Brubaker}\ \emph {et~al.}(2017)\citenamefont
  {Brubaker} \emph {et~al.}}]{Brubaker:2016ktl}%
  \BibitemOpen
  \bibfield  {author} {\bibinfo {author} {\bibfnamefont {B.~M.}\ \bibnamefont
  {Brubaker}} \emph {et~al.},\ }\href {\doibase 10.1103/PhysRevLett.118.061302}
  {\bibfield  {journal} {\bibinfo  {journal} {Phys. Rev. Lett.}\ }\textbf
  {\bibinfo {volume} {118}},\ \bibinfo {pages} {061302} (\bibinfo {year}
  {2017})},\ \Eprint {http://arxiv.org/abs/1610.02580} {arXiv:1610.02580
  [astro-ph.CO]} \BibitemShut {NoStop}%
\bibitem [{\citenamefont {Chung}(2016)}]{Chung:2016ysi}%
  \BibitemOpen
  \bibfield  {author} {\bibinfo {author} {\bibfnamefont {W.}~\bibnamefont
  {Chung}},\ }\bibfield  {booktitle} {\emph {\bibinfo {booktitle}
  {{Proceedings, 15th Hellenic School and Workshops on Elementary Particle
  Physics and Gravity (CORFU2015): Corfu, Greece, September 1-25, 2015}}},\
  }\href@noop {} {\bibfield  {journal} {\bibinfo  {journal} {PoS}\ }\textbf
  {\bibinfo {volume} {CORFU2015}},\ \bibinfo {pages} {047} (\bibinfo {year}
  {2016})}\BibitemShut {NoStop}%
\bibitem [{\citenamefont {Caldwell}\ \emph {et~al.}(2017)\citenamefont
  {Caldwell}, \citenamefont {Dvali}, \citenamefont {Majorovits}, \citenamefont
  {Millar}, \citenamefont {Raffelt}, \citenamefont {Redondo}, \citenamefont
  {Reimann}, \citenamefont {Simon},\ and\ \citenamefont
  {Steffen}}]{TheMADMAXWorkingGroup:2016hpc}%
  \BibitemOpen
  \bibfield  {author} {\bibinfo {author} {\bibfnamefont {A.}~\bibnamefont
  {Caldwell}}, \bibinfo {author} {\bibfnamefont {G.}~\bibnamefont {Dvali}},
  \bibinfo {author} {\bibfnamefont {B.}~\bibnamefont {Majorovits}}, \bibinfo
  {author} {\bibfnamefont {A.}~\bibnamefont {Millar}}, \bibinfo {author}
  {\bibfnamefont {G.}~\bibnamefont {Raffelt}}, \bibinfo {author} {\bibfnamefont
  {J.}~\bibnamefont {Redondo}}, \bibinfo {author} {\bibfnamefont
  {O.}~\bibnamefont {Reimann}}, \bibinfo {author} {\bibfnamefont
  {F.}~\bibnamefont {Simon}}, \ and\ \bibinfo {author} {\bibfnamefont
  {F.}~\bibnamefont {Steffen}} (\bibinfo {collaboration} {MADMAX Working
  Group}),\ }\href {\doibase 10.1103/PhysRevLett.118.091801} {\bibfield
  {journal} {\bibinfo  {journal} {Phys. Rev. Lett.}\ }\textbf {\bibinfo
  {volume} {118}},\ \bibinfo {pages} {091801} (\bibinfo {year} {2017})},\
  \Eprint {http://arxiv.org/abs/1611.05865} {arXiv:1611.05865
  [physics.ins-det]} \BibitemShut {NoStop}%
\bibitem [{\citenamefont {Arvanitaki}\ and\ \citenamefont
  {Geraci}(2014)}]{Arvanitaki:2014dfa}%
  \BibitemOpen
  \bibfield  {author} {\bibinfo {author} {\bibfnamefont {A.}~\bibnamefont
  {Arvanitaki}}\ and\ \bibinfo {author} {\bibfnamefont {A.~A.}\ \bibnamefont
  {Geraci}},\ }\href {\doibase 10.1103/PhysRevLett.113.161801} {\bibfield
  {journal} {\bibinfo  {journal} {Phys. Rev. Lett.}\ }\textbf {\bibinfo
  {volume} {113}},\ \bibinfo {pages} {161801} (\bibinfo {year} {2014})},\
  \Eprint {http://arxiv.org/abs/1403.1290} {arXiv:1403.1290 [hep-ph]}
  \BibitemShut {NoStop}%
\bibitem [{\citenamefont {Vogel}\ \emph {et~al.}(2013)\citenamefont {Vogel}
  \emph {et~al.}}]{Vogel:2013bta}%
  \BibitemOpen
  \bibfield  {author} {\bibinfo {author} {\bibfnamefont {J.~K.}\ \bibnamefont
  {Vogel}} \emph {et~al.},\ }in\ \href
  {http://lss.fnal.gov/archive/2013/pub/fermilab-pub-13-699-a.pdf} {\emph
  {\bibinfo {booktitle} {{8th Patras Workshop on Axions, WIMPs and WISPs
  (AXION-WIMP 2012) Chicago, Illinois, July 18-22, 2012}}}}\ (\bibinfo {year}
  {2013})\ \Eprint {http://arxiv.org/abs/1302.3273} {arXiv:1302.3273
  [physics.ins-det]} \BibitemShut {NoStop}%
\bibitem [{\citenamefont {Budker}\ \emph {et~al.}(2014)\citenamefont {Budker},
  \citenamefont {Graham}, \citenamefont {Ledbetter}, \citenamefont
  {Rajendran},\ and\ \citenamefont {Sushkov}}]{Budker:2013hfa}%
  \BibitemOpen
  \bibfield  {author} {\bibinfo {author} {\bibfnamefont {D.}~\bibnamefont
  {Budker}}, \bibinfo {author} {\bibfnamefont {P.~W.}\ \bibnamefont {Graham}},
  \bibinfo {author} {\bibfnamefont {M.}~\bibnamefont {Ledbetter}}, \bibinfo
  {author} {\bibfnamefont {S.}~\bibnamefont {Rajendran}}, \ and\ \bibinfo
  {author} {\bibfnamefont {A.}~\bibnamefont {Sushkov}},\ }\href {\doibase
  10.1103/PhysRevX.4.021030} {\bibfield  {journal} {\bibinfo  {journal} {Phys.
  Rev.}\ }\textbf {\bibinfo {volume} {X4}},\ \bibinfo {pages} {021030}
  (\bibinfo {year} {2014})},\ \Eprint {http://arxiv.org/abs/1306.6089}
  {arXiv:1306.6089 [hep-ph]} \BibitemShut {NoStop}%
\bibitem [{\citenamefont {Irastorza}\ and\ \citenamefont
  {Redondo}(2018)}]{Irastorza:2018dyq}%
  \BibitemOpen
  \bibfield  {author} {\bibinfo {author} {\bibfnamefont {I.~G.}\ \bibnamefont
  {Irastorza}}\ and\ \bibinfo {author} {\bibfnamefont {J.}~\bibnamefont
  {Redondo}},\ }\href {\doibase 10.1016/j.ppnp.2018.05.003} {\bibfield
  {journal} {\bibinfo  {journal} {Prog. Part. Nucl. Phys.}\ }\textbf {\bibinfo
  {volume} {102}},\ \bibinfo {pages} {89} (\bibinfo {year} {2018})},\ \Eprint
  {http://arxiv.org/abs/1801.08127} {arXiv:1801.08127 [hep-ph]} \BibitemShut
  {NoStop}%
\bibitem [{\citenamefont {Marsh}(2011)}]{Marsh:2011gr}%
  \BibitemOpen
  \bibfield  {author} {\bibinfo {author} {\bibfnamefont {D.~J.~E.}\
  \bibnamefont {Marsh}},\ }\href {\doibase 10.1103/PhysRevD.83.123526}
  {\bibfield  {journal} {\bibinfo  {journal} {Phys. Rev.}\ }\textbf {\bibinfo
  {volume} {D83}},\ \bibinfo {pages} {123526} (\bibinfo {year} {2011})},\
  \Eprint {http://arxiv.org/abs/1102.4851} {arXiv:1102.4851 [astro-ph.CO]}
  \BibitemShut {NoStop}%
\bibitem [{\citenamefont {Cicoli}\ \emph {et~al.}(2012)\citenamefont {Cicoli},
  \citenamefont {Goodsell},\ and\ \citenamefont {Ringwald}}]{Cicoli:2012sz}%
  \BibitemOpen
  \bibfield  {author} {\bibinfo {author} {\bibfnamefont {M.}~\bibnamefont
  {Cicoli}}, \bibinfo {author} {\bibfnamefont {M.}~\bibnamefont {Goodsell}}, \
  and\ \bibinfo {author} {\bibfnamefont {A.}~\bibnamefont {Ringwald}},\ }\href
  {\doibase 10.1007/JHEP10(2012)146} {\bibfield  {journal} {\bibinfo  {journal}
  {JHEP}\ }\textbf {\bibinfo {volume} {10}},\ \bibinfo {pages} {146} (\bibinfo
  {year} {2012})},\ \Eprint {http://arxiv.org/abs/1206.0819} {arXiv:1206.0819
  [hep-th]} \BibitemShut {NoStop}%
\bibitem [{\citenamefont {Marsh}\ \emph {et~al.}(2013)\citenamefont {Marsh},
  \citenamefont {Grin}, \citenamefont {Hlozek},\ and\ \citenamefont
  {Ferreira}}]{Marsh:2013taa}%
  \BibitemOpen
  \bibfield  {author} {\bibinfo {author} {\bibfnamefont {D.~J.~E.}\
  \bibnamefont {Marsh}}, \bibinfo {author} {\bibfnamefont {D.}~\bibnamefont
  {Grin}}, \bibinfo {author} {\bibfnamefont {R.}~\bibnamefont {Hlozek}}, \ and\
  \bibinfo {author} {\bibfnamefont {P.~G.}\ \bibnamefont {Ferreira}},\ }\href
  {\doibase 10.1103/PhysRevD.87.121701} {\bibfield  {journal} {\bibinfo
  {journal} {Phys. Rev.}\ }\textbf {\bibinfo {volume} {D87}},\ \bibinfo {pages}
  {121701} (\bibinfo {year} {2013})},\ \Eprint {http://arxiv.org/abs/1303.3008}
  {arXiv:1303.3008 [astro-ph.CO]} \BibitemShut {NoStop}%
\bibitem [{\citenamefont {Tashiro}\ \emph {et~al.}(2013)\citenamefont
  {Tashiro}, \citenamefont {Silk},\ and\ \citenamefont
  {Marsh}}]{Tashiro:2013yea}%
  \BibitemOpen
  \bibfield  {author} {\bibinfo {author} {\bibfnamefont {H.}~\bibnamefont
  {Tashiro}}, \bibinfo {author} {\bibfnamefont {J.}~\bibnamefont {Silk}}, \
  and\ \bibinfo {author} {\bibfnamefont {D.~J.~E.}\ \bibnamefont {Marsh}},\
  }\href {\doibase 10.1103/PhysRevD.88.125024} {\bibfield  {journal} {\bibinfo
  {journal} {Phys. Rev.}\ }\textbf {\bibinfo {volume} {D88}},\ \bibinfo {pages}
  {125024} (\bibinfo {year} {2013})},\ \Eprint {http://arxiv.org/abs/1308.0314}
  {arXiv:1308.0314 [astro-ph.CO]} \BibitemShut {NoStop}%
\bibitem [{\citenamefont {Yoshino}\ and\ \citenamefont
  {Kodama}(2015{\natexlab{a}})}]{Yoshino:2014wwa}%
  \BibitemOpen
  \bibfield  {author} {\bibinfo {author} {\bibfnamefont {H.}~\bibnamefont
  {Yoshino}}\ and\ \bibinfo {author} {\bibfnamefont {H.}~\bibnamefont
  {Kodama}},\ }\href {\doibase 10.1093/ptep/ptv067} {\bibfield  {journal}
  {\bibinfo  {journal} {PTEP}\ }\textbf {\bibinfo {volume} {2015}},\ \bibinfo
  {pages} {061E01} (\bibinfo {year} {2015}{\natexlab{a}})},\ \Eprint
  {http://arxiv.org/abs/1407.2030} {arXiv:1407.2030 [gr-qc]} \BibitemShut
  {NoStop}%
\bibitem [{\citenamefont {Kamionkowski}\ \emph {et~al.}(2014)\citenamefont
  {Kamionkowski}, \citenamefont {Pradler},\ and\ \citenamefont
  {Walker}}]{Kamionkowski:2014zda}%
  \BibitemOpen
  \bibfield  {author} {\bibinfo {author} {\bibfnamefont {M.}~\bibnamefont
  {Kamionkowski}}, \bibinfo {author} {\bibfnamefont {J.}~\bibnamefont
  {Pradler}}, \ and\ \bibinfo {author} {\bibfnamefont {D.~G.~E.}\ \bibnamefont
  {Walker}},\ }\href {\doibase 10.1103/PhysRevLett.113.251302} {\bibfield
  {journal} {\bibinfo  {journal} {Phys. Rev. Lett.}\ }\textbf {\bibinfo
  {volume} {113}},\ \bibinfo {pages} {251302} (\bibinfo {year} {2014})},\
  \Eprint {http://arxiv.org/abs/1409.0549} {arXiv:1409.0549 [hep-ph]}
  \BibitemShut {NoStop}%
\bibitem [{\citenamefont {Obata}\ \emph {et~al.}(2015)\citenamefont {Obata},
  \citenamefont {Miura},\ and\ \citenamefont {Soda}}]{Obata:2014loa}%
  \BibitemOpen
  \bibfield  {author} {\bibinfo {author} {\bibfnamefont {I.}~\bibnamefont
  {Obata}}, \bibinfo {author} {\bibfnamefont {T.}~\bibnamefont {Miura}}, \ and\
  \bibinfo {author} {\bibfnamefont {J.}~\bibnamefont {Soda}},\ }\href {\doibase
  10.1103/PhysRevD.95.109902, 10.1103/PhysRevD.92.063516} {\bibfield  {journal}
  {\bibinfo  {journal} {Phys. Rev.}\ }\textbf {\bibinfo {volume} {D92}},\
  \bibinfo {pages} {063516} (\bibinfo {year} {2015})},\ \bibinfo {note}
  {[Addendum: Phys. Rev.D95,no.10,109902(2017)]},\ \Eprint
  {http://arxiv.org/abs/1412.7620} {arXiv:1412.7620 [hep-ph]} \BibitemShut
  {NoStop}%
\bibitem [{\citenamefont {Yoshino}\ and\ \citenamefont
  {Kodama}(2015{\natexlab{b}})}]{Yoshino:2015nsa}%
  \BibitemOpen
  \bibfield  {author} {\bibinfo {author} {\bibfnamefont {H.}~\bibnamefont
  {Yoshino}}\ and\ \bibinfo {author} {\bibfnamefont {H.}~\bibnamefont
  {Kodama}},\ }\href {\doibase 10.1088/0264-9381/32/21/214001} {\bibfield
  {journal} {\bibinfo  {journal} {Class. Quant. Grav.}\ }\textbf {\bibinfo
  {volume} {32}},\ \bibinfo {pages} {214001} (\bibinfo {year}
  {2015}{\natexlab{b}})},\ \Eprint {http://arxiv.org/abs/1505.00714}
  {arXiv:1505.00714 [gr-qc]} \BibitemShut {NoStop}%
\bibitem [{\citenamefont {Daido}\ \emph {et~al.}(2015)\citenamefont {Daido},
  \citenamefont {Kitajima},\ and\ \citenamefont {Takahashi}}]{Daido:2015bva}%
  \BibitemOpen
  \bibfield  {author} {\bibinfo {author} {\bibfnamefont {R.}~\bibnamefont
  {Daido}}, \bibinfo {author} {\bibfnamefont {N.}~\bibnamefont {Kitajima}}, \
  and\ \bibinfo {author} {\bibfnamefont {F.}~\bibnamefont {Takahashi}},\ }\href
  {\doibase 10.1103/PhysRevD.92.063512} {\bibfield  {journal} {\bibinfo
  {journal} {Phys. Rev.}\ }\textbf {\bibinfo {volume} {D92}},\ \bibinfo {pages}
  {063512} (\bibinfo {year} {2015})},\ \Eprint
  {http://arxiv.org/abs/1505.07670} {arXiv:1505.07670 [hep-ph]} \BibitemShut
  {NoStop}%
\bibitem [{\citenamefont {Acharya}\ and\ \citenamefont
  {Pongkitivanichkul}(2016)}]{Acharya:2015zfk}%
  \BibitemOpen
  \bibfield  {author} {\bibinfo {author} {\bibfnamefont {B.~S.}\ \bibnamefont
  {Acharya}}\ and\ \bibinfo {author} {\bibfnamefont {C.}~\bibnamefont
  {Pongkitivanichkul}},\ }\href {\doibase 10.1007/JHEP04(2016)009} {\bibfield
  {journal} {\bibinfo  {journal} {JHEP}\ }\textbf {\bibinfo {volume} {04}},\
  \bibinfo {pages} {009} (\bibinfo {year} {2016})},\ \Eprint
  {http://arxiv.org/abs/1512.07907} {arXiv:1512.07907 [hep-ph]} \BibitemShut
  {NoStop}%
\bibitem [{\citenamefont {Emami}\ \emph {et~al.}(2016)\citenamefont {Emami},
  \citenamefont {Grin}, \citenamefont {Pradler}, \citenamefont {Raccanelli},\
  and\ \citenamefont {Kamionkowski}}]{Emami:2016mrt}%
  \BibitemOpen
  \bibfield  {author} {\bibinfo {author} {\bibfnamefont {R.}~\bibnamefont
  {Emami}}, \bibinfo {author} {\bibfnamefont {D.}~\bibnamefont {Grin}},
  \bibinfo {author} {\bibfnamefont {J.}~\bibnamefont {Pradler}}, \bibinfo
  {author} {\bibfnamefont {A.}~\bibnamefont {Raccanelli}}, \ and\ \bibinfo
  {author} {\bibfnamefont {M.}~\bibnamefont {Kamionkowski}},\ }\href {\doibase
  10.1103/PhysRevD.93.123005} {\bibfield  {journal} {\bibinfo  {journal} {Phys.
  Rev.}\ }\textbf {\bibinfo {volume} {D93}},\ \bibinfo {pages} {123005}
  (\bibinfo {year} {2016})},\ \Eprint {http://arxiv.org/abs/1603.04851}
  {arXiv:1603.04851 [astro-ph.CO]} \BibitemShut {NoStop}%
\bibitem [{\citenamefont {Karwal}\ and\ \citenamefont
  {Kamionkowski}(2016)}]{Karwal:2016vyq}%
  \BibitemOpen
  \bibfield  {author} {\bibinfo {author} {\bibfnamefont {T.}~\bibnamefont
  {Karwal}}\ and\ \bibinfo {author} {\bibfnamefont {M.}~\bibnamefont
  {Kamionkowski}},\ }\href {\doibase 10.1103/PhysRevD.94.103523} {\bibfield
  {journal} {\bibinfo  {journal} {Phys. Rev.}\ }\textbf {\bibinfo {volume}
  {D94}},\ \bibinfo {pages} {103523} (\bibinfo {year} {2016})},\ \Eprint
  {http://arxiv.org/abs/1608.01309} {arXiv:1608.01309 [astro-ph.CO]}
  \BibitemShut {NoStop}%
\bibitem [{\citenamefont {Gorbunov}\ and\ \citenamefont
  {Tokareva}(2017)}]{Gorbunov:2017ayg}%
  \BibitemOpen
  \bibfield  {author} {\bibinfo {author} {\bibfnamefont {D.}~\bibnamefont
  {Gorbunov}}\ and\ \bibinfo {author} {\bibfnamefont {A.}~\bibnamefont
  {Tokareva}},\ }\href {\doibase 10.1088/1475-7516/2017/06/016} {\bibfield
  {journal} {\bibinfo  {journal} {JCAP}\ }\textbf {\bibinfo {volume} {1706}},\
  \bibinfo {pages} {016} (\bibinfo {year} {2017})},\ \Eprint
  {http://arxiv.org/abs/1702.05924} {arXiv:1702.05924 [hep-ph]} \BibitemShut
  {NoStop}%
\bibitem [{\citenamefont {Yoshida}\ and\ \citenamefont
  {Soda}(2018)}]{Yoshida:2017cjl}%
  \BibitemOpen
  \bibfield  {author} {\bibinfo {author} {\bibfnamefont {D.}~\bibnamefont
  {Yoshida}}\ and\ \bibinfo {author} {\bibfnamefont {J.}~\bibnamefont {Soda}},\
  }\href {\doibase 10.1142/S0218271818500967} {\bibfield  {journal} {\bibinfo
  {journal} {Int. J. Mod. Phys.}\ }\textbf {\bibinfo {volume} {D27}},\ \bibinfo
  {pages} {1850096} (\bibinfo {year} {2018})},\ \Eprint
  {http://arxiv.org/abs/1708.09592} {arXiv:1708.09592 [gr-qc]} \BibitemShut
  {NoStop}%
\bibitem [{\citenamefont {Emami}\ \emph {et~al.}(2018)\citenamefont {Emami},
  \citenamefont {Broadhurst}, \citenamefont {Smoot}, \citenamefont {Chiueh},\
  and\ \citenamefont {Nhan}}]{Emami:2018rxq}%
  \BibitemOpen
  \bibfield  {author} {\bibinfo {author} {\bibfnamefont {R.}~\bibnamefont
  {Emami}}, \bibinfo {author} {\bibfnamefont {T.}~\bibnamefont {Broadhurst}},
  \bibinfo {author} {\bibfnamefont {G.}~\bibnamefont {Smoot}}, \bibinfo
  {author} {\bibfnamefont {T.}~\bibnamefont {Chiueh}}, \ and\ \bibinfo {author}
  {\bibfnamefont {L.~H.}\ \bibnamefont {Nhan}},\ }\href@noop {} {\  (\bibinfo
  {year} {2018})},\ \Eprint {http://arxiv.org/abs/1806.04518} {arXiv:1806.04518
  [astro-ph.CO]} \BibitemShut {NoStop}%
\bibitem [{\citenamefont {Bae}\ \emph {et~al.}(2015{\natexlab{a}})\citenamefont
  {Bae}, \citenamefont {Baer}, \citenamefont {Chun},\ and\ \citenamefont
  {Shin}}]{Bae:2014efa}%
  \BibitemOpen
  \bibfield  {author} {\bibinfo {author} {\bibfnamefont {K.~J.}\ \bibnamefont
  {Bae}}, \bibinfo {author} {\bibfnamefont {H.}~\bibnamefont {Baer}}, \bibinfo
  {author} {\bibfnamefont {E.~J.}\ \bibnamefont {Chun}}, \ and\ \bibinfo
  {author} {\bibfnamefont {C.~S.}\ \bibnamefont {Shin}},\ }\href {\doibase
  10.1103/PhysRevD.91.075011} {\bibfield  {journal} {\bibinfo  {journal} {Phys.
  Rev.}\ }\textbf {\bibinfo {volume} {D91}},\ \bibinfo {pages} {075011}
  (\bibinfo {year} {2015}{\natexlab{a}})},\ \Eprint
  {http://arxiv.org/abs/1410.3857} {arXiv:1410.3857 [hep-ph]} \BibitemShut
  {NoStop}%
\bibitem [{\citenamefont {Visinelli}(2015)}]{Visinelli:2014xsa}%
  \BibitemOpen
  \bibfield  {author} {\bibinfo {author} {\bibfnamefont {L.}~\bibnamefont
  {Visinelli}},\ }\href {\doibase 10.1007/s10714-015-1899-z} {\bibfield
  {journal} {\bibinfo  {journal} {Gen. Rel. Grav.}\ }\textbf {\bibinfo {volume}
  {47}},\ \bibinfo {pages} {62} (\bibinfo {year} {2015})},\ \Eprint
  {http://arxiv.org/abs/1410.1523} {arXiv:1410.1523 [gr-qc]} \BibitemShut
  {NoStop}%
\bibitem [{\citenamefont {Bae}\ \emph {et~al.}(2015{\natexlab{b}})\citenamefont
  {Bae}, \citenamefont {Baer}, \citenamefont {Lessa},\ and\ \citenamefont
  {Serce}}]{Bae:2015rra}%
  \BibitemOpen
  \bibfield  {author} {\bibinfo {author} {\bibfnamefont {K.~J.}\ \bibnamefont
  {Bae}}, \bibinfo {author} {\bibfnamefont {H.}~\bibnamefont {Baer}}, \bibinfo
  {author} {\bibfnamefont {A.}~\bibnamefont {Lessa}}, \ and\ \bibinfo {author}
  {\bibfnamefont {H.}~\bibnamefont {Serce}},\ }\href {\doibase
  10.3389/fphy.2015.00049} {\bibfield  {journal} {\bibinfo  {journal} {Front.in
  Phys.}\ }\textbf {\bibinfo {volume} {3}},\ \bibinfo {pages} {49} (\bibinfo
  {year} {2015}{\natexlab{b}})},\ \Eprint {http://arxiv.org/abs/1502.07198}
  {arXiv:1502.07198 [hep-ph]} \BibitemShut {NoStop}%
\bibitem [{\citenamefont {Baum}\ \emph {et~al.}(2017)\citenamefont {Baum},
  \citenamefont {Visinelli}, \citenamefont {Freese},\ and\ \citenamefont
  {Stengel}}]{Baum:2016oow}%
  \BibitemOpen
  \bibfield  {author} {\bibinfo {author} {\bibfnamefont {S.}~\bibnamefont
  {Baum}}, \bibinfo {author} {\bibfnamefont {L.}~\bibnamefont {Visinelli}},
  \bibinfo {author} {\bibfnamefont {K.}~\bibnamefont {Freese}}, \ and\ \bibinfo
  {author} {\bibfnamefont {P.}~\bibnamefont {Stengel}},\ }\href {\doibase
  10.1103/PhysRevD.95.043007} {\bibfield  {journal} {\bibinfo  {journal} {Phys.
  Rev.}\ }\textbf {\bibinfo {volume} {D95}},\ \bibinfo {pages} {043007}
  (\bibinfo {year} {2017})},\ \Eprint {http://arxiv.org/abs/1611.09665}
  {arXiv:1611.09665 [astro-ph.CO]} \BibitemShut {NoStop}%
\bibitem [{\citenamefont {Arkani-Hamed}\ \emph {et~al.}(2016)\citenamefont
  {Arkani-Hamed}, \citenamefont {Han}, \citenamefont {Mangano},\ and\
  \citenamefont {Wang}}]{Arkani-Hamed:2015vfh}%
  \BibitemOpen
  \bibfield  {author} {\bibinfo {author} {\bibfnamefont {N.}~\bibnamefont
  {Arkani-Hamed}}, \bibinfo {author} {\bibfnamefont {T.}~\bibnamefont {Han}},
  \bibinfo {author} {\bibfnamefont {M.}~\bibnamefont {Mangano}}, \ and\
  \bibinfo {author} {\bibfnamefont {L.-T.}\ \bibnamefont {Wang}},\ }\href
  {\doibase 10.1016/j.physrep.2016.07.004} {\bibfield  {journal} {\bibinfo
  {journal} {Phys. Rept.}\ }\textbf {\bibinfo {volume} {652}},\ \bibinfo
  {pages} {1} (\bibinfo {year} {2016})},\ \Eprint
  {http://arxiv.org/abs/1511.06495} {arXiv:1511.06495 [hep-ph]} \BibitemShut
  {NoStop}%
\bibitem [{\citenamefont {Pugnat}\ \emph {et~al.}(2008)\citenamefont {Pugnat}
  \emph {et~al.}}]{Pugnat:2007nu}%
  \BibitemOpen
  \bibfield  {author} {\bibinfo {author} {\bibfnamefont {P.}~\bibnamefont
  {Pugnat}} \emph {et~al.} (\bibinfo {collaboration} {OSQAR}),\ }\href
  {\doibase 10.1103/PhysRevD.78.092003} {\bibfield  {journal} {\bibinfo
  {journal} {Phys. Rev.}\ }\textbf {\bibinfo {volume} {D78}},\ \bibinfo {pages}
  {092003} (\bibinfo {year} {2008})},\ \Eprint {http://arxiv.org/abs/0712.3362}
  {arXiv:0712.3362 [hep-ex]} \BibitemShut {NoStop}%
\bibitem [{\citenamefont {Ballou}\ \emph {et~al.}(2015)\citenamefont {Ballou}
  \emph {et~al.}}]{Ballou:2015cka}%
  \BibitemOpen
  \bibfield  {author} {\bibinfo {author} {\bibfnamefont {R.}~\bibnamefont
  {Ballou}} \emph {et~al.} (\bibinfo {collaboration} {OSQAR}),\ }\href
  {\doibase 10.1103/PhysRevD.92.092002} {\bibfield  {journal} {\bibinfo
  {journal} {Phys. Rev.}\ }\textbf {\bibinfo {volume} {D92}},\ \bibinfo {pages}
  {092002} (\bibinfo {year} {2015})},\ \Eprint
  {http://arxiv.org/abs/1506.08082} {arXiv:1506.08082 [hep-ex]} \BibitemShut
  {NoStop}%
\bibitem [{\citenamefont {Ehret}\ \emph {et~al.}(2009)\citenamefont {Ehret}
  \emph {et~al.}}]{Ehret:2009sq}%
  \BibitemOpen
  \bibfield  {author} {\bibinfo {author} {\bibfnamefont {K.}~\bibnamefont
  {Ehret}} \emph {et~al.} (\bibinfo {collaboration} {ALPS}),\ }\href {\doibase
  10.1016/j.nima.2009.10.102} {\bibfield  {journal} {\bibinfo  {journal} {Nucl.
  Instrum. Meth.}\ }\textbf {\bibinfo {volume} {A612}},\ \bibinfo {pages} {83}
  (\bibinfo {year} {2009})},\ \Eprint {http://arxiv.org/abs/0905.4159}
  {arXiv:0905.4159 [physics.ins-det]} \BibitemShut {NoStop}%
\bibitem [{\citenamefont {Ehret}\ \emph {et~al.}(2010)\citenamefont {Ehret}
  \emph {et~al.}}]{Ehret:2010mh}%
  \BibitemOpen
  \bibfield  {author} {\bibinfo {author} {\bibfnamefont {K.}~\bibnamefont
  {Ehret}} \emph {et~al.},\ }\href {\doibase 10.1016/j.physletb.2010.04.066}
  {\bibfield  {journal} {\bibinfo  {journal} {Phys. Lett.}\ }\textbf {\bibinfo
  {volume} {B689}},\ \bibinfo {pages} {149} (\bibinfo {year} {2010})},\ \Eprint
  {http://arxiv.org/abs/1004.1313} {arXiv:1004.1313 [hep-ex]} \BibitemShut
  {NoStop}%
\bibitem [{\citenamefont {Sikivie}(1983)}]{Sikivie:1983ip}%
  \BibitemOpen
  \bibfield  {author} {\bibinfo {author} {\bibfnamefont {P.}~\bibnamefont
  {Sikivie}},\ }\bibfield  {booktitle} {\emph {\bibinfo {booktitle} {{Particle
  physics and cosmology: Dark matter}}},\ }\href {\doibase
  10.1103/PhysRevLett.51.1415, 10.1103/PhysRevLett.52.695.2} {\bibfield
  {journal} {\bibinfo  {journal} {Phys. Rev. Lett.}\ }\textbf {\bibinfo
  {volume} {51}},\ \bibinfo {pages} {1415} (\bibinfo {year} {1983})},\ \bibinfo
  {note} {[,321(1983)]}\BibitemShut {NoStop}%
\bibitem [{\citenamefont {Sikivie}(1985)}]{Sikivie:1985yu}%
  \BibitemOpen
  \bibfield  {author} {\bibinfo {author} {\bibfnamefont {P.}~\bibnamefont
  {Sikivie}},\ }\href {\doibase 10.1103/PhysRevD.36.974,
  10.1103/PhysRevD.32.2988} {\bibfield  {journal} {\bibinfo  {journal} {Phys.
  Rev.}\ }\textbf {\bibinfo {volume} {D32}},\ \bibinfo {pages} {2988} (\bibinfo
  {year} {1985})},\ \bibinfo {note} {[Erratum: Phys.
  Rev.D36,974(1987)]}\BibitemShut {NoStop}%
\bibitem [{\citenamefont {Asztalos}\ \emph {et~al.}(2001)\citenamefont
  {Asztalos} \emph {et~al.}}]{Asztalos:2001tf}%
  \BibitemOpen
  \bibfield  {author} {\bibinfo {author} {\bibfnamefont {S.~J.}\ \bibnamefont
  {Asztalos}} \emph {et~al.} (\bibinfo {collaboration} {ADMX}),\ }\href
  {\doibase 10.1103/PhysRevD.64.092003} {\bibfield  {journal} {\bibinfo
  {journal} {Phys. Rev.}\ }\textbf {\bibinfo {volume} {D64}},\ \bibinfo {pages}
  {092003} (\bibinfo {year} {2001})}\BibitemShut {NoStop}%
\bibitem [{\citenamefont {Al~Kenany}\ \emph {et~al.}(2017)\citenamefont
  {Al~Kenany} \emph {et~al.}}]{Kenany:2016tta}%
  \BibitemOpen
  \bibfield  {author} {\bibinfo {author} {\bibfnamefont {S.}~\bibnamefont
  {Al~Kenany}} \emph {et~al.},\ }\href {\doibase 10.1016/j.nima.2017.02.012}
  {\bibfield  {journal} {\bibinfo  {journal} {Nucl. Instrum. Meth.}\ }\textbf
  {\bibinfo {volume} {A854}},\ \bibinfo {pages} {11} (\bibinfo {year}
  {2017})},\ \Eprint {http://arxiv.org/abs/1611.07123} {arXiv:1611.07123
  [physics.ins-det]} \BibitemShut {NoStop}%
\bibitem [{\citenamefont {van Bibber}\ \emph {et~al.}(1989)\citenamefont {van
  Bibber}, \citenamefont {McIntyre}, \citenamefont {Morris},\ and\
  \citenamefont {Raffelt}}]{vanBibber:1988ge}%
  \BibitemOpen
  \bibfield  {author} {\bibinfo {author} {\bibfnamefont {K.}~\bibnamefont {van
  Bibber}}, \bibinfo {author} {\bibfnamefont {P.~M.}\ \bibnamefont {McIntyre}},
  \bibinfo {author} {\bibfnamefont {D.~E.}\ \bibnamefont {Morris}}, \ and\
  \bibinfo {author} {\bibfnamefont {G.~G.}\ \bibnamefont {Raffelt}},\ }\href
  {\doibase 10.1103/PhysRevD.39.2089} {\bibfield  {journal} {\bibinfo
  {journal} {Phys. Rev.}\ }\textbf {\bibinfo {volume} {D39}},\ \bibinfo {pages}
  {2089} (\bibinfo {year} {1989})}\BibitemShut {NoStop}%
\bibitem [{\citenamefont {Andriamonje}\ \emph {et~al.}(2007)\citenamefont
  {Andriamonje} \emph {et~al.}}]{Andriamonje:2007ew}%
  \BibitemOpen
  \bibfield  {author} {\bibinfo {author} {\bibfnamefont {S.}~\bibnamefont
  {Andriamonje}} \emph {et~al.} (\bibinfo {collaboration} {CAST}),\ }\href
  {\doibase 10.1088/1475-7516/2007/04/010} {\bibfield  {journal} {\bibinfo
  {journal} {JCAP}\ }\textbf {\bibinfo {volume} {0704}},\ \bibinfo {pages}
  {010} (\bibinfo {year} {2007})},\ \Eprint
  {http://arxiv.org/abs/hep-ex/0702006} {arXiv:hep-ex/0702006 [hep-ex]}
  \BibitemShut {NoStop}%
\bibitem [{\citenamefont {Anastassopoulos}\ \emph {et~al.}(2017)\citenamefont
  {Anastassopoulos} \emph {et~al.}}]{Anastassopoulos:2017ftl}%
  \BibitemOpen
  \bibfield  {author} {\bibinfo {author} {\bibfnamefont {V.}~\bibnamefont
  {Anastassopoulos}} \emph {et~al.} (\bibinfo {collaboration} {CAST}),\ }\href
  {\doibase 10.1038/nphys4109} {\bibfield  {journal} {\bibinfo  {journal}
  {Nature Phys.}\ }\textbf {\bibinfo {volume} {13}},\ \bibinfo {pages} {584}
  (\bibinfo {year} {2017})},\ \Eprint {http://arxiv.org/abs/1705.02290}
  {arXiv:1705.02290 [hep-ex]} \BibitemShut {NoStop}%
\bibitem [{\citenamefont {Payez}\ \emph {et~al.}(2015)\citenamefont {Payez},
  \citenamefont {Evoli}, \citenamefont {Fischer}, \citenamefont {Giannotti},
  \citenamefont {Mirizzi},\ and\ \citenamefont {Ringwald}}]{Payez:2014xsa}%
  \BibitemOpen
  \bibfield  {author} {\bibinfo {author} {\bibfnamefont {A.}~\bibnamefont
  {Payez}}, \bibinfo {author} {\bibfnamefont {C.}~\bibnamefont {Evoli}},
  \bibinfo {author} {\bibfnamefont {T.}~\bibnamefont {Fischer}}, \bibinfo
  {author} {\bibfnamefont {M.}~\bibnamefont {Giannotti}}, \bibinfo {author}
  {\bibfnamefont {A.}~\bibnamefont {Mirizzi}}, \ and\ \bibinfo {author}
  {\bibfnamefont {A.}~\bibnamefont {Ringwald}},\ }\href {\doibase
  10.1088/1475-7516/2015/02/006} {\bibfield  {journal} {\bibinfo  {journal}
  {JCAP}\ }\textbf {\bibinfo {volume} {1502}},\ \bibinfo {pages} {006}
  (\bibinfo {year} {2015})},\ \Eprint {http://arxiv.org/abs/1410.3747}
  {arXiv:1410.3747 [astro-ph.HE]} \BibitemShut {NoStop}%
\bibitem [{\citenamefont {Giannotti}\ \emph {et~al.}(2016)\citenamefont
  {Giannotti}, \citenamefont {Irastorza}, \citenamefont {Redondo},\ and\
  \citenamefont {Ringwald}}]{Giannotti:2015kwo}%
  \BibitemOpen
  \bibfield  {author} {\bibinfo {author} {\bibfnamefont {M.}~\bibnamefont
  {Giannotti}}, \bibinfo {author} {\bibfnamefont {I.}~\bibnamefont
  {Irastorza}}, \bibinfo {author} {\bibfnamefont {J.}~\bibnamefont {Redondo}},
  \ and\ \bibinfo {author} {\bibfnamefont {A.}~\bibnamefont {Ringwald}},\
  }\href {\doibase 10.1088/1475-7516/2016/05/057} {\bibfield  {journal}
  {\bibinfo  {journal} {JCAP}\ }\textbf {\bibinfo {volume} {1605}},\ \bibinfo
  {pages} {057} (\bibinfo {year} {2016})},\ \Eprint
  {http://arxiv.org/abs/1512.08108} {arXiv:1512.08108 [astro-ph.HE]}
  \BibitemShut {NoStop}%
\bibitem [{\citenamefont {Giannotti}\ \emph {et~al.}(2017)\citenamefont
  {Giannotti}, \citenamefont {Irastorza}, \citenamefont {Redondo},
  \citenamefont {Ringwald},\ and\ \citenamefont {Saikawa}}]{Giannotti:2017hny}%
  \BibitemOpen
  \bibfield  {author} {\bibinfo {author} {\bibfnamefont {M.}~\bibnamefont
  {Giannotti}}, \bibinfo {author} {\bibfnamefont {I.~G.}\ \bibnamefont
  {Irastorza}}, \bibinfo {author} {\bibfnamefont {J.}~\bibnamefont {Redondo}},
  \bibinfo {author} {\bibfnamefont {A.}~\bibnamefont {Ringwald}}, \ and\
  \bibinfo {author} {\bibfnamefont {K.}~\bibnamefont {Saikawa}},\ }\href
  {\doibase 10.1088/1475-7516/2017/10/010} {\bibfield  {journal} {\bibinfo
  {journal} {JCAP}\ }\textbf {\bibinfo {volume} {1710}},\ \bibinfo {pages}
  {010} (\bibinfo {year} {2017})},\ \Eprint {http://arxiv.org/abs/1708.02111}
  {arXiv:1708.02111 [hep-ph]} \BibitemShut {NoStop}%
\bibitem [{\citenamefont {Viaux}\ \emph {et~al.}(2013)\citenamefont {Viaux},
  \citenamefont {Catelan}, \citenamefont {Stetson}, \citenamefont {Raffelt},
  \citenamefont {Redondo}, \citenamefont {Valcarce},\ and\ \citenamefont
  {Weiss}}]{Viaux:2013lha}%
  \BibitemOpen
  \bibfield  {author} {\bibinfo {author} {\bibfnamefont {N.}~\bibnamefont
  {Viaux}}, \bibinfo {author} {\bibfnamefont {M.}~\bibnamefont {Catelan}},
  \bibinfo {author} {\bibfnamefont {P.~B.}\ \bibnamefont {Stetson}}, \bibinfo
  {author} {\bibfnamefont {G.}~\bibnamefont {Raffelt}}, \bibinfo {author}
  {\bibfnamefont {J.}~\bibnamefont {Redondo}}, \bibinfo {author} {\bibfnamefont
  {A.~A.~R.}\ \bibnamefont {Valcarce}}, \ and\ \bibinfo {author} {\bibfnamefont
  {A.}~\bibnamefont {Weiss}},\ }\href {\doibase 10.1103/PhysRevLett.111.231301}
  {\bibfield  {journal} {\bibinfo  {journal} {Phys. Rev. Lett.}\ }\textbf
  {\bibinfo {volume} {111}},\ \bibinfo {pages} {231301} (\bibinfo {year}
  {2013})},\ \Eprint {http://arxiv.org/abs/1311.1669} {arXiv:1311.1669
  [astro-ph.SR]} \BibitemShut {NoStop}%
\bibitem [{\citenamefont {Kobayashi}\ \emph {et~al.}(2017)\citenamefont
  {Kobayashi}, \citenamefont {Murgia}, \citenamefont {De~Simone}, \citenamefont
  {Ir{\v s}i{\v c}},\ and\ \citenamefont {Viel}}]{Kobayashi:2017jcf}%
  \BibitemOpen
  \bibfield  {author} {\bibinfo {author} {\bibfnamefont {T.}~\bibnamefont
  {Kobayashi}}, \bibinfo {author} {\bibfnamefont {R.}~\bibnamefont {Murgia}},
  \bibinfo {author} {\bibfnamefont {A.}~\bibnamefont {De~Simone}}, \bibinfo
  {author} {\bibfnamefont {V.}~\bibnamefont {Ir{\v s}i{\v c}}}, \ and\ \bibinfo
  {author} {\bibfnamefont {M.}~\bibnamefont {Viel}},\ }\href {\doibase
  10.1103/PhysRevD.96.123514} {\bibfield  {journal} {\bibinfo  {journal} {Phys.
  Rev.}\ }\textbf {\bibinfo {volume} {D96}},\ \bibinfo {pages} {123514}
  (\bibinfo {year} {2017})},\ \Eprint {http://arxiv.org/abs/1708.00015}
  {arXiv:1708.00015 [astro-ph.CO]} \BibitemShut {NoStop}%
\bibitem [{\citenamefont {Desjacques}\ \emph {et~al.}(2018)\citenamefont
  {Desjacques}, \citenamefont {Kehagias},\ and\ \citenamefont
  {Riotto}}]{Desjacques:2017fmf}%
  \BibitemOpen
  \bibfield  {author} {\bibinfo {author} {\bibfnamefont {V.}~\bibnamefont
  {Desjacques}}, \bibinfo {author} {\bibfnamefont {A.}~\bibnamefont
  {Kehagias}}, \ and\ \bibinfo {author} {\bibfnamefont {A.}~\bibnamefont
  {Riotto}},\ }\href {\doibase 10.1103/PhysRevD.97.023529} {\bibfield
  {journal} {\bibinfo  {journal} {Phys. Rev.}\ }\textbf {\bibinfo {volume}
  {D97}},\ \bibinfo {pages} {023529} (\bibinfo {year} {2018})},\ \Eprint
  {http://arxiv.org/abs/1709.07946} {arXiv:1709.07946 [astro-ph.CO]}
  \BibitemShut {NoStop}%
\bibitem [{\citenamefont {Poulin}\ \emph {et~al.}(2018)\citenamefont {Poulin},
  \citenamefont {Smith}, \citenamefont {Grin}, \citenamefont {Karwal},\ and\
  \citenamefont {Kamionkowski}}]{Poulin:2018dzj}%
  \BibitemOpen
  \bibfield  {author} {\bibinfo {author} {\bibfnamefont {V.}~\bibnamefont
  {Poulin}}, \bibinfo {author} {\bibfnamefont {T.~L.}\ \bibnamefont {Smith}},
  \bibinfo {author} {\bibfnamefont {D.}~\bibnamefont {Grin}}, \bibinfo {author}
  {\bibfnamefont {T.}~\bibnamefont {Karwal}}, \ and\ \bibinfo {author}
  {\bibfnamefont {M.}~\bibnamefont {Kamionkowski}},\ }\href@noop {} {\
  (\bibinfo {year} {2018})},\ \Eprint {http://arxiv.org/abs/1806.10608}
  {arXiv:1806.10608 [astro-ph.CO]} \BibitemShut {NoStop}%
\bibitem [{\citenamefont {Hlozek}\ \emph {et~al.}(2015)\citenamefont {Hlozek},
  \citenamefont {Grin}, \citenamefont {Marsh},\ and\ \citenamefont
  {Ferreira}}]{Hlozek:2014lca}%
  \BibitemOpen
  \bibfield  {author} {\bibinfo {author} {\bibfnamefont {R.}~\bibnamefont
  {Hlozek}}, \bibinfo {author} {\bibfnamefont {D.}~\bibnamefont {Grin}},
  \bibinfo {author} {\bibfnamefont {D.~J.~E.}\ \bibnamefont {Marsh}}, \ and\
  \bibinfo {author} {\bibfnamefont {P.~G.}\ \bibnamefont {Ferreira}},\ }\href
  {\doibase 10.1103/PhysRevD.91.103512} {\bibfield  {journal} {\bibinfo
  {journal} {Phys. Rev.}\ }\textbf {\bibinfo {volume} {D91}},\ \bibinfo {pages}
  {103512} (\bibinfo {year} {2015})},\ \Eprint {http://arxiv.org/abs/1410.2896}
  {arXiv:1410.2896 [astro-ph.CO]} \BibitemShut {NoStop}%
\bibitem [{\citenamefont {Hlozek}\ \emph {et~al.}(2018)\citenamefont {Hlozek},
  \citenamefont {Marsh},\ and\ \citenamefont {Grin}}]{Hlozek:2017zzf}%
  \BibitemOpen
  \bibfield  {author} {\bibinfo {author} {\bibfnamefont {R.}~\bibnamefont
  {Hlozek}}, \bibinfo {author} {\bibfnamefont {D.~J.~E.}\ \bibnamefont
  {Marsh}}, \ and\ \bibinfo {author} {\bibfnamefont {D.}~\bibnamefont {Grin}},\
  }\href {\doibase 10.1093/mnras/sty271} {\bibfield  {journal} {\bibinfo
  {journal} {Mon. Not. Roy. Astron. Soc.}\ }\textbf {\bibinfo {volume} {476}},\
  \bibinfo {pages} {3063} (\bibinfo {year} {2018})},\ \Eprint
  {http://arxiv.org/abs/1708.05681} {arXiv:1708.05681 [astro-ph.CO]}
  \BibitemShut {NoStop}%
\bibitem [{\citenamefont {Hoof}\ \emph {et~al.}(2018)\citenamefont {Hoof},
  \citenamefont {Kahlhoefer}, \citenamefont {Scott}, \citenamefont {Weniger},\
  and\ \citenamefont {White}}]{Hoof:2018ieb}%
  \BibitemOpen
  \bibfield  {author} {\bibinfo {author} {\bibfnamefont {S.}~\bibnamefont
  {Hoof}}, \bibinfo {author} {\bibfnamefont {F.}~\bibnamefont {Kahlhoefer}},
  \bibinfo {author} {\bibfnamefont {P.}~\bibnamefont {Scott}}, \bibinfo
  {author} {\bibfnamefont {C.}~\bibnamefont {Weniger}}, \ and\ \bibinfo
  {author} {\bibfnamefont {M.}~\bibnamefont {White}},\ }\href@noop {} {\
  (\bibinfo {year} {2018})},\ \Eprint {http://arxiv.org/abs/1810.07192}
  {arXiv:1810.07192 [hep-ph]} \BibitemShut {NoStop}%
\bibitem [{\citenamefont {Marsh}(2016)}]{Marsh:2015xka}%
  \BibitemOpen
  \bibfield  {author} {\bibinfo {author} {\bibfnamefont {D.~J.~E.}\
  \bibnamefont {Marsh}},\ }\href {\doibase 10.1016/j.physrep.2016.06.005}
  {\bibfield  {journal} {\bibinfo  {journal} {Phys. Rept.}\ }\textbf {\bibinfo
  {volume} {643}},\ \bibinfo {pages} {1} (\bibinfo {year} {2016})},\ \Eprint
  {http://arxiv.org/abs/1510.07633} {arXiv:1510.07633 [astro-ph.CO]}
  \BibitemShut {NoStop}%
\bibitem [{\citenamefont {Ade}\ \emph {et~al.}(2016{\natexlab{a}})\citenamefont
  {Ade} \emph {et~al.}}]{Ade:2015xua}%
  \BibitemOpen
  \bibfield  {author} {\bibinfo {author} {\bibfnamefont {P.~A.~R.}\
  \bibnamefont {Ade}} \emph {et~al.} (\bibinfo {collaboration} {Planck}),\
  }\href {\doibase 10.1051/0004-6361/201525830} {\bibfield  {journal} {\bibinfo
   {journal} {Astron. Astrophys.}\ }\textbf {\bibinfo {volume} {594}},\
  \bibinfo {pages} {A13} (\bibinfo {year} {2016}{\natexlab{a}})},\ \Eprint
  {http://arxiv.org/abs/1502.01589} {arXiv:1502.01589 [astro-ph.CO]}
  \BibitemShut {NoStop}%
\bibitem [{\citenamefont {Kobayashi}\ \emph {et~al.}(2013)\citenamefont
  {Kobayashi}, \citenamefont {Kurematsu},\ and\ \citenamefont
  {Takahashi}}]{Kobayashi:2013nva}%
  \BibitemOpen
  \bibfield  {author} {\bibinfo {author} {\bibfnamefont {T.}~\bibnamefont
  {Kobayashi}}, \bibinfo {author} {\bibfnamefont {R.}~\bibnamefont
  {Kurematsu}}, \ and\ \bibinfo {author} {\bibfnamefont {F.}~\bibnamefont
  {Takahashi}},\ }\href {\doibase 10.1088/1475-7516/2013/09/032} {\bibfield
  {journal} {\bibinfo  {journal} {JCAP}\ }\textbf {\bibinfo {volume} {1309}},\
  \bibinfo {pages} {032} (\bibinfo {year} {2013})},\ \Eprint
  {http://arxiv.org/abs/1304.0922} {arXiv:1304.0922 [hep-ph]} \BibitemShut
  {NoStop}%
\bibitem [{\citenamefont {Axenides}\ \emph {et~al.}(1983)\citenamefont
  {Axenides}, \citenamefont {Brandenberger},\ and\ \citenamefont
  {Turner}}]{Axenides:1983}%
  \BibitemOpen
  \bibfield  {author} {\bibinfo {author} {\bibfnamefont {M.}~\bibnamefont
  {Axenides}}, \bibinfo {author} {\bibfnamefont {R.}~\bibnamefont
  {Brandenberger}}, \ and\ \bibinfo {author} {\bibfnamefont {M.}~\bibnamefont
  {Turner}},\ }\href {\doibase http://dx.doi.org/10.1016/0370-2693(83)90586-5}
  {\bibfield  {journal} {\bibinfo  {journal} {Physics Letters B}\ }\textbf
  {\bibinfo {volume} {126}},\ \bibinfo {pages} {178 } (\bibinfo {year}
  {1983})}\BibitemShut {NoStop}%
\bibitem [{\citenamefont {Linde}(1985)}]{Linde:1985yf}%
  \BibitemOpen
  \bibfield  {author} {\bibinfo {author} {\bibfnamefont {A.~D.}\ \bibnamefont
  {Linde}},\ }\href {\doibase 10.1016/0370-2693(85)90436-8} {\bibfield
  {journal} {\bibinfo  {journal} {Phys. Lett.}\ }\textbf {\bibinfo {volume}
  {B158}},\ \bibinfo {pages} {375} (\bibinfo {year} {1985})}\BibitemShut
  {NoStop}%
\bibitem [{\citenamefont {Seckel}\ and\ \citenamefont
  {Turner}(1985)}]{Seckel:1985}%
  \BibitemOpen
  \bibfield  {author} {\bibinfo {author} {\bibfnamefont {D.}~\bibnamefont
  {Seckel}}\ and\ \bibinfo {author} {\bibfnamefont {M.~S.}\ \bibnamefont
  {Turner}},\ }\href {\doibase 10.1103/PhysRevD.32.3178} {\bibfield  {journal}
  {\bibinfo  {journal} {Phys. Rev. D}\ }\textbf {\bibinfo {volume} {32}},\
  \bibinfo {pages} {3178} (\bibinfo {year} {1985})}\BibitemShut {NoStop}%
\bibitem [{\citenamefont {Maldacena}(2003)}]{Maldacena:2002vr}%
  \BibitemOpen
  \bibfield  {author} {\bibinfo {author} {\bibfnamefont {J.~M.}\ \bibnamefont
  {Maldacena}},\ }\href {\doibase 10.1088/1126-6708/2003/05/013} {\bibfield
  {journal} {\bibinfo  {journal} {JHEP}\ }\textbf {\bibinfo {volume} {05}},\
  \bibinfo {pages} {013} (\bibinfo {year} {2003})},\ \Eprint
  {http://arxiv.org/abs/astro-ph/0210603} {arXiv:astro-ph/0210603 [astro-ph]}
  \BibitemShut {NoStop}%
\bibitem [{\citenamefont {Lyth}(1984)}]{Lyth:1984}%
  \BibitemOpen
  \bibfield  {author} {\bibinfo {author} {\bibfnamefont {D.}~\bibnamefont
  {Lyth}},\ }\href {\doibase http://dx.doi.org/10.1016/0370-2693(84)91391-1}
  {\bibfield  {journal} {\bibinfo  {journal} {Physics Letters B}\ }\textbf
  {\bibinfo {volume} {147}},\ \bibinfo {pages} {403 } (\bibinfo {year}
  {1984})}\BibitemShut {NoStop}%
\bibitem [{\citenamefont {Lyth}(1990)}]{Lyth:1990}%
  \BibitemOpen
  \bibfield  {author} {\bibinfo {author} {\bibfnamefont {D.~H.}\ \bibnamefont
  {Lyth}},\ }\href {\doibase http://dx.doi.org/10.1016/0370-2693(90)90374-F}
  {\bibfield  {journal} {\bibinfo  {journal} {Physics Letters B}\ }\textbf
  {\bibinfo {volume} {236}},\ \bibinfo {pages} {408 } (\bibinfo {year}
  {1990})}\BibitemShut {NoStop}%
\bibitem [{\citenamefont {Lyth}\ and\ \citenamefont
  {Stewart}(1992)}]{Lyth:1992yy}%
  \BibitemOpen
  \bibfield  {author} {\bibinfo {author} {\bibfnamefont {D.~H.}\ \bibnamefont
  {Lyth}}\ and\ \bibinfo {author} {\bibfnamefont {E.~D.}\ \bibnamefont
  {Stewart}},\ }\href {\doibase http://dx.doi.org/10.1016/0370-2693(92)90006-P}
  {\bibfield  {journal} {\bibinfo  {journal} {Physics Letters B}\ }\textbf
  {\bibinfo {volume} {283}},\ \bibinfo {pages} {189 } (\bibinfo {year}
  {1992})}\BibitemShut {NoStop}%
\bibitem [{\citenamefont {Ballesteros}\ \emph {et~al.}(2016)\citenamefont
  {Ballesteros}, \citenamefont {Redondo}, \citenamefont {Ringwald},\ and\
  \citenamefont {Tamarit}}]{Ballesteros:2016xej}%
  \BibitemOpen
  \bibfield  {author} {\bibinfo {author} {\bibfnamefont {G.}~\bibnamefont
  {Ballesteros}}, \bibinfo {author} {\bibfnamefont {J.}~\bibnamefont
  {Redondo}}, \bibinfo {author} {\bibfnamefont {A.}~\bibnamefont {Ringwald}}, \
  and\ \bibinfo {author} {\bibfnamefont {C.}~\bibnamefont {Tamarit}},\
  }\href@noop {} {\  (\bibinfo {year} {2016})},\ \Eprint
  {http://arxiv.org/abs/1610.01639} {arXiv:1610.01639 [hep-ph]} \BibitemShut
  {NoStop}%
\bibitem [{\citenamefont {Kitajima}\ and\ \citenamefont
  {Takahashi}(2015)}]{Kitajima:2014xla}%
  \BibitemOpen
  \bibfield  {author} {\bibinfo {author} {\bibfnamefont {N.}~\bibnamefont
  {Kitajima}}\ and\ \bibinfo {author} {\bibfnamefont {F.}~\bibnamefont
  {Takahashi}},\ }\href {\doibase 10.1088/1475-7516/2015/01/032} {\bibfield
  {journal} {\bibinfo  {journal} {JCAP}\ }\textbf {\bibinfo {volume} {1501}},\
  \bibinfo {pages} {032} (\bibinfo {year} {2015})},\ \Eprint
  {http://arxiv.org/abs/1411.2011} {arXiv:1411.2011 [hep-ph]} \BibitemShut
  {NoStop}%
\bibitem [{\citenamefont {Crotty}\ \emph {et~al.}(2003)\citenamefont {Crotty},
  \citenamefont {Garcia-Bellido}, \citenamefont {Lesgourgues},\ and\
  \citenamefont {Riazuelo}}]{Crotty:2003rz}%
  \BibitemOpen
  \bibfield  {author} {\bibinfo {author} {\bibfnamefont {P.}~\bibnamefont
  {Crotty}}, \bibinfo {author} {\bibfnamefont {J.}~\bibnamefont
  {Garcia-Bellido}}, \bibinfo {author} {\bibfnamefont {J.}~\bibnamefont
  {Lesgourgues}}, \ and\ \bibinfo {author} {\bibfnamefont {A.}~\bibnamefont
  {Riazuelo}},\ }\href {\doibase 10.1103/PhysRevLett.91.171301} {\bibfield
  {journal} {\bibinfo  {journal} {Phys. Rev. Lett.}\ }\textbf {\bibinfo
  {volume} {91}},\ \bibinfo {pages} {171301} (\bibinfo {year} {2003})},\
  \Eprint {http://arxiv.org/abs/astro-ph/0306286} {arXiv:astro-ph/0306286
  [astro-ph]} \BibitemShut {NoStop}%
\bibitem [{\citenamefont {Beltran}\ \emph {et~al.}(2005)\citenamefont
  {Beltran}, \citenamefont {Garcia-Bellido}, \citenamefont {Lesgourgues},
  \citenamefont {Liddle},\ and\ \citenamefont {Slosar}}]{Beltran:2005xd}%
  \BibitemOpen
  \bibfield  {author} {\bibinfo {author} {\bibfnamefont {M.}~\bibnamefont
  {Beltran}}, \bibinfo {author} {\bibfnamefont {J.}~\bibnamefont
  {Garcia-Bellido}}, \bibinfo {author} {\bibfnamefont {J.}~\bibnamefont
  {Lesgourgues}}, \bibinfo {author} {\bibfnamefont {A.~R.}\ \bibnamefont
  {Liddle}}, \ and\ \bibinfo {author} {\bibfnamefont {A.}~\bibnamefont
  {Slosar}},\ }\href {\doibase 10.1103/PhysRevD.71.063532} {\bibfield
  {journal} {\bibinfo  {journal} {Phys. Rev.}\ }\textbf {\bibinfo {volume}
  {D71}},\ \bibinfo {pages} {063532} (\bibinfo {year} {2005})},\ \Eprint
  {http://arxiv.org/abs/astro-ph/0501477} {arXiv:astro-ph/0501477 [astro-ph]}
  \BibitemShut {NoStop}%
\bibitem [{\citenamefont {Beltran}\ \emph {et~al.}(2007)\citenamefont
  {Beltran}, \citenamefont {Garcia-Bellido},\ and\ \citenamefont
  {Lesgourgues}}]{Beltran:2006sq}%
  \BibitemOpen
  \bibfield  {author} {\bibinfo {author} {\bibfnamefont {M.}~\bibnamefont
  {Beltran}}, \bibinfo {author} {\bibfnamefont {J.}~\bibnamefont
  {Garcia-Bellido}}, \ and\ \bibinfo {author} {\bibfnamefont {J.}~\bibnamefont
  {Lesgourgues}},\ }\href {\doibase 10.1103/PhysRevD.75.103507} {\bibfield
  {journal} {\bibinfo  {journal} {Phys. Rev.}\ }\textbf {\bibinfo {volume}
  {D75}},\ \bibinfo {pages} {103507} (\bibinfo {year} {2007})},\ \Eprint
  {http://arxiv.org/abs/hep-ph/0606107} {arXiv:hep-ph/0606107 [hep-ph]}
  \BibitemShut {NoStop}%
\bibitem [{\citenamefont {Ade}\ \emph {et~al.}(2014{\natexlab{a}})\citenamefont
  {Ade} \emph {et~al.}}]{Ade:2013zuv}%
  \BibitemOpen
  \bibfield  {author} {\bibinfo {author} {\bibfnamefont {P.~A.~R.}\
  \bibnamefont {Ade}} \emph {et~al.} (\bibinfo {collaboration} {Planck}),\
  }\href {\doibase 10.1051/0004-6361/201321591} {\bibfield  {journal} {\bibinfo
   {journal} {Astron. Astrophys.}\ }\textbf {\bibinfo {volume} {571}},\
  \bibinfo {pages} {A16} (\bibinfo {year} {2014}{\natexlab{a}})},\ \Eprint
  {http://arxiv.org/abs/1303.5076} {arXiv:1303.5076 [astro-ph.CO]} \BibitemShut
  {NoStop}%
\bibitem [{\citenamefont {Ade}\ \emph {et~al.}(2014{\natexlab{b}})\citenamefont
  {Ade} \emph {et~al.}}]{Planck:2013jfk}%
  \BibitemOpen
  \bibfield  {author} {\bibinfo {author} {\bibfnamefont {P.~A.~R.}\
  \bibnamefont {Ade}} \emph {et~al.} (\bibinfo {collaboration} {Planck}),\
  }\href {\doibase 10.1051/0004-6361/201321569} {\bibfield  {journal} {\bibinfo
   {journal} {Astron. Astrophys.}\ }\textbf {\bibinfo {volume} {571}},\
  \bibinfo {pages} {A22} (\bibinfo {year} {2014}{\natexlab{b}})},\ \Eprint
  {http://arxiv.org/abs/1303.5082} {arXiv:1303.5082 [astro-ph.CO]} \BibitemShut
  {NoStop}%
\bibitem [{\citenamefont {Barkats}\ \emph {et~al.}(2014)\citenamefont {Barkats}
  \emph {et~al.}}]{Barkats:2013jfa}%
  \BibitemOpen
  \bibfield  {author} {\bibinfo {author} {\bibfnamefont {D.}~\bibnamefont
  {Barkats}} \emph {et~al.} (\bibinfo {collaboration} {BICEP1}),\ }\href
  {\doibase 10.1088/0004-637X/783/2/67} {\bibfield  {journal} {\bibinfo
  {journal} {Astrophys. J.}\ }\textbf {\bibinfo {volume} {783}},\ \bibinfo
  {pages} {67} (\bibinfo {year} {2014})},\ \Eprint
  {http://arxiv.org/abs/1310.1422} {arXiv:1310.1422 [astro-ph.CO]} \BibitemShut
  {NoStop}%
\bibitem [{\citenamefont {Ade}\ \emph {et~al.}(2015)\citenamefont {Ade} \emph
  {et~al.}}]{Ade:2015tva}%
  \BibitemOpen
  \bibfield  {author} {\bibinfo {author} {\bibfnamefont {P.~A.~R.}\
  \bibnamefont {Ade}} \emph {et~al.} (\bibinfo {collaboration} {BICEP2,
  Planck}),\ }\href {\doibase 10.1103/PhysRevLett.114.101301} {\bibfield
  {journal} {\bibinfo  {journal} {Phys. Rev. Lett.}\ }\textbf {\bibinfo
  {volume} {114}},\ \bibinfo {pages} {101301} (\bibinfo {year} {2015})},\
  \Eprint {http://arxiv.org/abs/1502.00612} {arXiv:1502.00612 [astro-ph.CO]}
  \BibitemShut {NoStop}%
\bibitem [{\citenamefont {Ade}\ \emph {et~al.}(2016{\natexlab{b}})\citenamefont
  {Ade} \emph {et~al.}}]{Ade:2015lrj}%
  \BibitemOpen
  \bibfield  {author} {\bibinfo {author} {\bibfnamefont {P.~A.~R.}\
  \bibnamefont {Ade}} \emph {et~al.} (\bibinfo {collaboration} {Planck}),\
  }\href {\doibase 10.1051/0004-6361/201525898} {\bibfield  {journal} {\bibinfo
   {journal} {Astron. Astrophys.}\ }\textbf {\bibinfo {volume} {594}},\
  \bibinfo {pages} {A20} (\bibinfo {year} {2016}{\natexlab{b}})},\ \Eprint
  {http://arxiv.org/abs/1502.02114} {arXiv:1502.02114 [astro-ph.CO]}
  \BibitemShut {NoStop}%
\bibitem [{\citenamefont {Hlo{\v z}ek}\ \emph {et~al.}(2017)\citenamefont
  {Hlo{\v z}ek}, \citenamefont {Marsh}, \citenamefont {Grin}, \citenamefont
  {Allison}, \citenamefont {Dunkley},\ and\ \citenamefont
  {Calabrese}}]{Hlozek:2016lzm}%
  \BibitemOpen
  \bibfield  {author} {\bibinfo {author} {\bibfnamefont {R.}~\bibnamefont
  {Hlo{\v z}ek}}, \bibinfo {author} {\bibfnamefont {D.~J.~E.}\ \bibnamefont
  {Marsh}}, \bibinfo {author} {\bibfnamefont {D.}~\bibnamefont {Grin}},
  \bibinfo {author} {\bibfnamefont {R.}~\bibnamefont {Allison}}, \bibinfo
  {author} {\bibfnamefont {J.}~\bibnamefont {Dunkley}}, \ and\ \bibinfo
  {author} {\bibfnamefont {E.}~\bibnamefont {Calabrese}},\ }\href {\doibase
  10.1103/PhysRevD.95.123511} {\bibfield  {journal} {\bibinfo  {journal} {Phys.
  Rev.}\ }\textbf {\bibinfo {volume} {D95}},\ \bibinfo {pages} {123511}
  (\bibinfo {year} {2017})},\ \Eprint {http://arxiv.org/abs/1607.08208}
  {arXiv:1607.08208 [astro-ph.CO]} \BibitemShut {NoStop}%
\bibitem [{\citenamefont {Polchinski}(1998)}]{polchinski_1998}%
  \BibitemOpen
  \bibfield  {author} {\bibinfo {author} {\bibfnamefont {J.}~\bibnamefont
  {Polchinski}},\ }\href {\doibase 10.1017/CBO9780511816079} {\emph {\bibinfo
  {title} {String Theory}}},\ \bibinfo {series} {Cambridge Monographs on
  Mathematical Physics}, Vol.~\bibinfo {volume} {1}\ (\bibinfo  {publisher}
  {Cambridge University Press},\ \bibinfo {year} {1998})\BibitemShut {NoStop}%
\bibitem [{\citenamefont {Conlon}(2006)}]{Conlon:2006tq}%
  \BibitemOpen
  \bibfield  {author} {\bibinfo {author} {\bibfnamefont {J.~P.}\ \bibnamefont
  {Conlon}},\ }\href {\doibase 10.1088/1126-6708/2006/05/078} {\bibfield
  {journal} {\bibinfo  {journal} {JHEP}\ }\textbf {\bibinfo {volume} {05}},\
  \bibinfo {pages} {078} (\bibinfo {year} {2006})},\ \Eprint
  {http://arxiv.org/abs/hep-th/0602233} {arXiv:hep-th/0602233 [hep-th]}
  \BibitemShut {NoStop}%
\bibitem [{\citenamefont {Svrcek}(2006)}]{Svrcek:2006hf}%
  \BibitemOpen
  \bibfield  {author} {\bibinfo {author} {\bibfnamefont {P.}~\bibnamefont
  {Svrcek}},\ }\href@noop {} {\bibfield  {journal} {\bibinfo  {journal}
  {Submitted to: JHEP}\ } (\bibinfo {year} {2006})},\ \Eprint
  {http://arxiv.org/abs/hep-th/0607086} {arXiv:hep-th/0607086 [hep-th]}
  \BibitemShut {NoStop}%
\bibitem [{\citenamefont {Beasley}\ and\ \citenamefont
  {Witten}(2005)}]{Beasley:2004ys}%
  \BibitemOpen
  \bibfield  {author} {\bibinfo {author} {\bibfnamefont {C.}~\bibnamefont
  {Beasley}}\ and\ \bibinfo {author} {\bibfnamefont {E.}~\bibnamefont
  {Witten}},\ }\href {\doibase 10.1088/1126-6708/2005/01/056} {\bibfield
  {journal} {\bibinfo  {journal} {JHEP}\ }\textbf {\bibinfo {volume} {01}},\
  \bibinfo {pages} {056} (\bibinfo {year} {2005})},\ \Eprint
  {http://arxiv.org/abs/hep-th/0409149} {arXiv:hep-th/0409149 [hep-th]}
  \BibitemShut {NoStop}%
\bibitem [{\citenamefont {Easther}\ and\ \citenamefont
  {McAllister}(2006)}]{Easther:2005zr}%
  \BibitemOpen
  \bibfield  {author} {\bibinfo {author} {\bibfnamefont {R.}~\bibnamefont
  {Easther}}\ and\ \bibinfo {author} {\bibfnamefont {L.}~\bibnamefont
  {McAllister}},\ }\href {\doibase 10.1088/1475-7516/2006/05/018} {\bibfield
  {journal} {\bibinfo  {journal} {JCAP}\ }\textbf {\bibinfo {volume} {0605}},\
  \bibinfo {pages} {018} (\bibinfo {year} {2006})},\ \Eprint
  {http://arxiv.org/abs/hep-th/0512102} {arXiv:hep-th/0512102 [hep-th]}
  \BibitemShut {NoStop}%
\bibitem [{\citenamefont {Audren}\ \emph {et~al.}(2013)\citenamefont {Audren},
  \citenamefont {Lesgourgues}, \citenamefont {Benabed},\ and\ \citenamefont
  {Prunet}}]{Audren:2013}%
  \BibitemOpen
  \bibfield  {author} {\bibinfo {author} {\bibfnamefont {B.}~\bibnamefont
  {Audren}}, \bibinfo {author} {\bibfnamefont {J.}~\bibnamefont {Lesgourgues}},
  \bibinfo {author} {\bibfnamefont {K.}~\bibnamefont {Benabed}}, \ and\
  \bibinfo {author} {\bibfnamefont {S.}~\bibnamefont {Prunet}},\ }\href
  {http://stacks.iop.org/1475-7516/2013/i=02/a=001} {\bibfield  {journal}
  {\bibinfo  {journal} {Journal of Cosmology and Astroparticle Physics}\
  }\textbf {\bibinfo {volume} {2013}},\ \bibinfo {pages} {001} (\bibinfo {year}
  {2013})}\BibitemShut {NoStop}%
\bibitem [{\citenamefont {Vagnozzi}\ \emph {et~al.}(2017)\citenamefont
  {Vagnozzi}, \citenamefont {Giusarma}, \citenamefont {Mena}, \citenamefont
  {Freese}, \citenamefont {Gerbino}, \citenamefont {Ho},\ and\ \citenamefont
  {Lattanzi}}]{Vagnozzi:2017ovm}%
  \BibitemOpen
  \bibfield  {author} {\bibinfo {author} {\bibfnamefont {S.}~\bibnamefont
  {Vagnozzi}}, \bibinfo {author} {\bibfnamefont {E.}~\bibnamefont {Giusarma}},
  \bibinfo {author} {\bibfnamefont {O.}~\bibnamefont {Mena}}, \bibinfo {author}
  {\bibfnamefont {K.}~\bibnamefont {Freese}}, \bibinfo {author} {\bibfnamefont
  {M.}~\bibnamefont {Gerbino}}, \bibinfo {author} {\bibfnamefont
  {S.}~\bibnamefont {Ho}}, \ and\ \bibinfo {author} {\bibfnamefont
  {M.}~\bibnamefont {Lattanzi}},\ }\href {\doibase 10.1103/PhysRevD.96.123503}
  {\bibfield  {journal} {\bibinfo  {journal} {Phys. Rev.}\ }\textbf {\bibinfo
  {volume} {D96}},\ \bibinfo {pages} {123503} (\bibinfo {year} {2017})},\
  \Eprint {http://arxiv.org/abs/1701.08172} {arXiv:1701.08172 [astro-ph.CO]}
  \BibitemShut {NoStop}%
\bibitem [{\citenamefont {Klaer}\ and\ \citenamefont
  {Moore}(2017)}]{Klaer:2017ond}%
  \BibitemOpen
  \bibfield  {author} {\bibinfo {author} {\bibfnamefont {V.~B.}\ \bibnamefont
  {Klaer}}\ and\ \bibinfo {author} {\bibfnamefont {G.~D.}\ \bibnamefont
  {Moore}},\ }\href {\doibase 10.1088/1475-7516/2017/11/049} {\bibfield
  {journal} {\bibinfo  {journal} {JCAP}\ }\textbf {\bibinfo {volume} {1711}},\
  \bibinfo {pages} {049} (\bibinfo {year} {2017})},\ \Eprint
  {http://arxiv.org/abs/1708.07521} {arXiv:1708.07521 [hep-ph]} \BibitemShut
  {NoStop}%
\bibitem [{\citenamefont {Schive}\ \emph {et~al.}(2014)\citenamefont {Schive},
  \citenamefont {Chiueh},\ and\ \citenamefont {Broadhurst}}]{Schive:2014dra}%
  \BibitemOpen
  \bibfield  {author} {\bibinfo {author} {\bibfnamefont {H.-Y.}\ \bibnamefont
  {Schive}}, \bibinfo {author} {\bibfnamefont {T.}~\bibnamefont {Chiueh}}, \
  and\ \bibinfo {author} {\bibfnamefont {T.}~\bibnamefont {Broadhurst}},\
  }\href {\doibase 10.1038/nphys2996} {\bibfield  {journal} {\bibinfo
  {journal} {Nature Phys.}\ }\textbf {\bibinfo {volume} {10}},\ \bibinfo
  {pages} {496} (\bibinfo {year} {2014})},\ \Eprint
  {http://arxiv.org/abs/1406.6586} {arXiv:1406.6586 [astro-ph.GA]} \BibitemShut
  {NoStop}%
\bibitem [{\citenamefont {Schive}\ \emph {et~al.}(2016)\citenamefont {Schive},
  \citenamefont {Chiueh}, \citenamefont {Broadhurst},\ and\ \citenamefont
  {Huang}}]{Schive:2015kza}%
  \BibitemOpen
  \bibfield  {author} {\bibinfo {author} {\bibfnamefont {H.-Y.}\ \bibnamefont
  {Schive}}, \bibinfo {author} {\bibfnamefont {T.}~\bibnamefont {Chiueh}},
  \bibinfo {author} {\bibfnamefont {T.}~\bibnamefont {Broadhurst}}, \ and\
  \bibinfo {author} {\bibfnamefont {K.-W.}\ \bibnamefont {Huang}},\ }\href
  {\doibase 10.3847/0004-637X/818/1/89} {\bibfield  {journal} {\bibinfo
  {journal} {Astrophys. J.}\ }\textbf {\bibinfo {volume} {818}},\ \bibinfo
  {pages} {89} (\bibinfo {year} {2016})},\ \Eprint
  {http://arxiv.org/abs/1508.04621} {arXiv:1508.04621 [astro-ph.GA]}
  \BibitemShut {NoStop}%
\bibitem [{\citenamefont {Chen}\ \emph {et~al.}(2017)\citenamefont {Chen},
  \citenamefont {Schive},\ and\ \citenamefont {Chiueh}}]{Chen:2016unw}%
  \BibitemOpen
  \bibfield  {author} {\bibinfo {author} {\bibfnamefont {S.-R.}\ \bibnamefont
  {Chen}}, \bibinfo {author} {\bibfnamefont {H.-Y.}\ \bibnamefont {Schive}}, \
  and\ \bibinfo {author} {\bibfnamefont {T.}~\bibnamefont {Chiueh}},\ }\href
  {\doibase 10.1093/mnras/stx449} {\bibfield  {journal} {\bibinfo  {journal}
  {Mon. Not. Roy. Astron. Soc.}\ }\textbf {\bibinfo {volume} {468}},\ \bibinfo
  {pages} {1338} (\bibinfo {year} {2017})},\ \Eprint
  {http://arxiv.org/abs/1606.09030} {arXiv:1606.09030 [astro-ph.GA]}
  \BibitemShut {NoStop}%
\bibitem [{\citenamefont {Schive}\ and\ \citenamefont
  {Chiueh}(2018)}]{Schive:2017biq}%
  \BibitemOpen
  \bibfield  {author} {\bibinfo {author} {\bibfnamefont {H.-Y.}\ \bibnamefont
  {Schive}}\ and\ \bibinfo {author} {\bibfnamefont {T.}~\bibnamefont
  {Chiueh}},\ }\href {\doibase 10.1093/mnrasl/slx159} {\bibfield  {journal}
  {\bibinfo  {journal} {Mon. Not. Roy. Astron. Soc.}\ }\textbf {\bibinfo
  {volume} {473}},\ \bibinfo {pages} {L36} (\bibinfo {year} {2018})},\ \Eprint
  {http://arxiv.org/abs/1706.03723} {arXiv:1706.03723 [astro-ph.CO]}
  \BibitemShut {NoStop}%
\bibitem [{\citenamefont {Athron}\ \emph {et~al.}(2017)\citenamefont {Athron}
  \emph {et~al.}}]{Athron:2017ard}%
  \BibitemOpen
  \bibfield  {author} {\bibinfo {author} {\bibfnamefont {P.}~\bibnamefont
  {Athron}} \emph {et~al.} (\bibinfo {collaboration} {GAMBIT}),\ }\href
  {\doibase 10.1140/epjc/s10052-017-5513-2, 10.1140/epjc/s10052-017-5321-8}
  {\bibfield  {journal} {\bibinfo  {journal} {Eur. Phys. J.}\ }\textbf
  {\bibinfo {volume} {C77}},\ \bibinfo {pages} {784} (\bibinfo {year}
  {2017})},\ \bibinfo {note} {[Addendum: Eur. Phys. J.C78,no.2,98(2018)]},\
  \Eprint {http://arxiv.org/abs/1705.07908} {arXiv:1705.07908 [hep-ph]}
  \BibitemShut {NoStop}%
\bibitem [{\citenamefont {Bringmann}\ \emph {et~al.}(2017)\citenamefont
  {Bringmann} \emph {et~al.}}]{Workgroup:2017lvb}%
  \BibitemOpen
  \bibfield  {author} {\bibinfo {author} {\bibfnamefont {T.}~\bibnamefont
  {Bringmann}} \emph {et~al.} (\bibinfo {collaboration} {The GAMBIT Dark Matter
  Workgroup}),\ }\href {\doibase 10.1140/epjc/s10052-017-5155-4} {\bibfield
  {journal} {\bibinfo  {journal} {Eur. Phys. J.}\ }\textbf {\bibinfo {volume}
  {C77}},\ \bibinfo {pages} {831} (\bibinfo {year} {2017})},\ \Eprint
  {http://arxiv.org/abs/1705.07920} {arXiv:1705.07920 [hep-ph]} \BibitemShut
  {NoStop}%
\bibitem [{\citenamefont {Amendola}\ and\ \citenamefont
  {Barbieri}(2006)}]{Amendola:2005ad}%
  \BibitemOpen
  \bibfield  {author} {\bibinfo {author} {\bibfnamefont {L.}~\bibnamefont
  {Amendola}}\ and\ \bibinfo {author} {\bibfnamefont {R.}~\bibnamefont
  {Barbieri}},\ }\href {\doibase 10.1016/j.physletb.2006.08.069} {\bibfield
  {journal} {\bibinfo  {journal} {Phys. Lett.}\ }\textbf {\bibinfo {volume}
  {B642}},\ \bibinfo {pages} {192} (\bibinfo {year} {2006})},\ \Eprint
  {http://arxiv.org/abs/hep-ph/0509257} {arXiv:hep-ph/0509257 [hep-ph]}
  \BibitemShut {NoStop}%
\end{thebibliography}
\end{document}